\documentclass[10pt,a4paper]{article}
\usepackage[dvipsnames]{xcolor}
\usepackage[T1]{fontenc}
\usepackage{graphicx}
\usepackage{amsmath}
\usepackage{xspace}
\usepackage[numbers,sort&compress]{natbib}
\usepackage{caption}
\usepackage
[subrefformat=parens,position=top,skip=-15pt,margin=15pt,justification=justified,singlelinecheck=false]
{subcaption}
\usepackage[shortlabels]{enumitem}
\usepackage{authblk}

\usepackage{pstricks}
\usepackage{graphicx}
\usepackage{xspace}
\usepackage[pdfborder={0 0 0}]{hyperref}
\hypersetup{
  colorlinks   = true, 
  urlcolor     = blue, 
  linkcolor    = blue, 
  citecolor    = blue, 
  linktocpage  = true,
  pdftitle     = {Precise predictions for V+2 jet backgrounds in searches for invisible Higgs decays}
}
\usepackage{multirow}
\usepackage[left = 2.5cm, right = 2.5cm, top = 2.5cm, bottom = 2.5cm]{geometry}

\newcommand{\vecy}{\mathbf{y}}
\newcommand{\Mt}{m_{\Pt}}

\newcommand{\Mb}{m_{\Pb}}

\newcommand{\LO}{\text{LO}\xspace}
\newcommand{\NLO}{\ensuremath{\text{NLO}}\xspace}

\newcommand{\QCD}{\ensuremath{\text{QCD}}\xspace}
\newcommand{\interf}{\ensuremath{\text{interf}}\xspace}

\newcommand{\EW}{\ensuremath{\text{EW}}\xspace}

\newcommand{\QCDpEW}{\text{\QCD{}+\EW}\xspace}

\newcommand{\QCDtEW}{\QCD{}\ensuremath{\times}\EW\xspace}

\newcommand{\Z}{\ensuremath{Z}\xspace}
\newcommand{\W}{\ensuremath{W}\xspace}
\newcommand{\ZW}{\ensuremath{Z/W}\xspace}

\newcommand{\Pt}{\ensuremath{\mathrm{t}}\xspace}
\newcommand{\Pb}{\ensuremath{\mathrm{b}}\xspace}

\newcommand{\PW}{\ensuremath{\mathrm{W}}\xspace}

\newcommand{\PZ}{\ensuremath{\mathrm{Z}}\xspace}
\newcommand{\PH}{\ensuremath{\mathrm{H}}\xspace}

\newcommand{\jet}{\mathrm{j}}
\newcommand{\Vjj}{\ensuremath{V+2}\,\text{jet}\xspace}
\newcommand{\ppVjj}{\ensuremath{pp\to V+2}\,jet\xspace}

\newcommand{\Zjj}{\ensuremath{Z+2}\,\text{jet}\xspace}

\newcommand{\Wjj}{\ensuremath{W+2}\,\text{jet}\xspace}

\newcommand{\pscuts}[1]{\theta^{#1}_{\mathrm{cuts}}(\vecy)}
\def\mode{M}

\newcommand{\Znn}{\ensuremath{Z(\nu\bar\nu) }\xspace}

\newcommand{\Wen}{\ensuremath{W^{\pm}(\ell^{\pm}\nu})\xspace}

\newcommand{\Hinv}{\ensuremath{H\to {\rm inv}}\xspace}

\newcommand{\Znnjj}{\ensuremath{Z(\nu\bar\nu)+2}\,\text{jets}\xspace}
\newcommand{\Zlljj}{\ensuremath{Z(\ell^+\ell^-)+2}\,\text{jets}\xspace}
\newcommand{\Wenjj}{\ensuremath{W^{\pm}(\ell^{\pm}\nu)+2}\,\text{jets}\xspace}
\newcommand{\ppWenjj}{\ensuremath{pp\to W^{\pm}(\ell^{\pm}\nu)+2}\,\text{jets}\xspace}
\newcommand{\ppZnnjj}{\ensuremath{pp\to Z(\nu\bar\nu)+2}\,\text{jets}\,\xspace}
\newcommand{\ppZlljj}{\ensuremath{pp\to Z(\ell^+\ell^-)+2}\,\text{jets}\,\xspace}

\newcommand{\ri}{\mathrm{i}}
\newcommand{\rF}{\mathrm{F}}
\newcommand{\rR}{\mathrm{R}}

\newcommand{\rT}{\mathrm{T}}
\newcommand{\rd}{\mathrm{d}}
\newcommand{\rs}{\mathrm{s}}

\newcommand{\mur}{\mu_{\rR}}
\newcommand{\muf}{\mu_{\rF}}

\newcommand{\MeV}{\text{MeV}\xspace}
\newcommand{\GeV}{\text{GeV}\xspace}

\newcommand{\alphaS}{\alpha_{\rs}}
\newcommand{\ord}{\mathcal{O}}

\newcommand{\pTjone}{\ensuremath{p_{\mathrm{T},\jet_1}}}

\newcommand{\pTV}{\ensuremath{p_{\mathrm{T,V}}}\xspace}

\newcommand{\deltaphijj}{\ensuremath{\Delta\phi_{\jet_1\jet_2}}}

\newcommand{\deltaetajj}{\ensuremath{\Delta\eta_{\jet_1\jet_2}}}

\newcommand{\mjj}{\ensuremath{m_{\mathrm{\jet_1\jet_2}}}\xspace}

\newcommand{\half}{\frac{1}{2}}

\newcommand{\beqar}{\begin{eqnarray}}
\newcommand{\eeqar}{\end{eqnarray}}
\newcommand{\beq}{\begin{equation}}
\newcommand{\eeq}{\end{equation}}
\newcommand{\bit}{\begin{itemize}}
\newcommand{\eit}{\end{itemize}}

\def\refeq#1{\mbox{(\ref{#1})}}

\def\refta#1{\mbox{Table~\ref{#1}}}

\def\refse#1{\mbox{Section~\ref{#1}}}

\def\ie{i.e.\ }

\def\max{\mathrm{max}}

\def\thetaw{\theta_{\mathrm{w}}}

\def\parx{\frac{\rm d}{{\rm d}x}}
\def\pary{\frac{\rm d}{{\rm d}\vecy}}

\def\eps{\varepsilon}
\def\MC{\mathrm{MC}}
\def\TH{\mathrm{TH}}
\def\vepsTH{\vec\eps_{\TH}}

\newcommand{\GF}{{G_\mu}}
\newcommand{\shortequal}{\ensuremath{\!\!\!=\!\!\!}}
\newcommand{\MW}{M_\PW}
\newcommand{\MZ}{M_\PZ}
\newcommand{\MH}{M_\PH}
\newcommand{\GW}{\Gamma_{\PW}}
\newcommand{\GZ}{\Gamma_{\PZ}}
\newcommand{\GH}{\Gamma_{\PH}}
\newcommand{\Gb}{\Gamma_{\Pb}}
\newcommand{\Gt}{\Gamma_{\Pt}}

\usepackage[left,mathlines]{lineno} 
\usepackage{soul} 

\def\bnote{\begin{notes}}
\def\enote{\end{notes}}
\usepackage{tcolorbox}
\newcounter{notescounter}
\newenvironment{notes}{\stepcounter{notescounter}
\begin{center} 
\begin{minipage}[t]{\textwidth} 
\begin{tcolorbox}[colback=Apricot,colframe=Apricot]
[\thenotescounter]\;}{ 
\end{tcolorbox} 
\end{minipage}  
\end{center}}
\newcommand{\plotsep}{\hspace*{0.05\textwidth}}
\newcommand{\diagwidth}{0.24\textwidth}

\title{Precise predictions for V+2 jet backgrounds\\ in searches for invisible Higgs decays}

\author[1]{J. M.~Lindert}
\author[2]{S. Pozzorini}
\author[3]{M. Sch\"onherr}

\affil[1]{\footnotesize
Department of Physics and Astronomy, University of Sussex, Brighton BN1 9QH, UK
\normalsize}
\affil[2]{\footnotesize
Physik-Institut, Universit\"at Z\"urich, CH-8057 Z\"urich, Switzerland \normalsize}
\affil[3]{\footnotesize Institute for Particle Physics Phenomenology, Department of Physics, Durham University, Durham, DH1 3LE, UK \normalsize}

\date{}

\begin{document}

\maketitle

\begin{abstract}\noindent
We present next-to-leading order QCD and electroweak (EW) theory predictions
for \Vjj production, with $V=Z,W^{\pm}$, considering both the QCD and EW
production modes and their interference.  We focus on phase-space regions
where \Vjj production is dominated by vector-boson fusion,
and where these processes yield the dominant irreducible backgrounds in
searches for invisible Higgs boson decays.  Predictions at parton level are
provided together with 
detailed prescriptions
for their implementation in
experimental analyses based on the reweighting of Monte Carlo samples. 
The key idea is that, exploiting accurate data 
for \Wjj production in combination with a theory-driven extrapolation 
to the \Zjj process can lead to a determination of the irreducible
background at the few-percent level.
Particular attention is devoted to the estimate of 
the residual theoretical uncertainties due to unknown higher-order 
QCD and EW effects
and their correlation
between the different \Vjj processes, which is key to improve the
sensitivity to invisible Higgs decays.

\end{abstract}

\tableofcontents

\section{Introduction}

Along the main objectives of current and future runs of the Large Hadron Collider (LHC) will be a further detailed investigation of the Higgs sector and the search for physics beyond the Standard Model (BSM).
In fact, these two objects are linked,
since very precise measurements of Higgs couplings and properties might reveal hints of BSM physics. A prime example of this is given by the branching ratio of the Higgs boson into   
invisible particles. In the Standard Model (SM), the only invisible decay mode of the Higgs boson proceeds via $H \to ZZ^* \to 4 \nu$, with a branching ratio of only about $10^{-3}$~\cite{Denner:2019fcr}. In various extensions of the SM this 
invisible branching ratio
can be strongly enhanced~\cite{Shrock:1982kd,Choudhury:1993hv,Dominici:2009pq}, in particular in scenarios where the Higgs boson can decay
into a pair of weakly interacting massive particles -- prime candidates of particle dark matter~\cite{Belanger:2001am,Djouadi:2011aa,Baek:2012se,Calibbi:2013poa,Beniwal:2015sdl,Butter:2015fqa} (for a recent review see Ref.~\cite{Argyropoulos:2021sav}). Therefore, experimental limits on invisible Higgs decays ($H\to {\rm inv}$) can be used to exclude regions of parameter space of these models. At the LHC any production mode where the Higgs boson is produced in association with visible SM particles can in principle be used in order to search for $H\to {\rm inv}$. Most stringent bounds have been obtained combining searches in Higgs production
via vector-boson fusion (VBF) and Higgs production in association with a vector boson (VH) performed by both ATLAS~\cite{ATLAS:2017nyv,ATLAS:2018bnv,ATLAS:2018nda,ATLAS:2019cid} and CMS~\cite{CMS:2014gab,CMS:2016dhk,CMS:2018yfx}. These searches yield as currently best limit on the invisible Higgs branching ratio ${\rm Br}(H \to \rm{inv}) <  0.19$ at 95\%
confidence level~\cite{CMS:2018yfx}. The sensitivity in these searches is dominated by the VBF channel, i.e. the signature of two forward jets with large invariant mass together with sizeable missing transverse
energy. 
This signature receives large contributions from irreducible SM backgrounds, originating in particular from $Z$-boson production and decay into neutrinos in association with two jets.
Significant sensitivity improvements in $H\to {\rm inv}$ searches can  be 
achieved by controlling these backgrounds at the percent level. 
This in turn becomes possible via a theory assisted data-driven strategy,
where precision measurements 
are combined with state-of-the-art theoretical predictions 
for $Z$+2\,jet and $W$+2\,jet distributions and for 
their ratios.
Using this approach for the $V+$jet backgrounds to 
monojet signals~\cite{Lindert:2017olm} made it possible to 
enhance the sensitivity of dark-matter searches at the LHC in a very significant
way~\cite{ATLAS:2021kxv,CMS:2021far}.

Besides controlling backgrounds in \Hinv searches, 
\Vjj
production is of
importance and relevance in its own right. It serves as a laboratory for QCD dynamics and can be used to derive stringent bounds on anomalous triple gauge boson couplings and corresponding dimension-6 effective field theory coefficients~\cite{ATLAS:2014sjq,CMS:2016myt,ATLAS:2017nei,ATLAS:2017luz,CMS:2017dmo}.  In regard of the former, VBF production of vector
bosons, which contributes to 
\Vjj production at large dijet invariant mass and/or rapidity separation, 
can 
provide important insights
for the understanding of 
the
QCD dynamics in vector boson scattering (VBS) processes.

In this paper we present new theory predictions for 
\Vjj production, with $V=Z,W^{\pm}$, including higher-order QCD and electroweak (EW) corrections together with detailed recommendations for their implementation for 
improving \Vjj
backgrounds in searches for invisible Higgs decays. To be precise, we consider \Vjj production at next-to-leading order (NLO) QCD and
EW.
At the leading order (LO), these processes receive
three perturbative contributions. The leading one in the strong coupling constant $\alphaS$ is customary denoted as QCD production mode, while the contribution with the lowest order in $\alphaS$ is denoted as EW production mode. The third LO contribution corresponds to the interference between the QCD and the EW modes. The EW mode receives contributions from VBF-type production as well as from diboson production with subsequent semi-leptonic decays, 
and in the case of \Wjj also 
from single-top production with leptonic decays. At the NLO-level four perturbative contributions emerge, of which only the highest and lowest order in $\alphaS$ can unambiguously be denoted as NLO QCD corrections to the QCD modes and NLO EW corrections to the EW
mode, respectively. The remaining two contributions formally receive both $\ord(\alphaS)$ and $\ord(\alpha)$ corrections and partly overlap. In this study we present predictions for all of these LO and NLO
contributions, considering 
$pp\to W^{\pm
}+2$\,jets and $pp\to Z+2$\,jets
 including off-shell leptonic decays and invisible decays in the case of $Z+2$ jet production. We critically investigate remaining higher-order uncertainties at the NLO level and their correlation between the different \Vjj processes. To this end we consider besides remaining QCD and EW uncertainties also uncertainties due
to missing mixed QCD-EW corrections and due to the matching to parton showers (PS). 
For the implementation of these theoretical predictions in the
framework of invisible Higgs searches 
we propose a procedure based on the 
reweighting of Monte Carlo samples, providing also 
detailed prescriptions for the 
estimate of theoretical uncertainties including
correlations between 
the \Zjj and \Wjj processes.

The NLO QCD corrections to the \Vjj QCD production modes are widely
available~\cite{Campbell:2002tg,FebresCordero:2006nvf,Campbell:2008hh}
(for
$pp \to V+ n$ jets with $n > 2$ see
e.g.~\cite{Berger:2009zg,Ellis:2009zw,Ellis:2009zyy,Berger:2009ep,Berger:2010zx,Bern:2013gka,Anger:2017nkq})
and even next-to-next-to leading order (NNLO) corrections are within
reach~\cite{Badger:2021nhg,Abreu:2021asb}.  
The NLO
QCD corrections to the QCD
modes are readily available within general purpose shower Monte Carlo  (SMC)
programs~\cite{Re:2012zi,Alwall:2014hca,Sherpa:2019gpd,Frederix:2015eii},
where they typically enter Monte Carlo samples when NLO predictions for
$V+0,1,2$ jets production are merged and combined with parton showers at
NLO~\cite{Hoeche:2012yf,Gehrmann:2012yg,Lonnblad:2012ix,Frederix:2012ps}.
Additionally, logarithmically enhanced corrections  beyond fixed-order NLO 
due to wide-angle QCD emissions are available~\cite{Andersen:2012gk,Andersen:2016vkp,Andersen:2020yax}.
NLO EW corrections to the QCD modes of \Vjj production are known at fixed-order~\cite{Denner:2014ina,Kallweit:2014xda,Kallweit:2015dum,Chiesa:2015mya} and have also been combined with a QCD+QED parton shower 
using an approximation where only subleading 
QED effects are
 neglected~\cite{Kallweit:2015dum}. 
The QCD corrections to the EW modes 
 are only known in the so-called \textit{VBF approximation}, 
where the VBF subprocess alone is considered, and 
the cross-talk between quark lines is neglected 
in the higher-order
 corrections~\cite{Oleari:2003tc,Jager:2010aj}. 
Within this approximation NLO QCD corrections to the EW modes have been
matched to parton showers~\cite{Jager:2012xk,Schissler:2013nga}.  
The NLO EW
corrections to the EW production modes are currently not known and 
presented here for the first time.

The paper is organised as following. In Section~\ref{sec:modes} we discuss the structure of the NLO corrections to \Vjj production considering both the QCD and EW production modes and their interference. In Section~\ref{se:reweighting} we propose a reweighting procedure for the incorporation of the higher-order corrections into Monte Carlo samples. 
Theoretical predictions and uncertainties are 
presented in Section~\ref{sec:results}, 
and our
conclusions can be found in Section~\ref{sec:conclusions}.

\section{\texorpdfstring{\Vjj}{V+2jet} QCD and EW production modes at NLO}
\label{sec:modes}
\begin{figure}[tb]
\centering
\includegraphics[width=0.5\textwidth]{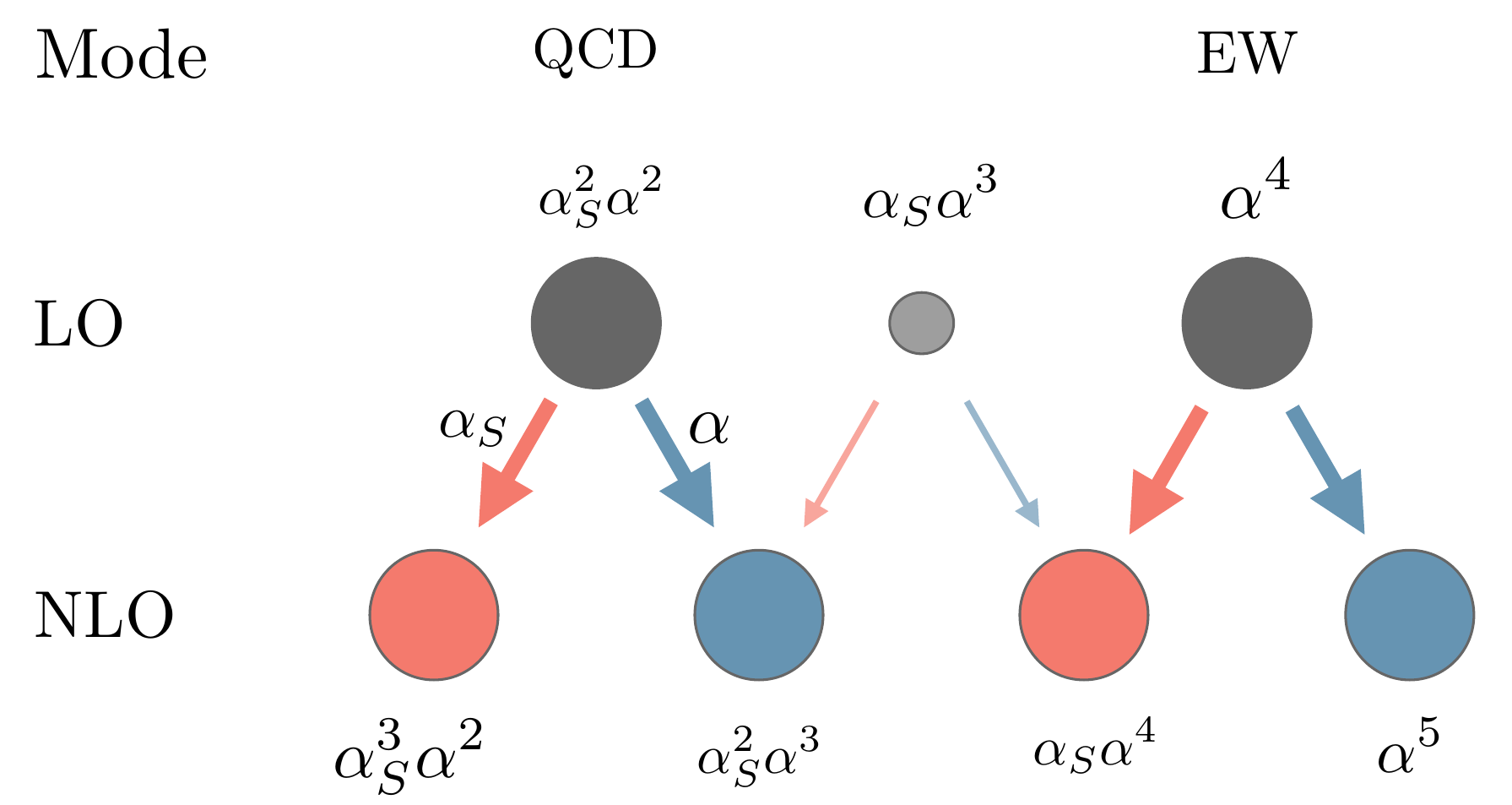}
  \caption{Tower of perturbative contributions to \Vjj production at LO and NLO considered and evaluated in this study. In the presented counting the $\ord(\alpha)$ 
vector-boson decays 
are included.}
  \label{fig:perturbativeorders}
\end{figure}

\begin{figure}[tb]
\centering
  \captionsetup[subfigure]{justification=centering}
  \begin{subfigure}{\diagwidth}
  \centering
    \includegraphics[width=\textwidth]{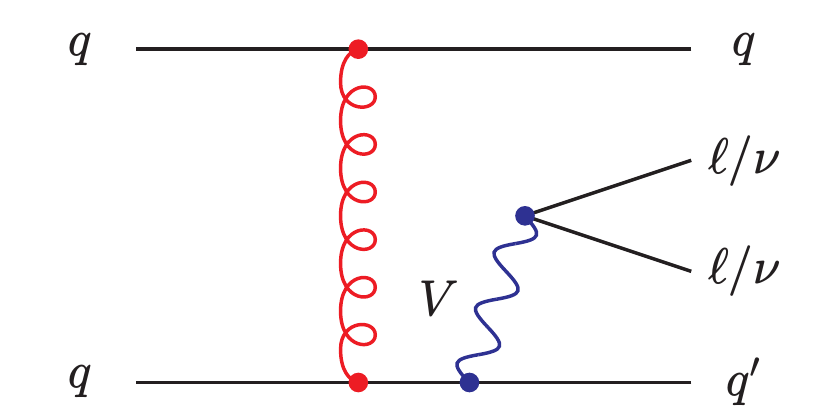}
    \caption{}
    \label{fig:LOdiags:O22A}
  \end{subfigure}
  \begin{subfigure}{\diagwidth}
    \includegraphics[width=\textwidth]{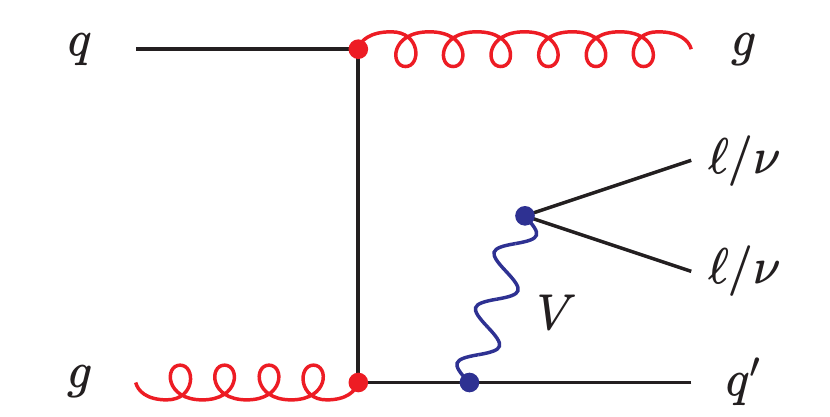}
    \caption{}
    \label{fig:LOdiags:O22B}
  \end{subfigure}
  \begin{subfigure}{\diagwidth}
    \includegraphics[width=\textwidth]{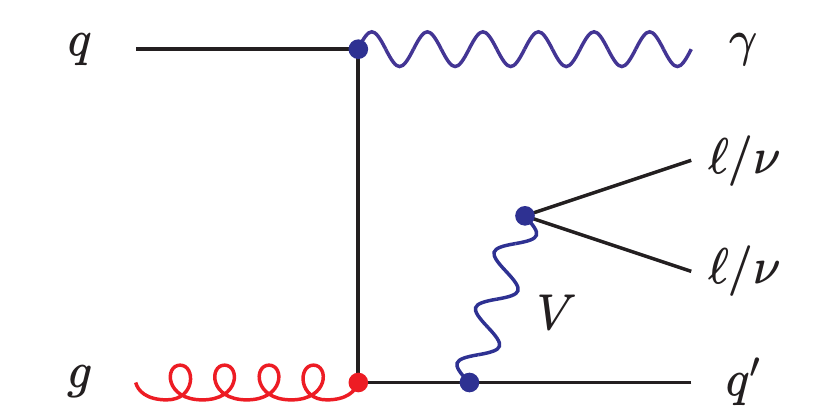}
    \caption{}
    \label{fig:LOdiags:O13A}
  \end{subfigure}
  \begin{subfigure}{\diagwidth}
    \includegraphics[width=\textwidth]{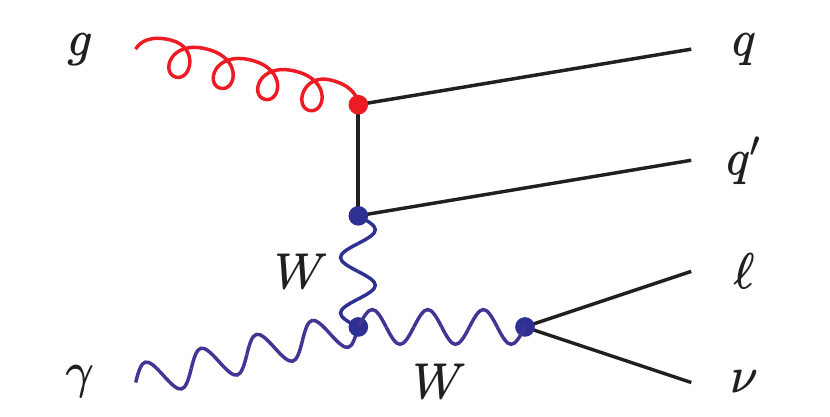}
    \caption{}
    \label{fig:LOdiags:O13B}
  \end{subfigure}\\[6mm]
  \begin{subfigure}{\diagwidth}
  \centering
    \includegraphics[width=\textwidth]{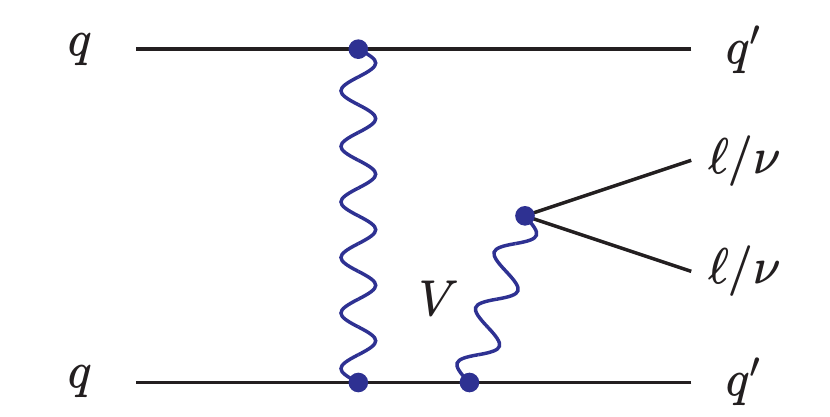}
    \caption{}
    \label{fig:LOdiags:O04A}
  \end{subfigure}
  \begin{subfigure}{\diagwidth}
    \includegraphics[width=\textwidth]{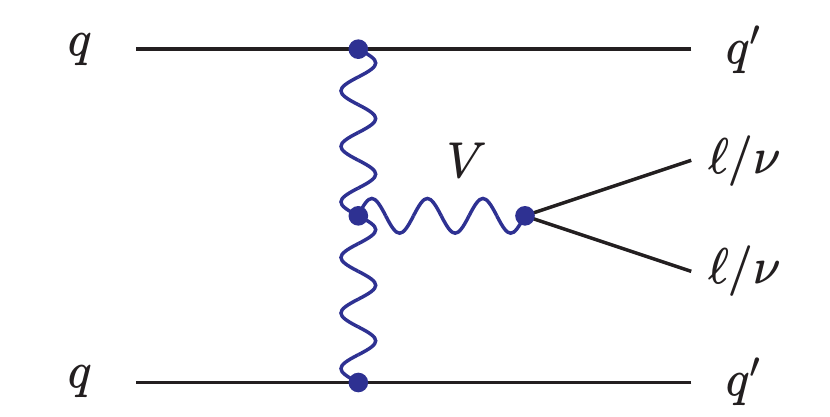}
    \caption{}
    \label{fig:LOdiags:O04VBFA}
  \end{subfigure}
  \begin{subfigure}{\diagwidth}
    \includegraphics[width=\textwidth]{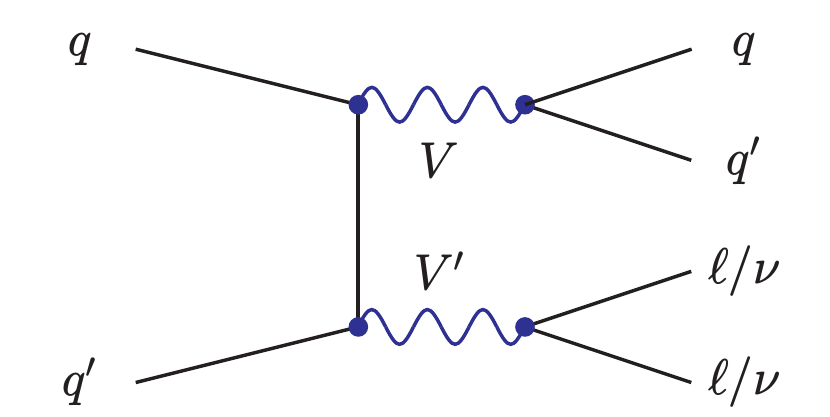}
    \caption{}
    \label{fig:LOdiags:O04VVA}
  \end{subfigure}
  \begin{subfigure}{\diagwidth}
    \includegraphics[width=\textwidth]{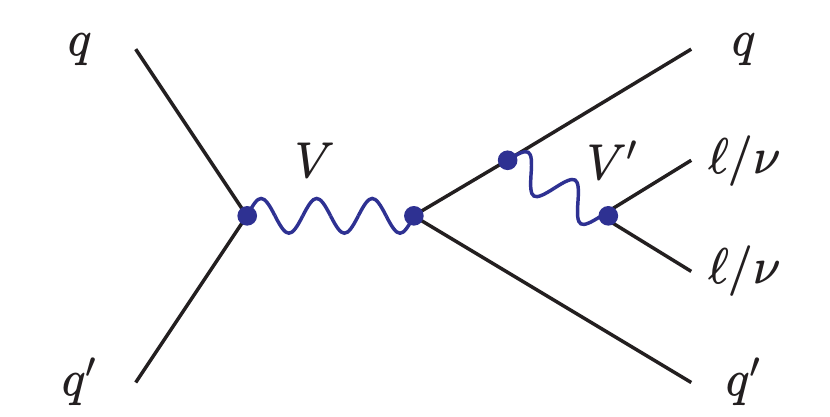}
    \caption{}
    \label{fig:LOdiags:O04VA}
  \end{subfigure}\\[6mm]
  \begin{subfigure}{\diagwidth}
  \centering
    \includegraphics[width=\textwidth]{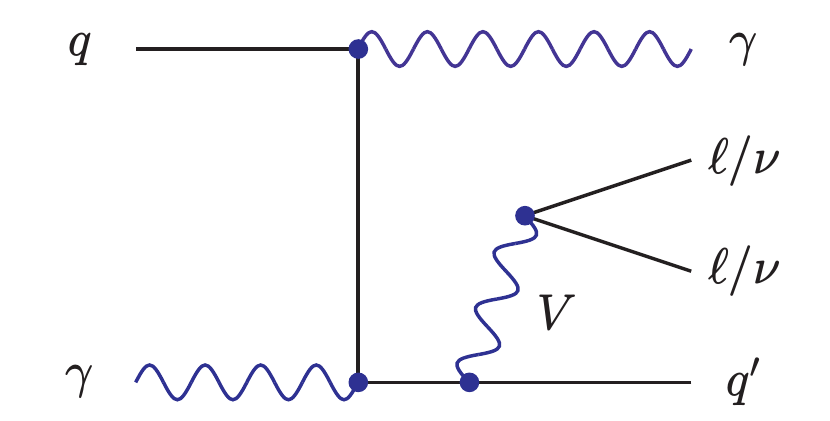}
    \caption{}
    \label{fig:LOdiags:O04B}
  \end{subfigure}
  \begin{subfigure}{\diagwidth}
  \centering
    \includegraphics[width=\textwidth]{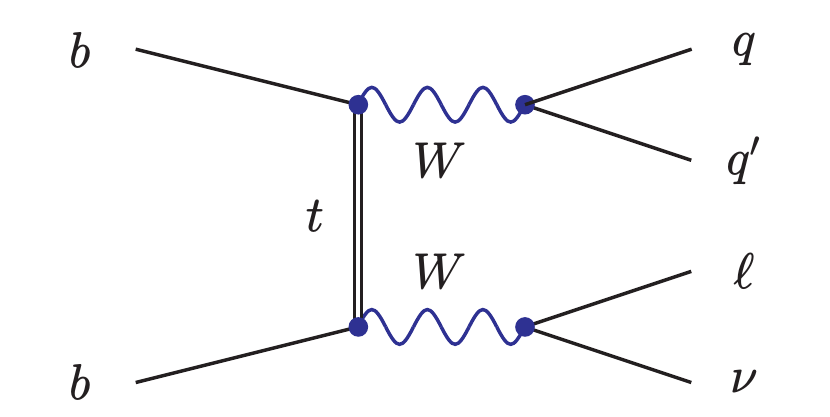}
    \caption{}
    \label{fig:LOdiags:O04tC}
  \end{subfigure}
  \begin{subfigure}{\diagwidth}
    \includegraphics[width=\textwidth]{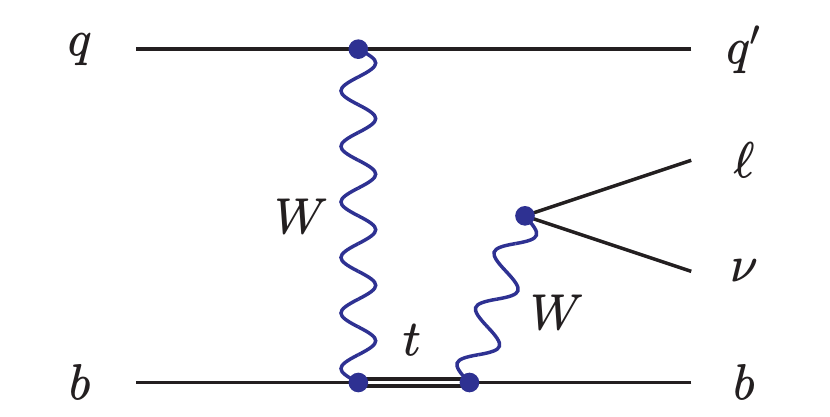}
    \caption{}
    \label{fig:LOdiags:O04tB}
  \end{subfigure}
  \begin{subfigure}{\diagwidth}
    \includegraphics[width=\textwidth]{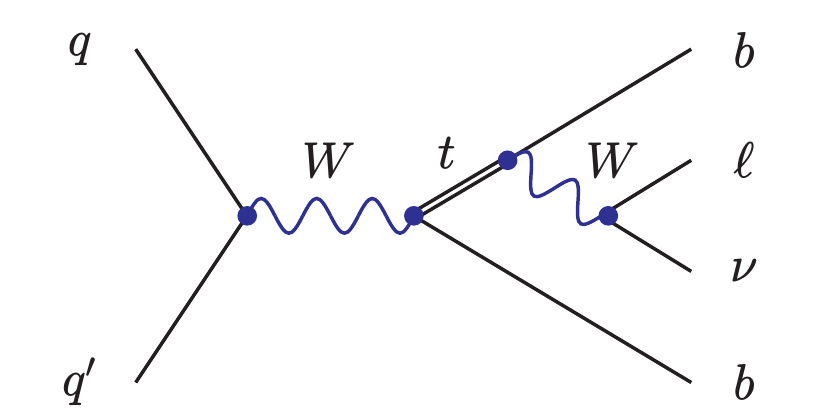}
    \caption{}
    \label{fig:LOdiags:O04tA}
  \end{subfigure}\\[3mm]
  \caption{
    Example LO diagrams at 
    $\ord(g_s^2e^2)$ (a,b),
    $\ord(g_se^3)$ (c,d), and 
    $\ord(e^4)$ (e-l). The square of $\ord(g_s^2e^2)$ diagrams yields the $\ord(\alphaS^2\alpha^2)$ QCD LO amplitude, while the square of the $\ord(e^4)$ diagrams yields the $\ord(\alpha^4)$ EW LO amplitude. The $\ord(\alphaS\alpha^3)$ perturbative contribution emerges as square of $\ord(g_se^3)$ diagrams, or due to the interference between $\ord(g_s^2e^2)$  and $\ord(e^4)$ diagrams. 
  }
  \label{fig:lodiagrams}
\end{figure}

At LO the process $pp\to\Vjj$, with 
\beq
\label{eq:Vmodes}
V\,=\,
\begin{cases}
Z^{\nu} & \mbox{for} \quad  \ppZnnjj\\
Z^{\ell} & \mbox{for}\quad   \ppZlljj\\
W^{\pm}        & \mbox{for} \quad  \ppWenjj  
\end{cases}
\eeq
receives three perturbative contributions as illustrated in the top row of Fig.~\ref{fig:perturbativeorders}.
Thus, the total LO differential cross section in a certain observable $x$ can be written as  
\begin{align}
\parx \sigma^{V}_{\rm LO} = \parx  \sigma^{V,\QCD}_{\rm LO}  +\parx  \sigma^{V,\EW}_{\rm LO} +\parx  \sigma^{V,\interf}_{\rm LO} \,.
\end{align}
The \textit{QCD mode} contributes at $\ord(\alphaS^2\alpha^2)$ and 
consists of absolute squares of the coherent sum of diagrams of 
$\ord(g_s^2e^2)$, exemplified by Figs~\ref{fig:LOdiags:O22A} and \ref{fig:LOdiags:O22B}. 
In this counting vector-boson decays
($\ell^{+}\ell^{-}$/$\nu\bar\nu$/$\ell^{\pm}\nu$) are included.
The \textit{EW mode}, on the other hand, contributes at $\ord(\alpha^4)$ 
and comprises the absolute square 
of the coherent sum of all diagrams of $\ord(e^4)$, 
see Figs.~\ref{fig:LOdiags:O04A}-\ref{fig:LOdiags:O04tA} for example diagrams. 
Their \textit{interference} contribution at $\ord(\alphaS\alpha^3)$ 
then is mostly comprised of the 
interference of $\ord(g_s^2e^2)$ diagrams with $\ord(e^4)$ diagrams. 
It, however, also contains genuine contributions consisting of absolute 
squares of $\ord(g_se^3)$ diagrams, for an example see 
Figs.~\ref{fig:LOdiags:O13A} and \ref{fig:LOdiags:O13B}, typically containing an external gluon and 
an external photon.

The contributions to the EW mode (and consequently also to the interference) deserve some closer inspection.
Diagrams illustrated in Figs.~\ref{fig:LOdiags:O04A} and \ref{fig:LOdiags:O04VBFA}, 
contribute to VBF-type production, while diagrams as in Figs.~\ref{fig:LOdiags:O04VVA} and \ref{fig:LOdiags:O04VA} contribute to (off-shell) diboson production with one vector boson decaying hadronically and the other leptonically. In the literature these are often denoted as $t$-channel and $s$-channel contributions, respectively. At LO these $t$- and $s$-channel contributions can easily be separated in a gauge invariant way in the well known VBF approximation. For
example, requiring at least one $t$-channel vector boson propagator and omitting $t$-$u$-channel interferences selects the VBF process, which includes contributions where the leptonically decaying vector boson couples directly to one of the external quark lines, as shown in Fig.~\ref{fig:LOdiags:O04A}. 
In addition, the EW mode also features photon-induced processes, see Fig.~\ref{fig:LOdiags:O04B}.
Since we employ the five-flavour (5F) number scheme throughout in the PDFs,
$b$-quarks are treated as massless partons, and channels with initial-state
$b$-quarks are taken into account for all processes and perturbative orders.
In the 5F scheme, the process $pp\to W+2\,$jets
includes partonic channels of type $q b\to q' b W$ that 
involve EW topologies corresponding to $t$-channel single-top production,
$q b\to q' t (bW)$, as illustrated in Fig.~\ref{fig:LOdiags:O04tB}.
Top resonances occur also in light-flavour channels of type
$q \bar q' \to \bar b b W$, which 
receive contributions from 
$s$-channel single-top production,
$q \bar q' \to \bar b t (bW)$, illustrated in Fig.~\ref{fig:LOdiags:O04tA}.
All these single-top contributions are consistently included in our predictions. 
At small \mjj their numerical impact can yield a substantial fraction of the total EW \Wjj
cross section at LO. 
For example, the combined $t$-channel and $s$-channel $pp \to t j$ processes yields around $25\%$ of the total EW \Wjj process at $\mjj=500\,$GeV. 
At higher \mjj the impact of the single-top modes is increasingly suppressed,
and for $\mjj > 2.5\,$TeV it is below $1\%$ of the EW \Wjj process.
More details on the impact of single-top contributions can be found in Sect.~\ref{res:interference}.

The LO interferences between QCD and EW modes that contribute at $\ord(\alphaS\alpha^3)$ are largely colour suppressed and yield very small contributions. This in particular holds in the VBF phase space, i.e.\ with large dijet invariant masses and large rapidity separation of the leading jets.

As illustrated in Fig.~\ref{fig:perturbativeorders} (bottom row) at NLO four perturbative contributions emerge. Out of these only the contribution with the highest and lowest power of $\alphaS$, i.e. the ones of $\ord(\alphaS^3\alpha^2)$ and $\ord(\alpha^5)$, can unambiguously be considered
as,
respectively, QCD corrections to the QCD mode and EW corrections to the EW mode. The former are well known in the literature~\cite{Campbell:2002tg,FebresCordero:2006nvf,Campbell:2008hh} while the latter are considered here for the first time. The contribution of $\ord(\alphaS^2\alpha^3)$ can 
be seen as NLO EW correction to the QCD mode; however, it also receives QCD-like corrections with respect to the LO interference. Corresponding results have been presented in~\cite{Denner:2014ina,Kallweit:2014xda,Kallweit:2015dum,Chiesa:2015mya}. Similarly, the contribution of  $\ord(\alphaS\alpha^4)$ can 
be seen as NLO QCD correction to the EW mode which likewise also receives EW-like corrections with respect to the LO interference. So far QCD corrections to the EW mode are known in the literature only in the VBF approximation, where  on top of the LO requirement of at least one $t$-channel vector-boson propagator (see above) QCD interactions between different quark lines are   
not allowed~\cite{Oleari:2003tc}. In this approximation the mentioned interference contributions do not enter. 
In this paper we present the first complete
computation of the $\ord(\alphaS\alpha^4)$ contribution, i.e.~the complete NLO QCD corrections to the EW
mode,  which go beyond the VBF approximation. This computation takes into account any contribution at the given perturbative order, i.e.\ it entails cross-talk between different quark lines, $t$-, $u$-, and $s$-channel
contributions and their interference, interference effects between the QCD and EW modes, as well as $s$-channel and $t$-channel single-top contribution in the case of EW \Wenjj production. Therefore, this computation can be seen as a unified NLO description of VBF vector-boson production, vector-boson pair-production with semi-leptonic decays, and (in the case of \Wenjj production) $t$-channel plus $s$-channel single-top production. 

In \Wjj production at $\ord(\alphaS \alpha^4)$, 
top resonances occur, besides in $t$-channel and $s$-channel configurations, also in channels of type $g b \to W b q \bar q'$,
which involve $Wt$-channel single-top production,
$g b \to Wt(b q\bar q')$. We have verified that these contributions are always at or significantly below the $1\%$ level with respect to the EW LO mode for all considered observables. For $\mjj > 2\,$TeV these contributions are suppressed to below the permil level.
In all perturbative orders we consider QCD partons (quarks and gluons) on the same footing as
photons, i.e.\ in the process definition $pp\to Vjj$ we have $j\in \{q,\bar q, g,
\gamma\}$, and everywhere
photon-induced production modes are included. However, we have verified that photon-induced production modes are at or below the $1\%$ level for all considered observables.

The total differential NLO cross section for \ppVjj production in a certain observable $x$ can be written as
\begin{align}
\parx \sigma^{V}_{\rm NLO} = \parx  \sigma^{V,\QCD}_{\rm NLO\, QCD+EW}  +\parx  \sigma^{V,\EW}_{\rm NLO\, QCD+EW} +\parx  \sigma^{V,\interf}_{\rm LO} \,,
\end{align}
where
\begin{align}
\parx \sigma^{V,\mode}_{\rm NLO\, QCD+EW} = \parx  \sigma^{V,\mode}_{\rm LO}  +\parx  \delta\sigma^{V,\mode}_{\rm NLO\, QCD}  +\parx  \delta\sigma^{V,\mode}_{\rm NLO\, EW}\,,
\end{align}
and $\mode = \{\QCD, \EW\}$ identifies the corresponding production mode. The NLO QCD and NLO EW 
corrections $\delta\sigma^{V,\mode}_{\rm NLO\, QCD}$ and  $\delta\sigma^{V,\mode}_{\rm NLO\, EW}$ correspond to the perturbative contributions of $\ord(\alphaS^3\alpha^2)$ and $\ord(\alphaS^2\alpha^3)$ for $\mode=\QCD$, and of $\ord(\alphaS\alpha^4)$ and $\ord(\alpha^5)$ for $\mode=\EW$. 
For later convenience we also define pure NLO QCD predictions 
without EW corrections,
\begin{align}
\label{eq:nloqcddef}
\parx \sigma^{V,\mode}_{\rm NLO\, QCD} \,=\, 
\parx  \sigma^{V,\mode}_{\rm LO}  
+\parx  \delta\sigma^{V,\mode}_{\rm NLO\, QCD}\,,  
\end{align}
and pure NLO EW predictions without QCD corrections,
\begin{align}
\label{eq:nloewdef}
\parx \sigma^{V,\mode}_{\rm NLO\, EW} \,=\, 
\parx  \sigma^{V,\mode}_{\rm LO}  
+\parx  \delta\sigma^{V,\mode}_{\rm NLO\, EW}\,.
\end{align}

As a natural
approximation of mixed QCD--EW higher-order corrections we also define a factorised combination 
of NLO QCD and NLO EW corrections,
\begin{align}
\parx \sigma^{V,\mode}_{\rm NLO\, \QCDtEW} \,=\, 
\parx  \sigma^{V,\mode}_{\rm NLO\, QCD
} 
\left(1+\kappa^{V,\mode}_{\EW}(x)\right)\,,
\end{align}
with the NLO EW correction factors
\beq 
\kappa^{V,\mode}_{\EW}(x)=\frac{\parx\delta\sigma^{V,\mode}_{\NLO\,\EW}}{\parx\sigma^{V,\mode}_{\LO}}\,.
\eeq


\section{Reweighting of Monte Carlo samples}
\label{se:reweighting}

The reweighting of MC samples is a natural way of combining (N)LO MC simulations with (N)NLO QCD+EW perturbative calculations and to account for the respective uncertainties and correlations 
in a systematic way. In the following we define a 
Monte Carlo reweighting procedure
for the individual 
QCD and EW production modes in \ppVjj.

For practical purposes the reweighting has to be performed based on a
one-dimensional distribution in a certain observable $x$.
To be precise, 
the relevant  
higher-order theory (TH) predictions for the observable at hand 
are defined as 
\beq
\label{eq:singlx}
\parx \sigma^{V,\mode}_{\rm TH}\left(\vec\eps^{\,V,\mode}_{\rm TH}\right)
\,=\,
\int\rd \vecy\, \pscuts{V}\,\parx \pary\, 
\sigma^{V,\mode}_{\rm TH}\left(\vec\eps^{\,V,\mode}_{\rm TH}\right)\,,
\eeq
where $V$ indicates the specific \Vjj process in Eq.~\eqref{eq:Vmodes}, and $\mode=\QCD, \EW$ 
identifies the corresponding production mode.
%
As reweighting observable  for the case at hand, i.e.\ $V+$\,multijet production in the VBF
phase-space, we choose the dijet invariant mass, 
\beq
x=\mjj\,,
\eeq
which is defined in more detail in Sect.~\ref{se:cutsnadobs}.
The integration on the r.h.s.\ of Eq.~\refeq{eq:singlx} involves all 
degrees of freedom $\vecy$ that are independent of $x$.
Such degrees of freedom include the fully differential kinematic dependence on the
vector-boson decay products 
and the two leading jets,
as well as the 
QED and QCD radiation that accompanies the VBF production process, \ie
extra jets and photons, and also possible extra leptons and neutrinos from hadron
decays.  %

The function $\pscuts{V}$ on the r.h.s.\ of Eq.~\refeq{eq:singlx} 
describes selection cuts for
\mbox{$pp\to$\,\Vjj},
and the details of its definition (see Sects.~\ref{se:cutsnadobs}--\ref{se:objects})
play 
an important role for the 
consistent implementation of the MC 
reweighting procedure.
Such cuts are typically chosen in a very similar way for
$V=Z,W$, but are not necessarily identical. For instance, in the case
$V=W$ the QED radiation from the lepton stemming from the $W\to\ell\nu$ decay 
is typically subject to a dressing prescription, while 
dressing is irrelevant for $Z\to \nu\nu$ decays. 
Note also that 
the cuts that are applied to the theoretical calculations
in Eq.~\refeq{eq:singlx}
do not need to be identical to the ones employed in the
experimental analysis.  They are typically rather similar to the actual
experimental cuts but more inclusive.%
\footnote{This is not a necessary prerequisite, i.e.\ the theoretical cuts
$\pscuts{V}$ may be also more exclusive then experimental cuts.  The crucial
prerequisite is that the MC samples that are going to be reweighted
with Eq.~\refeq{eq:singlx} and applied to the experimental analyis should extend
over the full phase-space regions 
that are covered by the theoretical calculations and by the
experimental analyses.
}

%
Theory uncertainties in Eq.~\eqref{eq:singlx} are 
parametrised through sets of nuisance parameters $\vec\eps^{\, V,\mode}_\TH$, and variations of individual nuisance parameters in the range 
\beq 
\label{eq:nuisancepar}
 \eps^{\, V,\mode}_{i,\TH}\;\in\,[-1,1]
\eeq 
should be understood as $1\sigma$ Gaussian uncertainties.
%



In a similar way as was proposed for monojet dark matter searches~\cite{Lindert:2017olm}, the theory predictions for the 
$\Vjj$ $x$-distributions can be embodied into the corresponing MC
simulations through a one-dimensional reweighting procedure. In this 
approach, the reweighted MC samples are defined as
\beq
\label{eq:rewa}
\parx\pary\, \sigma^{V,\mode}(\vec\eps^{\, V,\mode}_{\rm MC},
\vec\eps^{\,V,\mode}_{\rm TH}) 
:= 
\left[\frac{\parx\sigma^{V,\mode}_{\TH}(\vec\eps^{\,V,\mode}_{\rm TH})}
{\parx\sigma^{V,\mode}_{\MC}(\vec\eps^{\,V,\mode}_{\rm MC})}\right]
\,
\parx\pary\, \sigma^{V,\mode}_{\rm MC}(\vec\eps^{\,V,\mode}_{\rm MC}) \,.
\eeq
On the r.h.s., $\sigma^{V,\mode}_{\rm MC}$ with $\mode$= QCD, EW correspond to the fully differential
$V+2$\,jet
Monte Carlo samples before reweighting, and the 
$\sigma^{V,\mode}_{\rm MC}$ terms in the 
numerator and denominator 
must correspond to the same MC samples used in the experimental analysis.
Monte Carlo uncertainties, described by $\vec\eps^{\,V,\mode}_{\rm MC}$,
must be correlated in the numerator and denominator, 
while they can be kept uncorrelated across different processes, apart from 
$Z(\nu\bar\nu)+$\,jets and $Z(\ell\ell)+$\,jets.
As for the 
$\parx\sigma^{V,\mode}_{\TH}/\parx\sigma^{V,\mode}_{\MC}$
ratio on the r.h.s.\ of Eq.~\eqref{eq:rewa},
it is crucial that the numerator and the denomiator are 
determined using the same definition of the 
$x$-distribution, which is provided in
Sects.~\ref{se:cutsnadobs}--\ref{se:objects}.

The method proposed in~\cite{Lindert:2017olm} foresees the separate 
reweighting of the various $V+$jet processes, while the correlations 
between different processes and different $x$-regions is encoded into 
the corresponding correlations between nuisance parameters.
In this paper we adopt a simplified approach, which is designed for the case
where experimental analyses do not exploit theoretical 
information on the shape of the $x$-distribution, but only on the 
correlation between different processes at fixed $x$.
In this case, the relevant information can be encoded into the
$Z/W$ ratio
\beq 
\label{eq:ratios}
R^{\ZW,\mode}_{\TH}
(x,\vepsTH^{\,Z,\mode},\vepsTH^{\,W,\mode})
%
=\frac{\parx\sigma^{Z,\mode}_{\TH}(\vec\eps^{\,Z,\mode}_{\rm TH})}{\parx\sigma^{W,\mode}_{\TH}(\vec\eps^{\,W,\mode}_{\rm
TH})}\,,
\eeq  
where $Z=Z^{\nu}\, {\rm or}\, Z^{\ell}$, 
and $W\equiv \W^+ + W^-$.
In this ratio theory uncertainties largely cancel due to the very similar
dynamics of the $Z+2$\,jet and $W+2$\,jet processes.  This in particular holds for the
uncertainties related to higher-order QCD effects.
Such cancellations depend on the amount of correlation between the uncertainties of 
the individual distributions, which 
are encoded into the corresponding nuisance parameters.
%
%
Our theory predictions to be used for MC reweighting are provided directly at the level of the ratio of
Eq.~\refeq{eq:ratios}.

This ratio makes it possible to translate the MC prediction 
for the $x$-distribution in $W+2$\,jet into a correposnding 
$Z+2$\,jet prediction,
\beq
\label{eq:rewb}
\parx\, \sigma^{Z,\mode}(\vec\eps^{\, W,\mode}_{\rm MC},
\vec\eps^{\,Z,\mode}_{\rm TH},
\vec\eps^{\,W,\mode}_{\rm TH}) 
:= 
R^{\ZW,\mode}_{\TH}
(x,\vepsTH^{\,Z,\mode},\vepsTH^{\,W,\mode})
\,
\parx\, \sigma^{W,\mode}_{\MC}(\vec\eps^{\,W,\mode}_{\rm MC}) \,.
\eeq
Here the idea is that the MC uncertainties in $\sigma^{W,\mode}_{\MC}$ 
can be strongly constrained through data, while theory uncertainties 
are strongly reduced through cancellations in the ratio, which results into
an accurate prediction for the 
$x$-distribution in $Z+2$\,jets. The latter can be applied to the whole
$Z+$\,jets sample  via reweighting,
\beq
\label{eq:rewc}
\parx\pary\, \sigma^{\Z,\mode}(\vec\eps^{\, Z,\mode}_{\rm MC},
\vec\eps^{\, W,\mode}_{\rm MC},
\vec\eps^{\,Z,\mode}_{\rm TH},
\vec\eps^{\,W,\mode}_{\rm TH}) 
:= \left[
\frac{\parx\, \sigma^{Z,\mode}(\vec\eps^{\, W,\mode}_{\rm MC},
\vec\eps^{\,Z,\mode}_{\rm TH},
\vec\eps^{\,W,\mode}_{\rm TH}) 
}
{\parx\, \sigma^{\Z,\mode}_{\rm MC}(\vec\eps^{\,Z,\mode}_{\rm MC})
}
\right]
\parx\pary\, \sigma^{\Z,\mode}_{\rm MC}(\vec\eps^{\,Z,\mode}_{\rm MC}) \,.
\eeq
Note that the double reweighting procedure defined in
Eqs.~\eqref{eq:rewb}--\eqref{eq:rewc}
is equivalent to a single reweighting of the $Z+2$\,jet $x$-distribution, 
\beq
\label{eq:rew}
\parx\pary\, \sigma^{\Z,\mode}(\vec\eps^{\, Z,\mode}_{\rm MC},
\vec\eps^{\, W,\mode}_{\rm MC}
\vec\eps^{\,Z,\mode}_{\rm TH},
\vec\eps^{\,W,\mode}_{\rm TH}) 
:= \left[
\frac{R^{\Z/\W,\mode}_{\rm TH}
(x,\vepsTH^{\,Z,\mode},\vepsTH^{\,W,\mode})
}
{ R^{\Z/\W,\mode}_{\rm MC}(x, \vec\eps^{\, \Z,\mode}_{\rm MC}, \vec\eps^{\, \W,\mode}_{\rm MC})}\right] 
\parx\pary\, \sigma^{\Z,\mode}_{\rm MC}(\vec\eps^{\,Z,\mode}_{\rm MC}) \,.
\eeq
where $\sigma^{\Z,\mode}_{\rm MC}$ is the MC counterpart of the $Z/W$ ratio
defined in Eq.~\eqref{eq:ratios}.
%
%
As discussed above, the definition of the variable $x$ and the binning of its distribution 
need to be the same in all three terms on the r.h.s.~of~Eq.~\eqref{eq:rew}.
Instead,  acceptance cuts  must be identical in the numerator and denominator of the
double ratio, while particle-level MC predictions can be subject to more
exclusive or inclusive cuts in the experimental analysis.

In addition to the cancellation of theoretical uncertainties in 
the ratio $R^{\Z/\W,\mode}_{\rm TH}$ also correlated 
MC uncertainties tend to cancel in 
$R^{\Z/\W,\mode}_{\rm MC}$, thus the reweighting procedure~Eq.~\eqref{eq:rew}
turns a precise $W+2$\,jet measurement into a precise prediction for 
$Z+2$\,jets.
 

%

The reweighting in Eq.\ \eqref{eq:rew} can be applied to a \Znnjj as well as to
a \Zlljj MC sample, the former allows to constrain the irreducible backgrounds
in Higgs to invisible searches, while the latter allows for validation
against data in control regions.

\subsection{Reweighting observables and cuts}
\label{se:setup}

In this section we specify the observables, acceptance cuts,
and physics objects relevant for the reweighting in Eq.~\eqref{eq:rew}.
The theoretical calculations presented in Sect.~\ref{sec:results} 
are based on these definitions, which need to be adopted also for the MC predictions that enter in the
denominator of the double ratio on the r.h.s.\ of Eq.~\eqref{eq:rew}.
The details of this reweighting setup are designed such as to
take full advantage of the precision of perturbative calculations,
while excluding all effects 
that are better described by 
MC simulations
(e.g.~parton showering, hadronisation, and leptons or missing energy from hadron decays).


\subsection{Observables and cuts}
\label{se:cutsnadobs}

The reweighting in Eq.~\eqref{eq:rew} should be performed based on the rato of
the one-dimensional distribution in the 
dijet invariant mass $x=\mjj$, where $\jet_1, \jet_2$ are the two hardest jets.
The following binning is adopted for distributions in \mjj
\beqar
\label{eq:binning}\nonumber
\frac{\mjj}{\GeV} & \in  [500,550, \dots,950,1000,1100,\dots,1900,2000,2500,3000,3500,4000,6000,13000]\,.\nonumber\\
\eeqar
Theoretical predictions for the $\mjj$-distribution and their 
MC counterpart should be determined in the 
presence of the following cuts,
\beqar 
\label{eq:nominalcuts}
p_{\rT,\jet_{1}}&>&100\,\GeV\,,\qquad
p_{\rT,\jet_{2}}\,>\,50\,\GeV\,, \qquad
\mjj \,>\, 500 \,\GeV\,, \qquad
\Delta\eta_{\jet_{\rm 1}\jet_{\rm 2}} \,>\, 2.5\,, 
\nonumber\\[1mm]
p_{\rT,V}&>&150\,\GeV\,,
\eeqar 
for $V=W^\pm,Z$. The relevant definitions of jets and 
$p_{\rT,V}$ are discussed in \refse{se:objects}. Note that
only the reconstructed vector-boson momenta are subject to cuts, while 
no restriction is applied to the individual 
momenta of their decay products.
For $pp\to \ell\ell+\jet\jet$ the additional process-specific cut
\beqar 
m_{\ell \ell} > 40\,\GeV
\eeqar 
should be applied.

For a realistic assessment of theoretical uncertainties, one should also 
consider the fact that, 
whithin experimental analyses, VBF cuts can be supplemented by a
veto on additional jet radiation.
In this case we recommend to perform two alternative reweightings with and without jet
veto.
The difference between MC samples reweighted 
with jet veto and in the nominal setup of Eq.~\eqref{eq:nominalcuts} 
should be small and can be taken as an additional uncertainty.
In particular we consider an additional veto on jet radiation
\beqar
\label{eq:modveto}
p_{T,j_3} < p_{T,{\rm cut}}= \max(500\,\GeV, m_{jj})/20\,.
\eeqar
{We choose to employ a dynamic jet veto to minimise possible 
large logarithms that may spoil the perturbative convergence of our results.

Finally, in order to address the limitations of the proposed 
one-dimensional reweighting in \mjj,
we split the phase space into the following three $\deltaphijj$ regions,
where $\deltaphijj$ is the azimuthal-angle separation 
between the two leading jets,
\beqar
\label{eq:moddphi}
\Phi_1 = \{\deltaphijj < 1\}\,, \qquad
\Phi_2 = \{1 < \deltaphijj < 2\}\,, \qquad
\Phi_3 = \{2< \deltaphijj\}\,.
\eeqar
These $\deltaphijj$ bins are motivated by the fact 
that 
the higher-order corrections to the reweigthing ratios
$R^{\ZW,\mode}_{\TH}$, defined in Eq.~\refeq{eq:ratios},
feature a non-negligible dependence on $\deltaphijj$.
As discussed in Sect.~\ref{se:ratios}, 
this effect is taken into account through a
theoretical uncertainty that is derived from the 
differences between the $R^{\ZW,\mode}_{\TH}$ ratios
in the above $\deltaphijj$ regions.



\subsection{Definition of physics objects}
\label{se:objects}

In the following we define the various physics objects relevant for 
higher-order perturbative calculations and for the reweighting in the Monte Carlo counterparts in Eq.\ \refeq{eq:rew}.

\subsubsection*{Neutrinos}
\label{se:neutrinos}
In parton-level calculations of $pp\to \Vjj$,
neutrinos originate only from vector-boson decays, while
in Monte Carlo samples they can arise also from 
hadron decays.  In order to avoid any bias in the reweighting procedure, 
only neutrinos arising
from $Z$ and $W$ decays at Monte Carlo truth level should be considered.

\subsubsection*{Charged leptons}
\label{se:dressing}
Distributions in the lepton $p_\rT$ and other leptonic observables are known
to be highly sensitive to QED radiative corrections, and the differences in
the treatment of QED radiation on Monte Carlo and theory side can lead to a bias in
the reweighting procedure.  To avoid such a bias, dressed
leptons should be used, i.e.\ all leptons are combined with all nearly collinear photons that lie
within a cone of
\def\rrec{\Delta R_{\mathrm{rec}}}
\beq 
\Delta R_{\ell\gamma}=\sqrt{\Delta \phi^2_{\ell\gamma}
+\Delta \eta^2_{\ell\gamma}}<\rrec\,.
\eeq 
For the radius of the recombination cone we employ the standard value $\rrec=0.1$,
which allows one to capture the bulk of the collinear final-state radiation, 
while keeping contamination from large-angle photon radiation from other sources at a negligible level.
All lepton observables as well as the kinematics of the reconstructed $W$ and $Z$ bosons are defined in terms of dressed leptons,
and, in accordance with standard experimental practice, both muons and electrons should be
 dressed.  In this way differences between electrons and muons, $\ell=e,\mu$,
become negligible, and the reweighting function needs to be computed only
once for a generic lepton flavour $\ell$.

Similarly as for neutrinos, only charged leptons that arise from $Z$ and $W$
decays at Monte Carlo truth level should be considered. 
Concerning QCD radiation in the vicinity of leptons, 
no lepton isolation requirement should be imposed
in the context of the reweighting procedure. Instead,
in the experimental analysis
lepton isolation cuts can be applied in the usual manner.

\subsubsection*{$Z$ and $W$ bosons}
\label{se:ptVdef}
The off-shell four-momenta of $W$ and $Z$ bosons are defined as
\beqar
\label{eq:reconstruction}
p^\mu_{W^+}&=&p^\mu_{\ell^+}+p^\mu_{\nu_\ell},\qquad
p^\mu_{W^-}=p^\mu_{\ell^-}+p^\mu_{\bar\nu_\ell},\\
p^\mu_{Z}&=&p^\mu_{\ell^+}+p^\mu_{\ell^-},\qquad
p^\mu_{Z}=p^\mu_{\nu_\ell}+p^\mu_{\bar\nu_\ell},\nonumber
\eeqar
where the leptons and neutrinos that result from $Z$ and $W$ decays are defined as discussed 
above.

\subsubsection*{Jets}

Similarly as for the charged leptons, photons are recombined with collinear quarks within $\Delta R_{q\gamma} < \rrec$ prior to jet clustering. Subsequently, QCD partons (quarks and gluons) together with
the remaining photons are clustered into jets according to the anti-$k_\rT$ algorithm~\cite{Cacciari:2008gp} using R=0.4 and ordered by their transverse momentum.

%
%
%

\section{Theoretical predictions and uncertainties}
\label{sec:results}

In this section we present our theoretical input for 
invisible-Higgs searches. The relevant input parameters are 
documented in Sect.~\ref{sec:setup}, and in
Sect.~\ref{se:hopred} we discuss 
NLO QCD+EW predictions for $pp\to V+2$\,jets 
at parton level and matched to 
the parton shower.
Our main results for $Z/W$ ratios and their theoretical uncertainties are
presented in Sect.~\ref{se:ratios}.

All predictions presented in this paper have been obtained 
within the {\sc Sherpa+OpenLoops} framework, 
which supports fully automated NLO QCD+EW 
calculations at parton level~\cite{Kallweit:2014xda,Schonherr:2017qcj,Sherpa:2019gpd}
as well as matching~\cite{Frixione:2002ik,Hoeche:2011fd} to {\sc Sherpa}'s parton shower~\cite{Schumann:2007mg} and multi-jet
merging~\cite{Hoeche:2012yf} at NLO as implemented in the {\sc Sherpa} Monte Carlo framework~\cite{Krauss:2001iv,Gleisberg:2007md,Gleisberg:2008ta,Sherpa:2019gpd}.
In particular, {\sc Sherpa+OpenLoops} 
allows for the simulation of the entire tower of QCD and EW contributions of
 $\ord(\alphaS^n \alpha^m)$  that are
relevant for
multi-jet processes like $pp\to V+2$\,jets at LO and NLO.
All relevant renormalised virtual amplitudes are provided by the {\sc OpenLoops 2}
program \cite{Buccioni:2019sur} which implements the techniques
of~\cite{Cascioli:2011va,Buccioni:2017yxi} and is interfaced 
with {\sc Collier}~\cite{Denner:2016kdg} and {\sc OneLOop}~\cite{vanHameren:2010cp} 
for the calculation of scalar integrals.

\subsection{Definition of numerical setup}
\label{sec:setup}

In the following we specify input parameters and PDFs employed for 
theoretical predictions in this study.
As  discussed in \refse{se:reweighting},
Monte Carlo samples used in the experimental analyses 
do not need to be generated with the same input parameters and PDFs used 
for higher-order theoretical predictions.

\begin{table}
  \begin{center}
    \begin{tabular}{rclrcl}
      $\MW$ & \shortequal & $80.399~\GeV$\,  & $\GW$ & \shortequal & $2.085~\GeV$\, \\
      $\MZ$ & \shortequal & $91.1876~\GeV$\, & $\GZ$ & \shortequal & $2.495~\GeV$\, \\
      $\MH$ & \shortequal & $125~\GeV$\,     & $\GH$ & \shortequal & $4.07~\MeV$\, \\
      $\Mb$ & \shortequal & $0~\GeV$\,    & $\Gb$ & \shortequal & $0$\, \\
      $\Mt$ & \shortequal & $172.5~\GeV$\,   & $\Gt$ & \shortequal & $1.32~\GeV$\,, \\
      $\GF$ & \shortequal & $1.1663787\cdot 10^{-5}~\GeV^{-2}$\, & \qquad\qquad & &\\
    \end{tabular}
  \end{center}
  \caption{Values of the various physical input parameters. The value of
  $\Mb$ depends on the employed flavour-number scheme as discussed in the
  text.}
    \label{tab:inputs}
\end{table}

In the calculation of  $pp\to\nu\nu/\ell\nu/\ell\ell\,+2$\,jets we use the 
coupling constants, masses and widths as listed in \refta{tab:inputs}.
All unstable particles are treated in the complex-mass scheme~\cite{Denner:2005fg},
where width effects are absorbed into the complex-valued renormalised masses
\beqar\label{eq:complexmasses}
\mu^2_i=M_i^2-\ri\Gamma_iM_i \qquad\mbox{for}\;i=W,Z,t.
\eeqar 
The EW couplings are derived from the gauge-boson masses and the Fermi constant
$\GF$ using 
\beq\label{eq:defalpha}
\alpha=\left|\frac{\sqrt{2}\sin^2\thetaw\,\mu^2_W G_\mu}{\pi}\right|\,,
\eeq
and the weak mixing angle $\thetaw$. The latter is determined by 
\beq\label{eq:defsintheta}
\sin^2\thetaw=1-\cos^2\thetaw=1-\frac{\mu_W^2}{\mu_Z^2}
\eeq
in the complex-mass scheme.
The $G_\mu$-scheme guarantees an  optimal description of pure SU(2) interactions
at the 
EW
scale as it absorbs universal higher-order corrections to the weak 
mixing angle into the LO contribution already and, thus, minimises higher-order 
corrections. It is therefore the scheme of choice for
$W+$\,multijet production, and it provides a very good description of
$Z\,+$\,multijet production as well. 
The CKM matrix is assumed to be diagonal, and we checked at LO and NLO QCD that for $W$+multijet production the difference with respect to a non-diagonal CKM matrix  is always well
below 1\%.

As renormalisation scale $\mur$ and factorisation scale $\muf$ we set
\beq
\mu_{\rm R,F} = \xi_{\rm R,F}\mu_0\,,\quad {\rm with}\quad \mu_0=\frac{1}{2}H_{\rT}' \quad{\rm and} \quad\frac{1}{2} \leq \xi_{\rm R},\xi_{\rm F} \leq 2\,.
\eeq
Here $H_{\rm T}'$ is defined as the scalar sum of the transverse energy of all 
parton-level final-state objects,
\beq
H_{\rT}' = E_{\rT,V} + \sum\limits_{i \in \mathrm{partons}} p_{\rT,i}\,,\quad {\rm with}\quad E_{\rT,V}=\sqrt{m_V^2 + p_{\rT,V}^2}
\eeq
where $m_V$ and $p_{\rT,V}$ are, respectively, the invariant mass and the
transverse momentum of the reconstructed off-shell vector boson momenta as
defined in Eq.~\eqref{eq:reconstruction}, 
while the sum includes all final-state QCD and QED partons ($q, g, \gamma$) 
including those emitted at NLO.\footnote{This scale choice
corresponds to the scale setter {\tt DH\_Tp2} in {\sc Sherpa}.} Our default scale
choice corresponds to $\xi_{\rm R} = \xi_{\rm F} =1$, and theoretical QCD
scale uncertainties are assessed by applying the standard 7-point variations
$(\xi_{\rm R}, \xi_{\rm F}) = (2, 2), (2, 1), (1, 2), (1, 1), (1,\half),
(\half, 1), (\half,\half)$.

For the calculation of hadron-level cross sections at NLO(PS) QCD\,+\,NLO
EW we employ the
{\tt NNPDF31\_nlo\_as\_0118\_luxqed}
PDF set, which encodes QED effects via the LUXqed methodology of~\cite{Manohar:2016nzj}.
The same PDF set, and the related $\alphaS$ value,  is used throughout, i.e.\ also in the relevant LO and NLO 
ingredients used in the estimate of theoretical uncertainties.
Consistently with the 5F number scheme employed in the PDFs,
$b$-quarks are treated as massless partons, and channels with initial-state
$b$-quarks are taken into account for all processes and production modes.

In addition to fixed-order calculation including NLO QCD and EW corrections, 
we also match the NLO QCD corrections to the QCD mode to the parton shower. 
Here we set the scales according to the CKKW scale setting algorithm 
of \cite{Hoeche:2009rj,Hoeche:2012yf}, i.e.\ we interprete the given 
configuration using the inverse of the parton shower (using only its 
QCD splitting functions) to arrive at a core process and the reconstructed 
splitting scales $t_i$, 
\beq
  \alphaS^{n+k}(\mu_{\rm R}^2)=\alphaS^k(\mu_\text{core}^2)\,\prod\limits_{i=1}^n\alphaS(t_i)\,.
\eeq
We restrict ourselves to strongly ordered hierarchies only, i.e.\ 
$\mu_{\rm Q}>t_1>t_2>..>t_n$, as the parton shower would produce them 
in its regular evolution. 
In consequence, depending on the phase space point, possible core 
configurations are $pp\to V$, $pp\to V+j$, $pp\to V+jj$, and $pp\to V+jjj$.
Further, we set both the factorisation and the shower starting scale, 
$\mu_{\rm F}$ and $\mu_{\rm Q}$ respectively, 
to the scale $\mu_\text{core}=\tfrac{1}{2}\,H_\rT'$ defined on the 
reconstructed core process.
In our region of interest where the usual Sudakov factors are negligible, 
our NLOPS simulation is thus equivalent to the two-jet component of 
an inclusive NLO merged calculation in the MEPS@NLO algorithm without 
additional multiplicities merged on top of it.

%

\subsection{Higher-order QCD, EW and PS predictions for \texorpdfstring{\Vjj}{V+2jet}}
\label{se:hopred}

In this section we present LO and NLO QCD+EW predictions for 
\ppZnnjj and \ppWenjj including also parton-shower effects.
Each process is split into a QCD and EW 
production mode as discussed in Sect.~\ref{sec:modes}.


%

\begin{figure}[t]
\centering
	\includegraphics[width=0.45\textwidth]{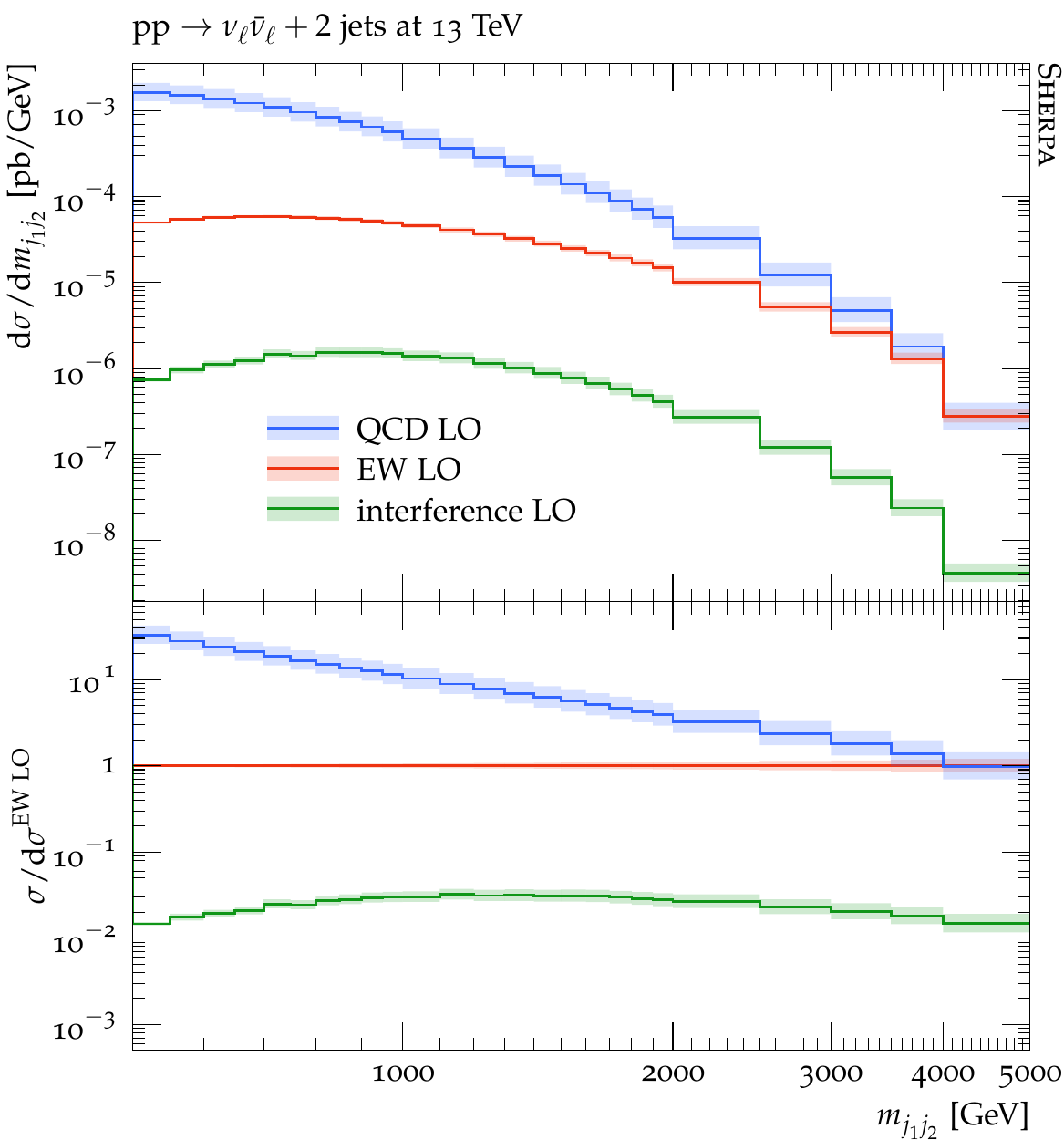}
	\plotsep
	\includegraphics[width=0.45\textwidth]{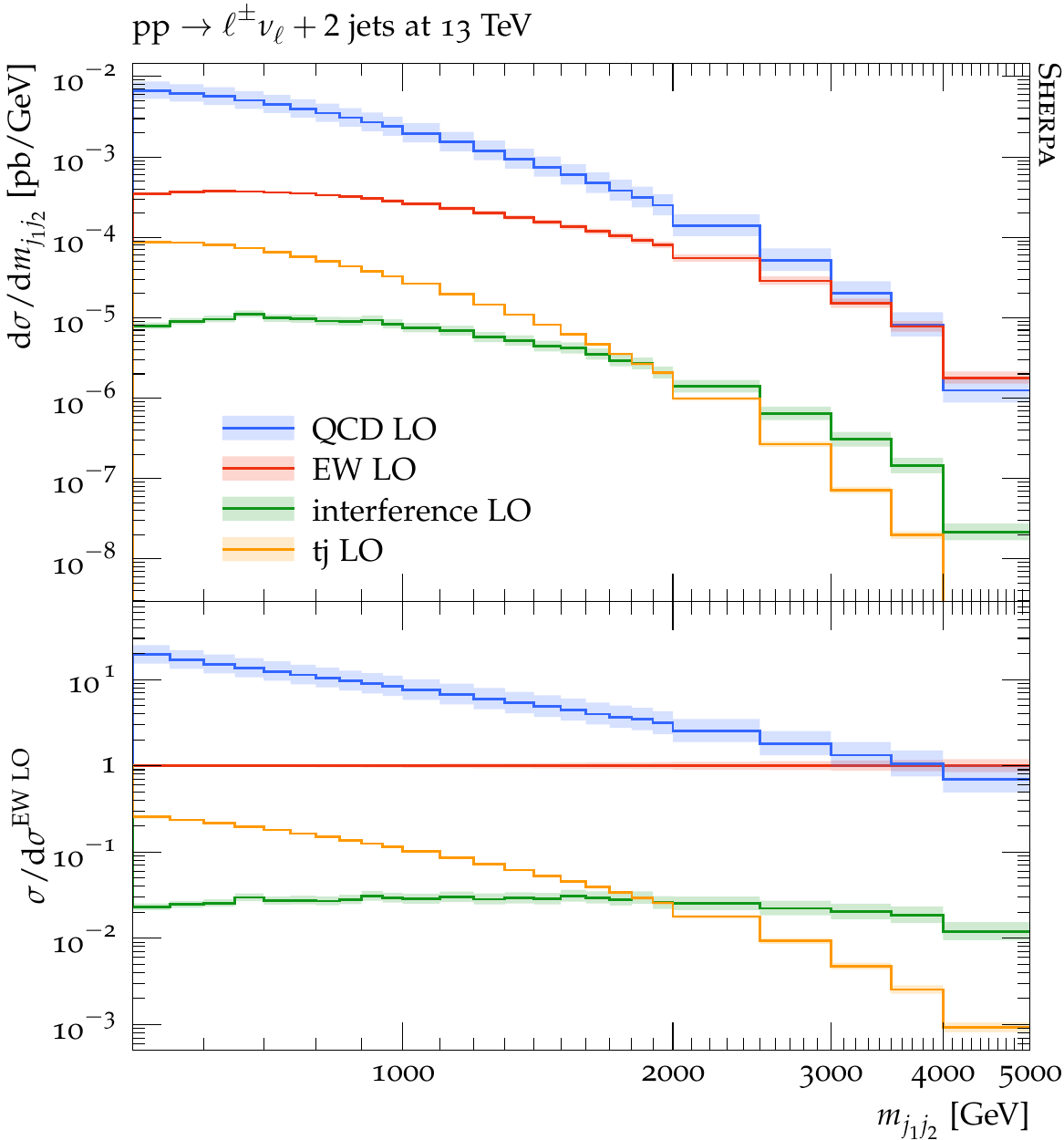}
\caption{Distribution in the invariant mass of the two hardest jets, \mjj, for
\ppZnnjj (left) and \ppWenjj (right) at LO.  The upper frame 
shows absolute predictions for the QCD (blue), EW (red),
and interference (green) production modes. 
For \ppWenjj we also show the LO $pp\to tj$
contributions (orange), which belong to the EW production mode
and include $t$-channel and $s$-channel single-top 
production.
The relative importance of the various contributions normalised to the
EW production mode is displayed in the lower frame.
The bands correspond to QCD scale variations, and in the case of ratios
only the numerator is varied.  
}
\label{fig:nom_LO_mjj}
\end{figure}

\subsubsection{LO contributions and interference}
\label{res:interference}

In Fig.~\ref{fig:nom_LO_mjj} we show LO predictions for \Zjj (left) and \Wjj (right) production considering the QCD and EW modes together with the LO interference. In the case of \Wjj production we also show the LO contribution due to $pp\to t j$ with leptonic on-shell decays of the top. The final-state jet can be a light jet or a bottom-quark jet, i.e.\ this process comprises $t$-channel and $s$-channel single-top production at LO.
The single-top processes are consistently included in the the off-shell matrix elements of the EW mode of \ppWenjj. For both \Zjj and \Wjj production the QCD mode largely dominates over the EW mode in the bulk of the phase-space; however, at large \mjj the EW mode becomes subsequently more and more important, eventually dominating over the QCD mode for about $\mjj > 4\,$TeV. For both considered processes the LO interference remains more or less constant with respect to the EW mode, at about $2-3\%$ relative to it over the entire \mjj range.
The $pp \to t j$ process yields around $25\%$ of the total EW \Wjj process at the lower end of the considered \mjj range.  At large \mjj the impact of the single-top modes is increasingly suppressed,
and for $\mjj > 2.5\,$TeV it drops below $1\%$ of the EW \Wjj process.

\subsubsection{QCD production}
\label{sec:nomQCD}

The NLO QCD and EW corrections to the production of $V+2$\,jets via 
QCD interactions are well known in the literature.
For example, Ref.~\cite{Kallweit:2015dum} presents a systematic investigation of QCD and EW correction
effects on high-energy observables. Here we focus on NLO corrections 
and correlations relevant for invisible-Higgs searches at large invariant masses 
of the two hardest jets. 
Besides fixed-order NLO corrections we also investigate the effect of parton-shower
matching at NLO QCD.

\begin{figure}[t]
\centering
	\includegraphics[width=0.45\textwidth]{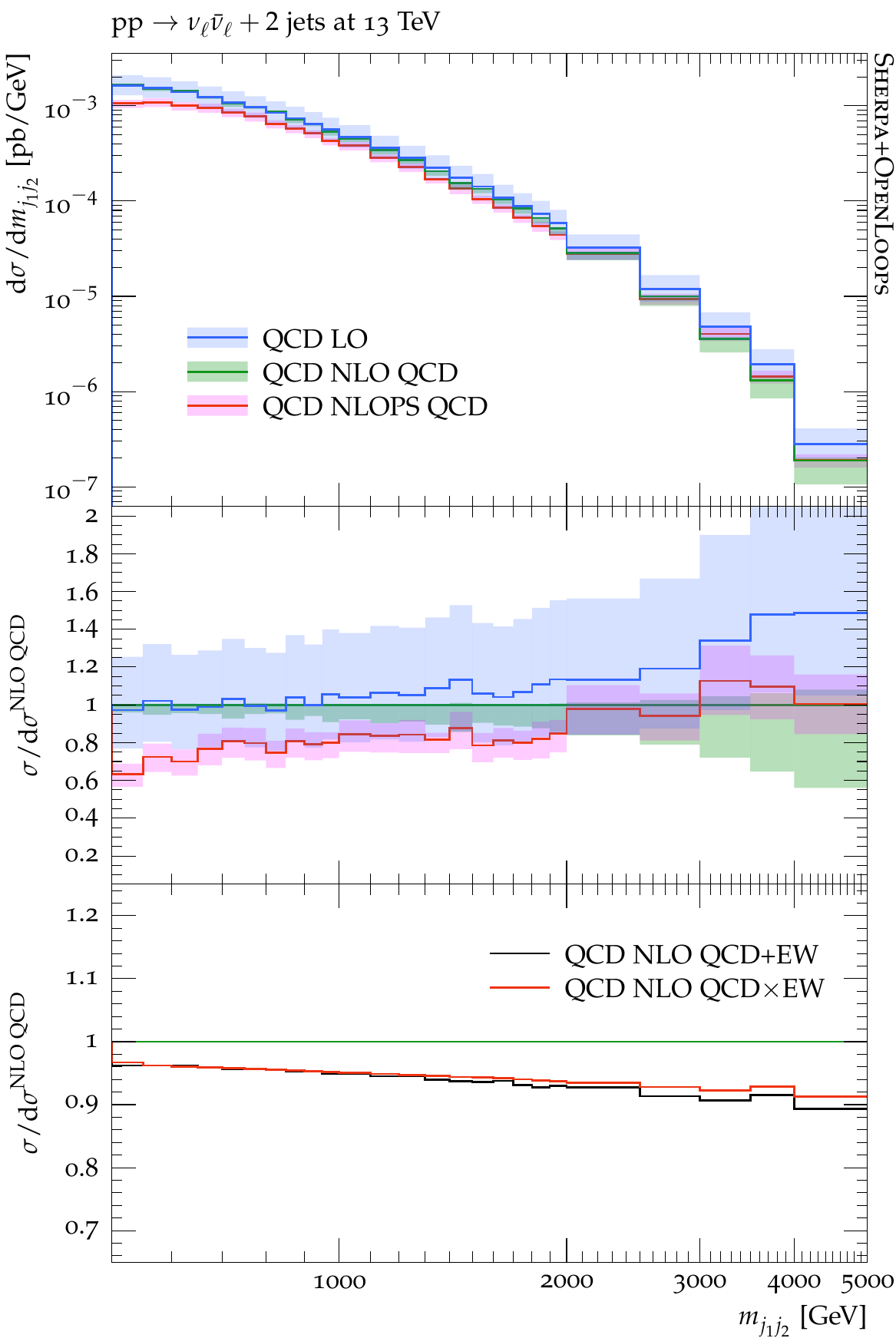}
	\plotsep
	\includegraphics[width=0.45\textwidth]{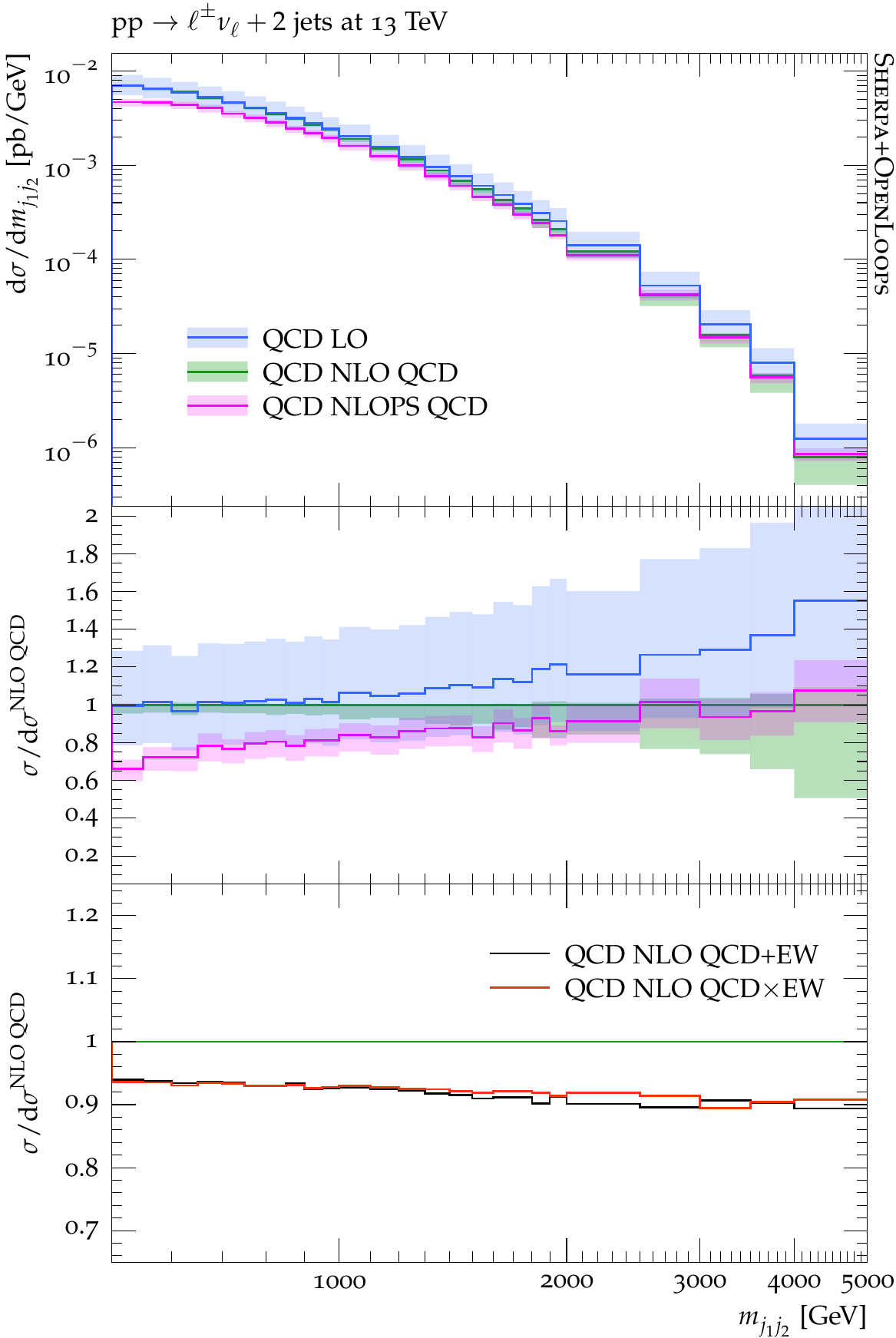}
\caption{Distribution in the invariant mass distribution of the two hardest
jets, \mjj, for QCD \ppZnnjj (left) and QCD \ppWenjj (right).  The upper
frame displays absolute LO QCD
(blue), NLO QCD (green), and NLO+PS QCD (magenta) predictions, and
ratios with respect to NLO QCD are presented in the central panel.  
The bands correspond to QCD scale
variations, and in the case of ratios only the numerator is varied.  
The lower panel shows the relative impact of NLO QCD+EW (black) and NLO QCD$\times$EW
(red) predictions normalised to NLO QCD.}
\label{fig:nom_QCD_mjj}
\end{figure}

Fig.~\ref{fig:nom_QCD_mjj} shows the distribution in \mjj for 
\ppWenjj and \ppZnnjj in various approximations.
Predictions and scale variations at LO QCD, NLO QCD and NLOPS
QCD accuracy are presented together with the additive and multiplicative combination of 
NLO QCD and EW corrections.
For both processes the effect of QCD, EW and shower corrections, as well as
the QCD scale variations is remarkably 
similar. 
 
The impact of QCD corrections is negative, and below 
1\,TeV it remains quite small, while 
in the $\mjj$ tail it becomes increasingly large,
reaching around $-20\%$ at 2--3\,TeV and $-50\%$
at 4\,TeV. 
Parton-shower corrections are at the percent level in the 
\mjj-tail, while below 2\,TeV their effect is more 
sizeable and negative, reaching $20$--$30\%$ around 500\,GeV.
Also the NLO EW corrections yield an increasingly negative 
contribution with rising $\mjj$.
Their impact, however, is rather mild
and reaches only about $-10\%$ in the multi-TeV region.

In Fig.~\ref{fig:nom_QCD_cjv_mjj} we show the 
same \mjj-distributions and theoretical predictions of 
Fig.~\ref{fig:nom_QCD_mjj} in the presence of 
the dynamic veto of Eq.~\eqref{eq:modveto} against a third jet.
At LO QCD, where only two jets are present, the veto has no effect, while the
NLO QCD and NLOPS QCD predictions are strongly reduced.
The maximal effect is observed at $\mjj=500$\,GeV, 
where the veto of Eq.~\eqref{eq:modveto} corresponds to $p_{T,{\rm cut}}=25$\,GeV, and
the NLO~\QCD cross section is suppressed by a factor four.
Above 500\,GeV the value of $p_{T,{\rm cut}}$ grows linearly 
with $\mjj$, and the effect of the veto on the cross section 
becomes less important.
%
In spite of the large NLO QCD corrections up to 1--2\,TeV, 
the small difference between NLO and NLOPS predictions 
suggests the absence of large higher-order 
effects beyond NLO.
Moreover, we observe that the 
pattern of suppression with respect to LO
and also the NLOPS corrections with respect to NLO QCD are highly universal
between \Zjj and \Wjj production.  The NLO EW corrections are almost 
identical to the inclusive selection.

%

\begin{figure}[t]
\centering
	\includegraphics[width=0.45\textwidth]{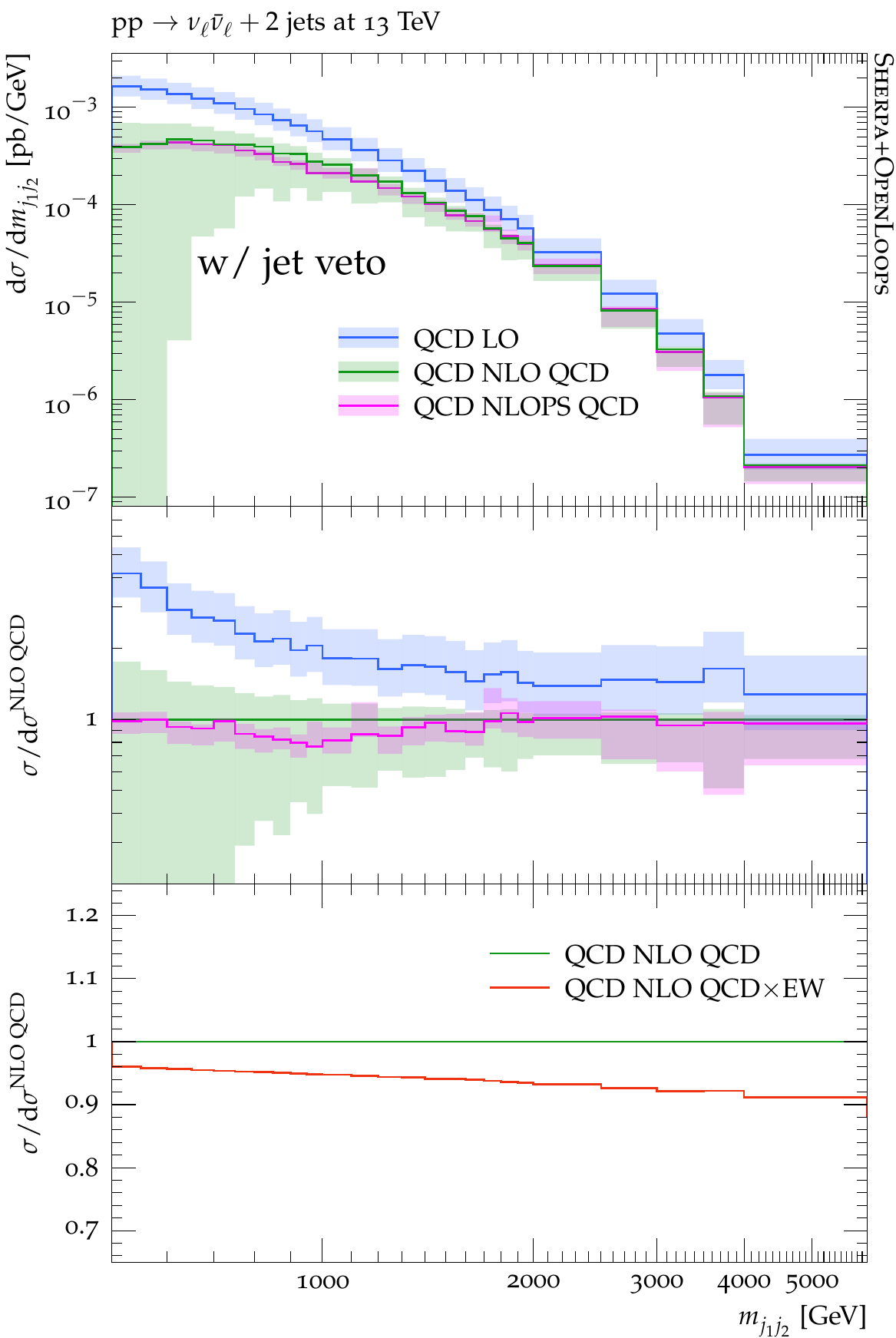}
	\plotsep
	\includegraphics[width=0.45\textwidth]{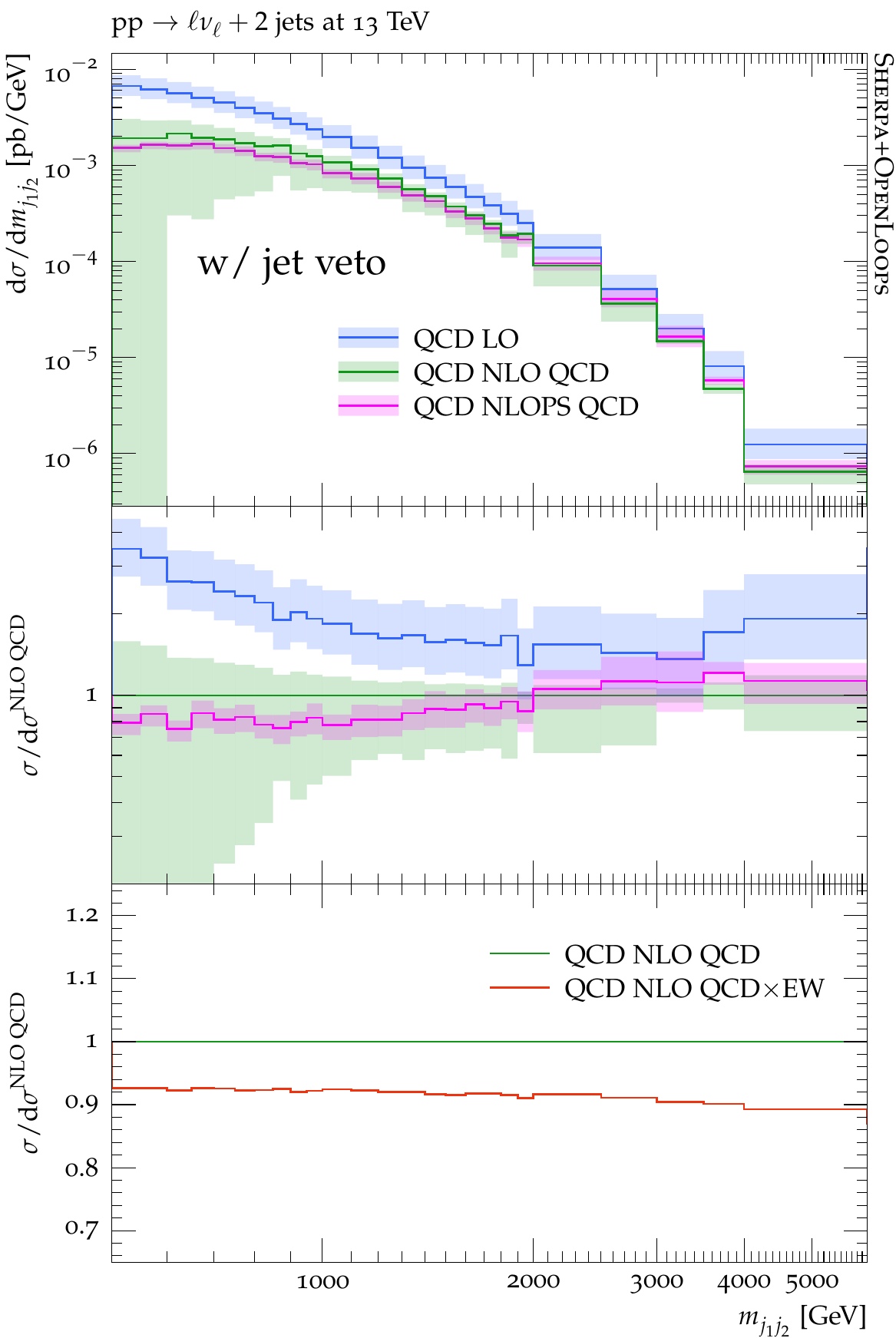}
\caption{Distribution in the invariant mass distribution of the two hardest
jets, \mjj, for QCD \ppZnnjj (left) and QCD \ppWenjj (right) 
subject to the dynamic veto of Eq.~\eqref{eq:modveto} against a third jet. 
Curves and bands as in Fig.~\ref{fig:nom_QCD_mjj} but without NLO QCD+EW predictions. 
}
\label{fig:nom_QCD_cjv_mjj}
\end{figure}

\subsubsection{EW production}
\label{sec:nomEW}

Numerical results for EW \Vjj production including QCD and EW corrections are shown in Figs.~\ref{fig:nom_EW_pTV}-\ref{fig:nom_EW_deta}. We remind the reader that here we present the first complete computation of the QCD corrections to the EW production modes, i.e.\ without resorting to the VBF approximation, and also the first computation of the EW corrections to the EW modes.

\begin{figure}[t!]
\centering
	\includegraphics[width=0.45\textwidth]{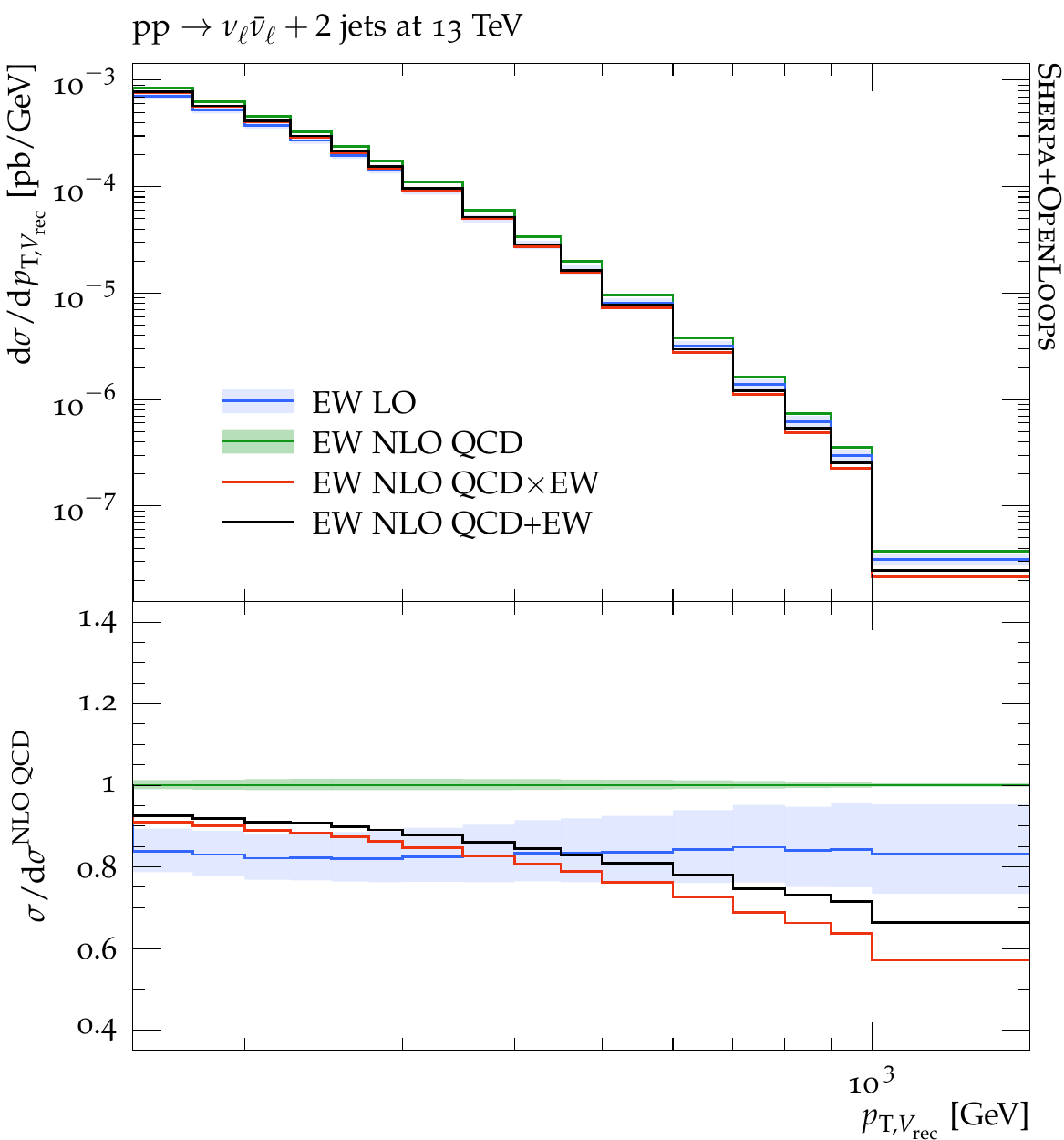}
	\plotsep
	\includegraphics[width=0.45\textwidth]{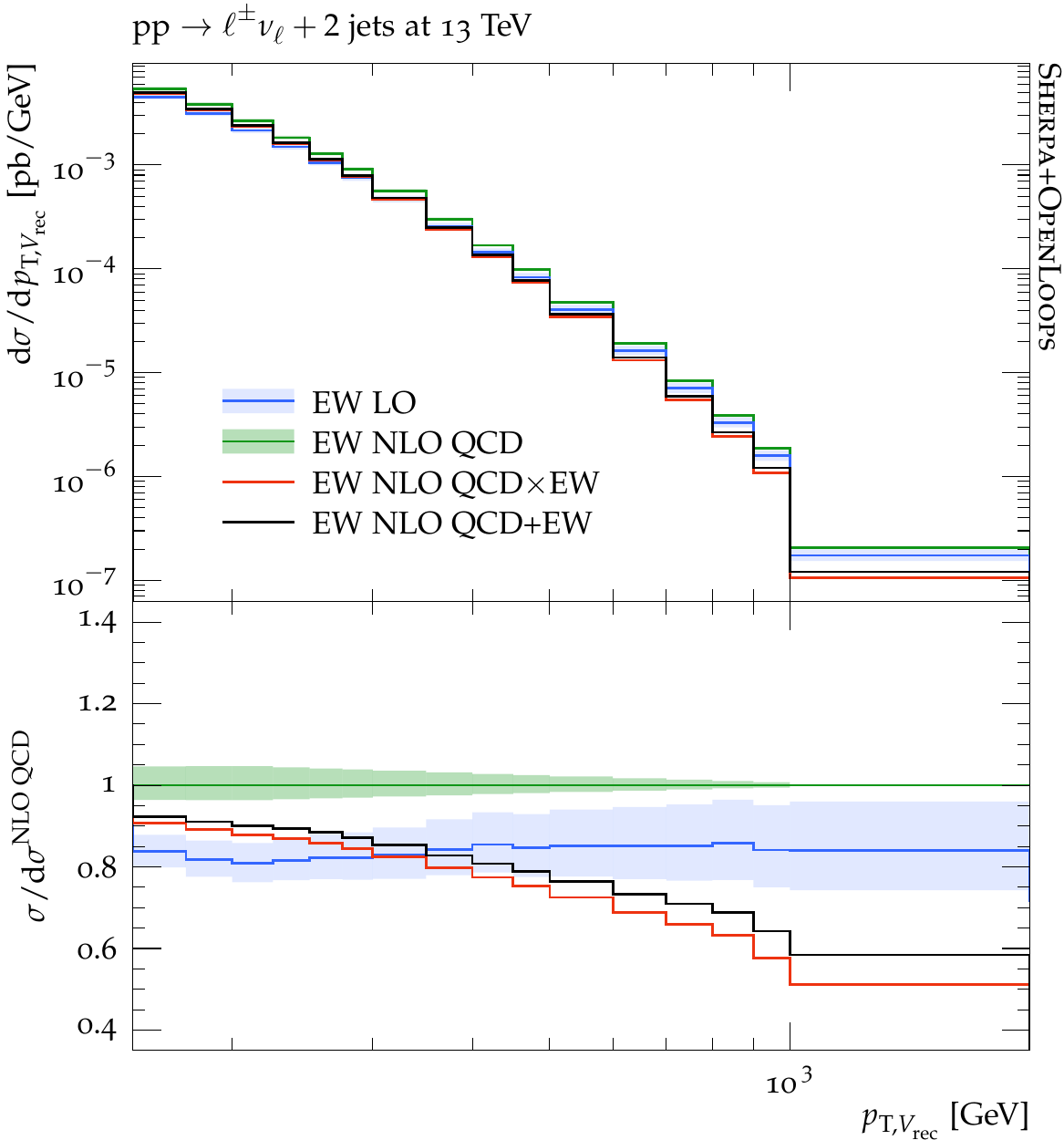}
\caption{Distribution in the reconstructed transverse momentum of the
off-shell vector boson, \pTV, for EW \ppZnnjj (left) and EW \ppWenjj
(right).  Absolute EW LO (blue), NLO QCD (green), NLO QCD+EW (black) and NLO
QCD$\times$EW (red) predictions are shown in the upper panel.
Here NLO QCD and NLO EW corrections should be understood as 
$\ord(\alphaS)$ and $\ord(\alpha)$ effects wrt to the EW LO.
The same predictions normalised to NLO QCD are shown in the lower panel.
The bands correspond to QCD scale variations, and in the case of ratios only the numerator is varied.}
\label{fig:nom_EW_pTV}
\end{figure}

In Fig.~\ref{fig:nom_EW_pTV} differential predictions in the transverse
momentum of the (reconstructed) vector bosons, $p_{\rT,V}$, are shown.  We observe that 
the NLO QCD corrections increase the LO EW cross section 
by about $20\%$ showing
hardly any \pTV{} dependence.  
QCD scale uncertainties at LO are at $10\%$ for small $p_{\rT,V}$ and increase
up to  $20\%$ 
in the tail of the \pTV{} distribution.
The QCD scale
uncertainties at NLO QCD are only at the level of a few percent and decrease to
negligible levels in the tail.  This is consistent 
with the computation of NLO QCD corrections for the \Vjj processes
in the VBF approximation, where residual scale uncertainties are at the $2\%$
level~\cite{Oleari:2003tc}.  
Here we note that, given the rather large size of the NLO QCD corrections, 
such small scale uncertainties cannot be regarded as a 
reliable estimate of unknown higher-order effects.
In the \pTV
distribution the EW corrections display a typical behaviour induced by the
dominance of EW Sudakov logarithms.  At 1\,TeV the EW
corrections reduce the NLO QCD cross section by $40$--$50\%$, 
with a spread of about $10\%$
between the additive and the multiplicative combinations.  Both QCD and EW
corrections are highly correlated between the two considered processes, i.e. 
the relative impact of these corrections is almost identical.

\begin{figure}[t!]
\centering
	\includegraphics[width=0.45\textwidth]{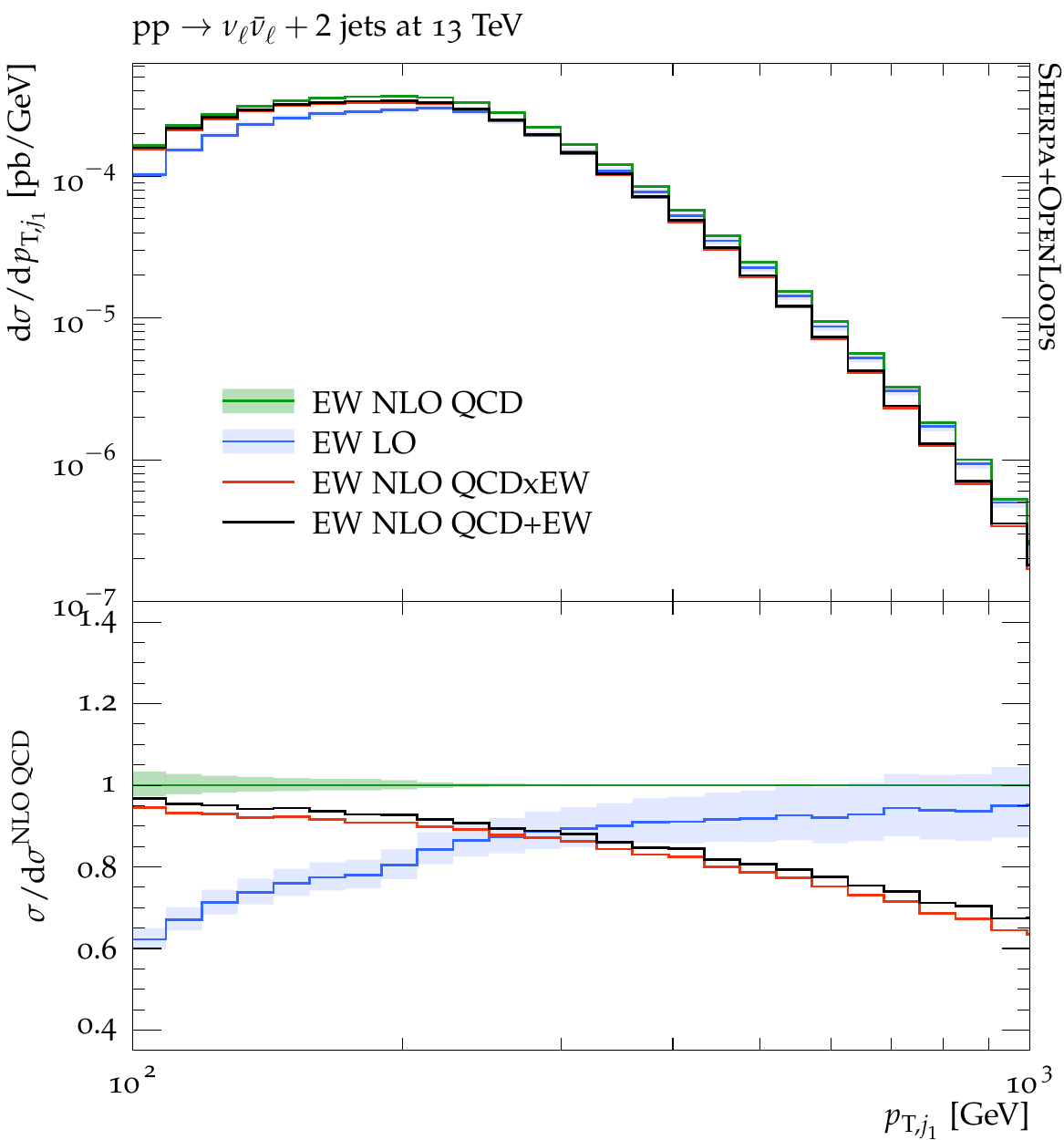}
	\plotsep
	\includegraphics[width=0.45\textwidth]{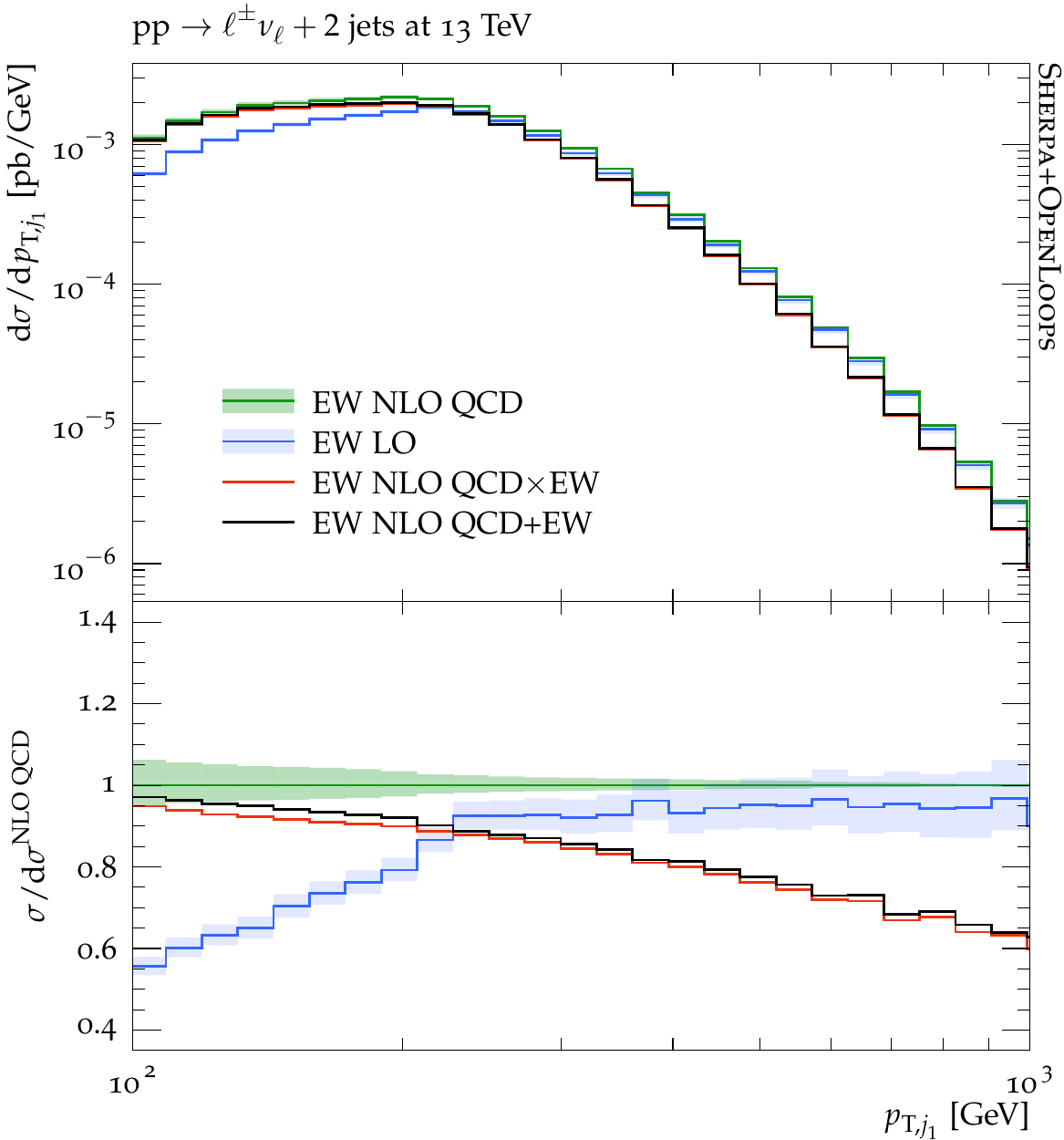}
\caption{Distribution in the transverse momentum of the hardest jet, \pTjone,
for
EW \ppZnnjj (left) 
and
EW \ppWenjj (right). 
Curves and bands as in Fig. \ref{fig:nom_EW_pTV}.}
\label{fig:nom_EW_pTj1}
\end{figure}

Higher-order QCD and EW corrections to the transverse momentum distribution of the hardest
jet, $p_{\rT,\jet_1}$, are shown in Fig. \ref{fig:nom_EW_pTj1}. Here the QCD corrections are largest at small \pTjone and decrease in the tail. For \pTjone above a few hundred GeV NLO QCD corrections drop below $10\%$.
The NLO EW corrections increase logarithmically at large \pTjone and reach $-40\%$ at 1\,TeV. Due to the smallness of the higher-order QCD corrections in the
tail, differences between additive and multiplicative combinations are negligible. Again a very high degree of correlation of the higher-order corrections is observed between the two processes.

In Figs.~\ref{fig:nom_EW_mjj} and \ref{fig:nom_EW_mjj_jv} we turn to
the distribution
in the invariant mass between of two leading jets, $\mjj$, defined 
inclusively and with an additional dynamic veto on central jet activity
as introduced in Sect.\ \ref{se:cutsnadobs}, respectively.
These distributions are crucial for background estimations in
invisible-Higgs searches.  For the jet-inclusive distributions higher-order
QCD and EW corrections are highly correlated between the two considered
processes with differences at the $5\%$ level for the QCD corrections at
small \mjj.  At LO QCD, scale uncertainties increase with
\mjj and reach 20--30\% in the multi-TeV range.  At NLO QCD, 
scale uncertainties are reduced to the $1\%$ level all the way up to the
multi-TeV regime.
Overall, the NLO QCD corrections have a marked impact on the shape of the
\mjj distribution, ranging from $+70\%$ at small \mjj to about $+5\%$ above
2\,TeV.  At the same time, NLO EW corrections are negative and increase
towards the \mjj tail.  However, they remain smaller compared to the
corresponding corrections in \pTV or \pTjone.  
This is due to the fact that, at very large \mjj,
the Mandelstam invariants $\hat t$ and $\hat u$ are 
much smaller as compared to $\hat s \sim \mjj$. As a consequence
the double Sudakov logarithms $\ln^2(|\hat r|/M_W^2)$ with $\hat r=\hat t,\hat u$
are significantly suppressed with respect to $\ln^2(\hat s/M_W^2)$.
%
At $\mjj=5$\, TeV the EW corrections amount to about $20\%$ and differences between an additive and a multiplicative combinations of QCD and EW corrections remain at $1\%$ level. 
The dynamic central jet veto has marked impact on the NLO QCD corrections, 
in particular in the small \mjj region. Here, the corrections are reduced to 
about $+20\%$ for \Zjj production, and turn negative to about $-20\%$ for \Wjj production. The jet veto has a much smaller effect in the TeV regime. Here the QCD corrections for both \Znn and \Wen production are at the percent level only.
Unsurprisingly, the EW corrections are hardly effected by the central jet veto.

\begin{figure}[t!]
\centering
	\includegraphics[width=0.45\textwidth]{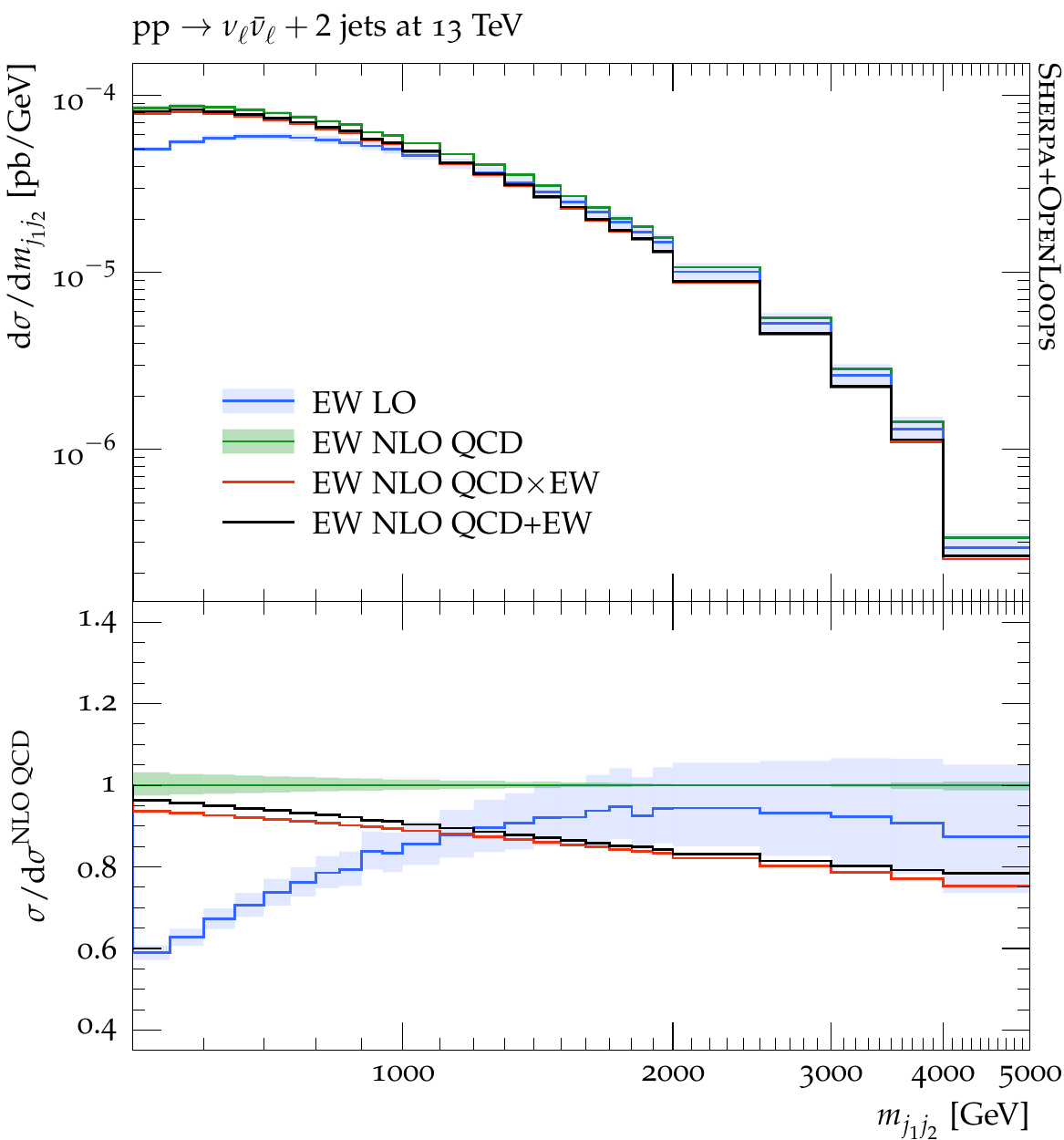}
	\plotsep
	\includegraphics[width=0.45\textwidth]{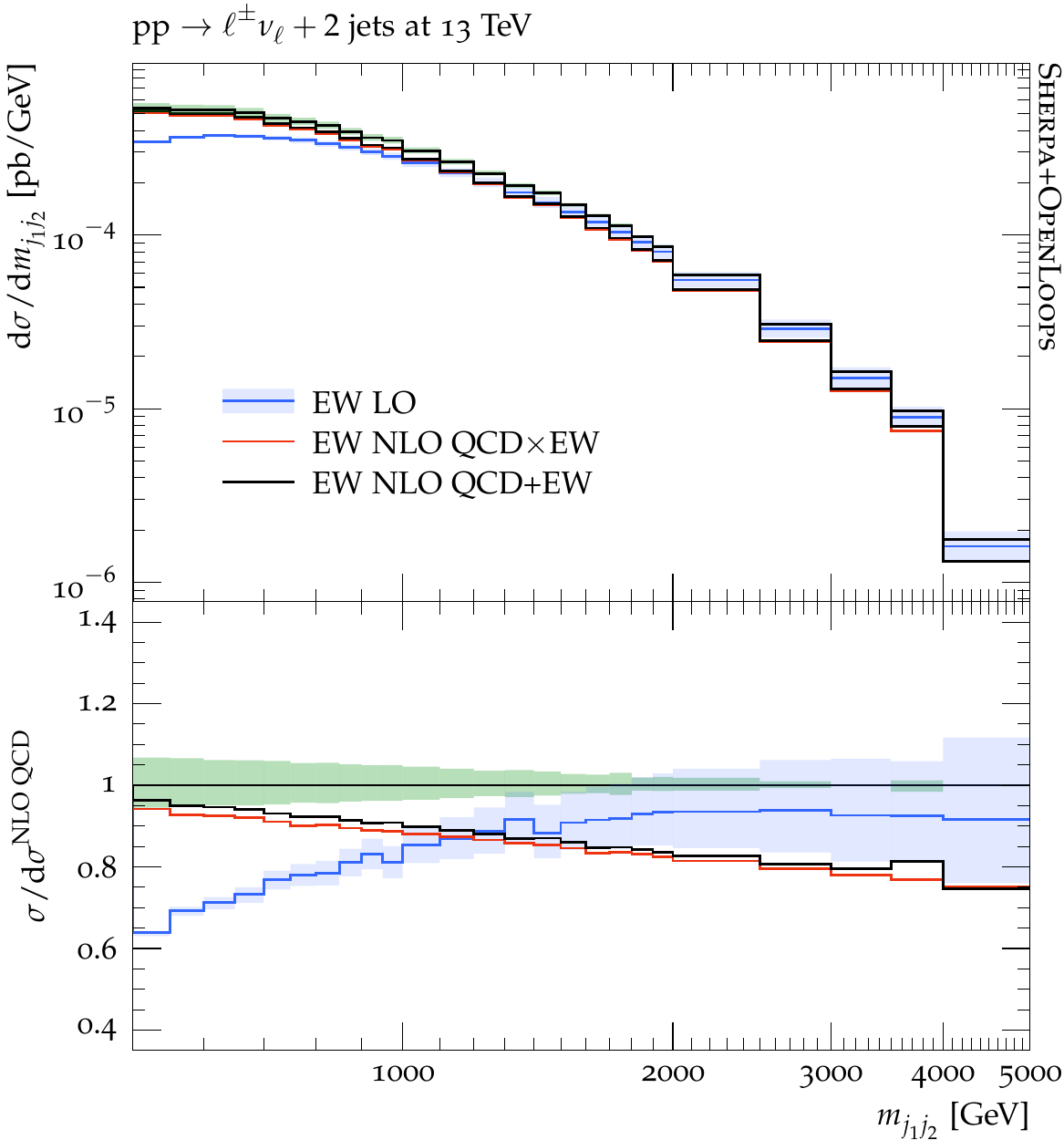}
\caption{Distribution in the invariant mass of the two hardest jets, \mjj, for 
EW \ppZnnjj (left) 
and
EW \ppWenjj (right). 
Curves and bands as in Fig. \ref{fig:nom_EW_pTV}.}
\label{fig:nom_EW_mjj}
\end{figure}

\begin{figure}[t!]
\centering
	\includegraphics[width=0.45\textwidth]{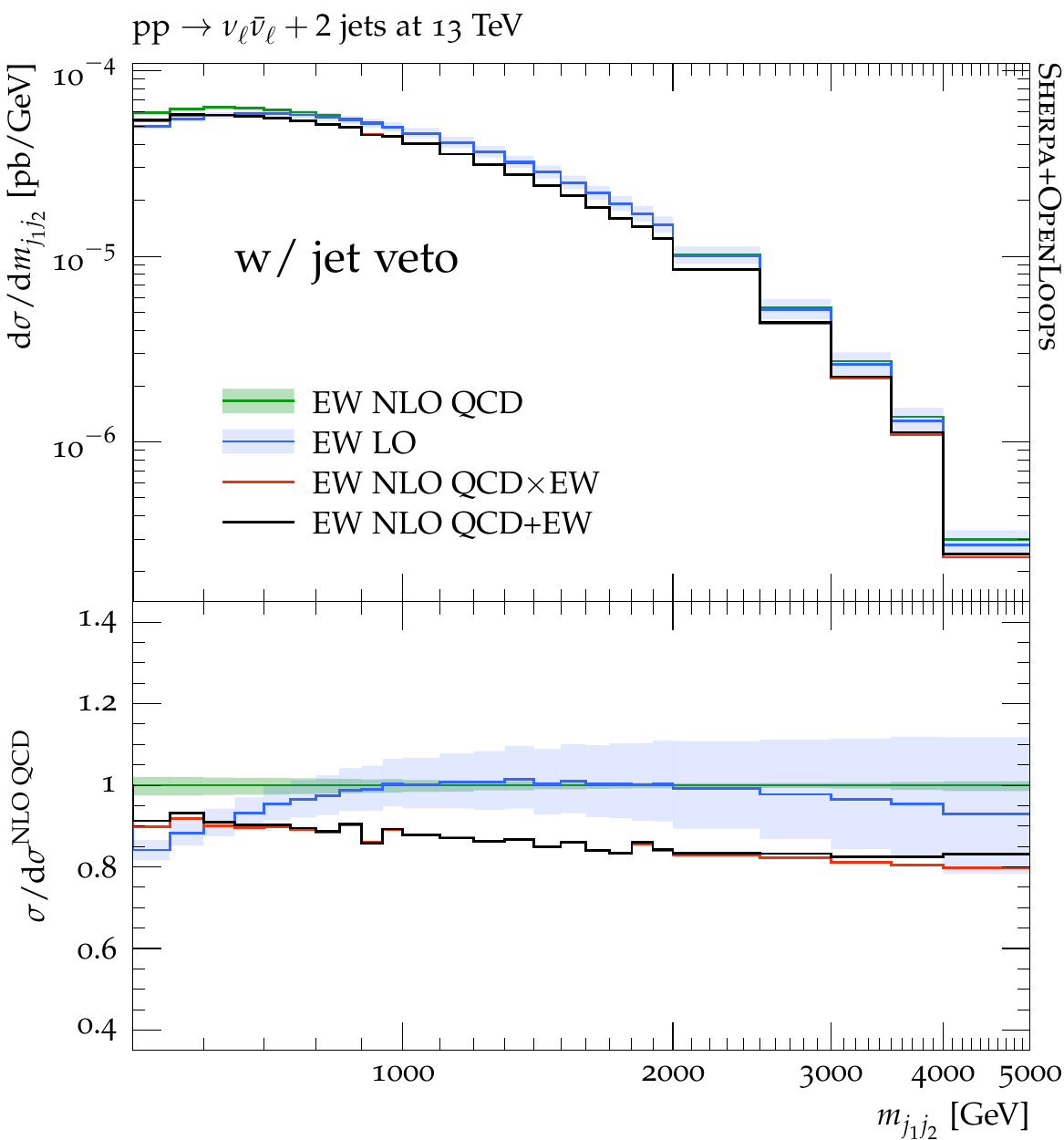}
	\plotsep
	\includegraphics[width=0.45\textwidth]{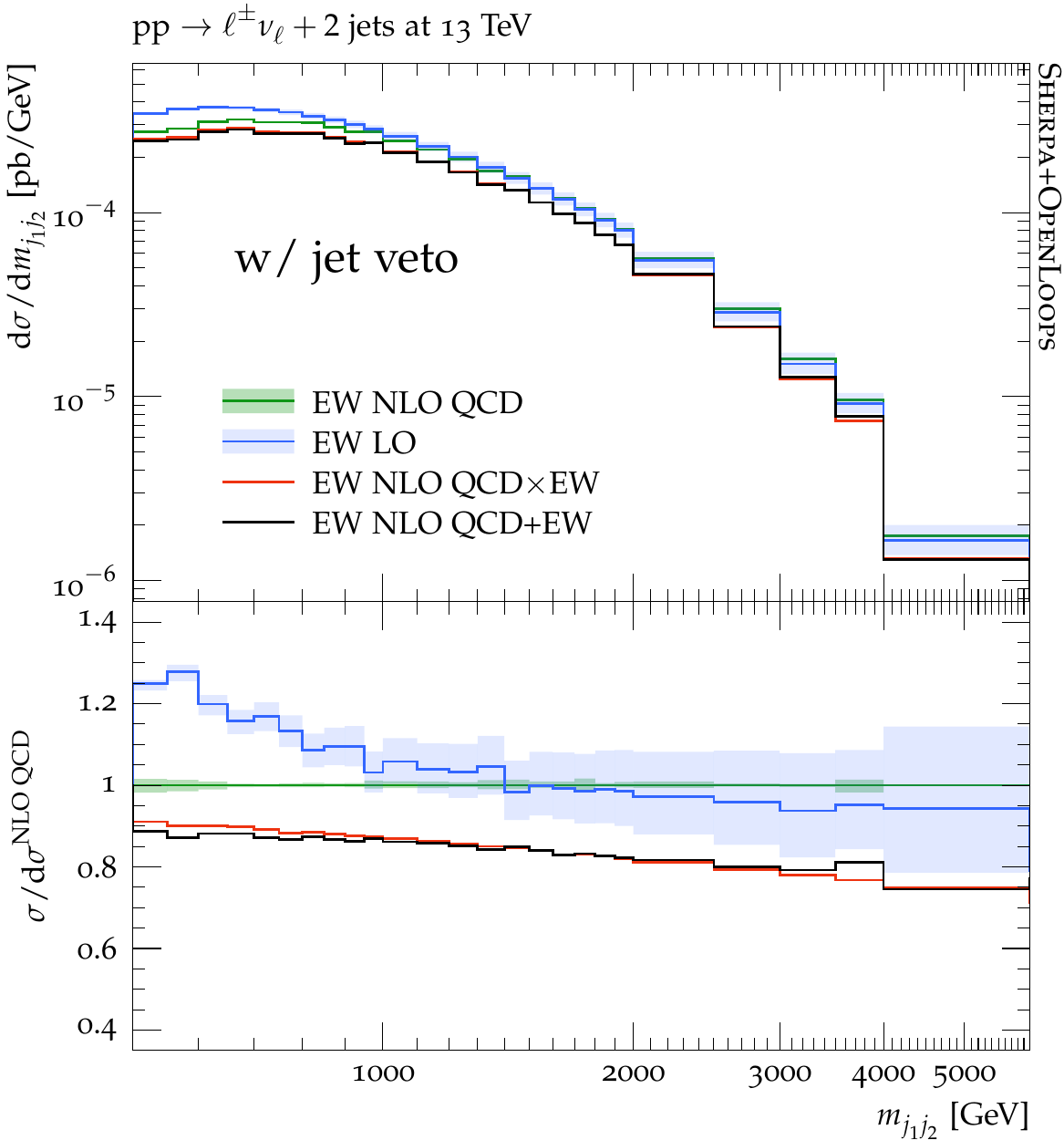}
\caption{Distribution in the invariant mass of the two hardest jets, \mjj, for 
EW \ppZnnjj (left) 
and
EW \ppWenjj (right) subject to the dynamic third jet veto of Eq.~\eqref{eq:modveto}. 
Curves and bands as in Fig. \ref{fig:nom_EW_pTV}.}
\label{fig:nom_EW_mjj_jv}
\end{figure}

\begin{figure}[t!]
\centering
	\includegraphics[width=0.45\textwidth]{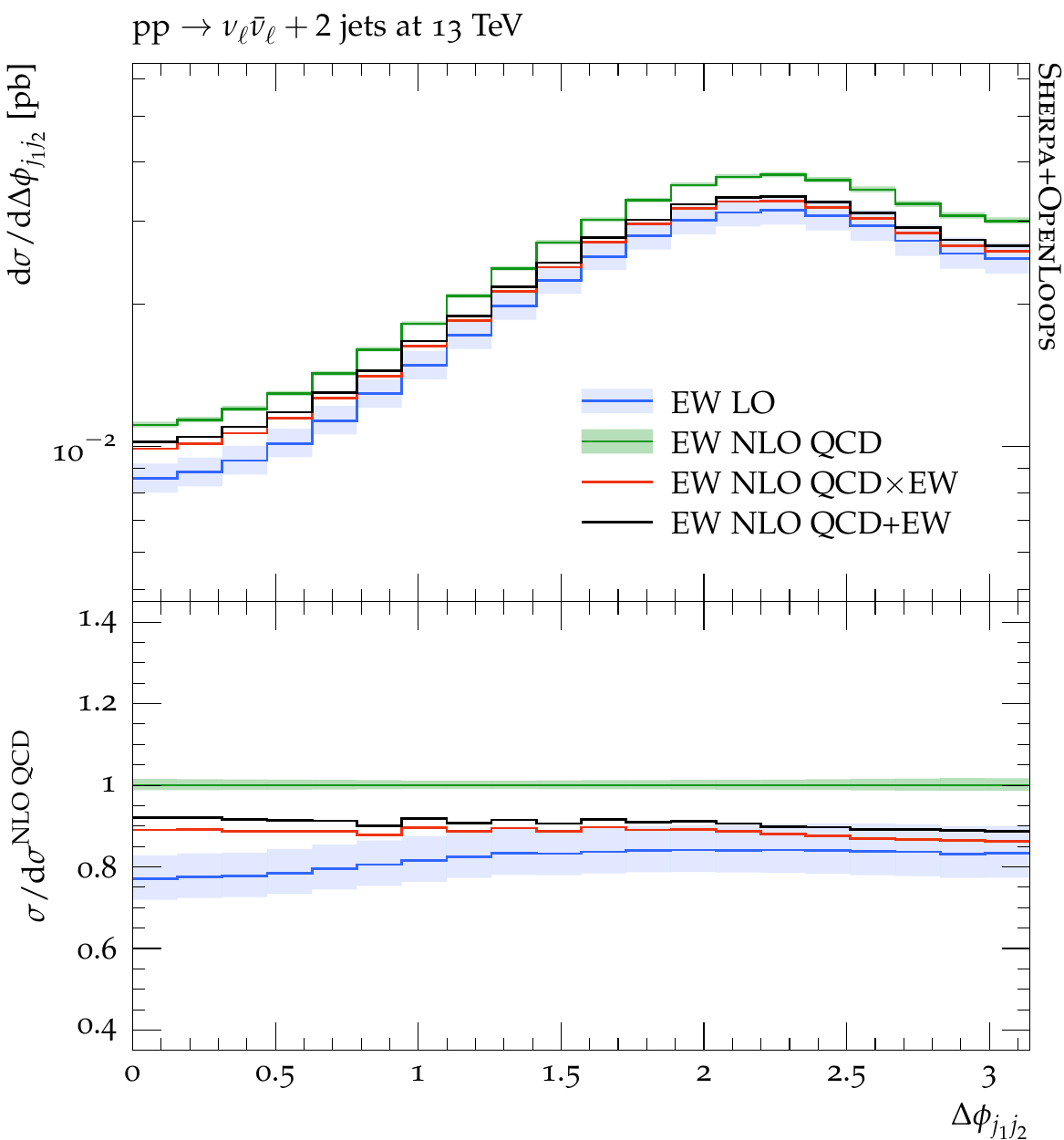}
	\plotsep
	\includegraphics[width=0.45\textwidth]{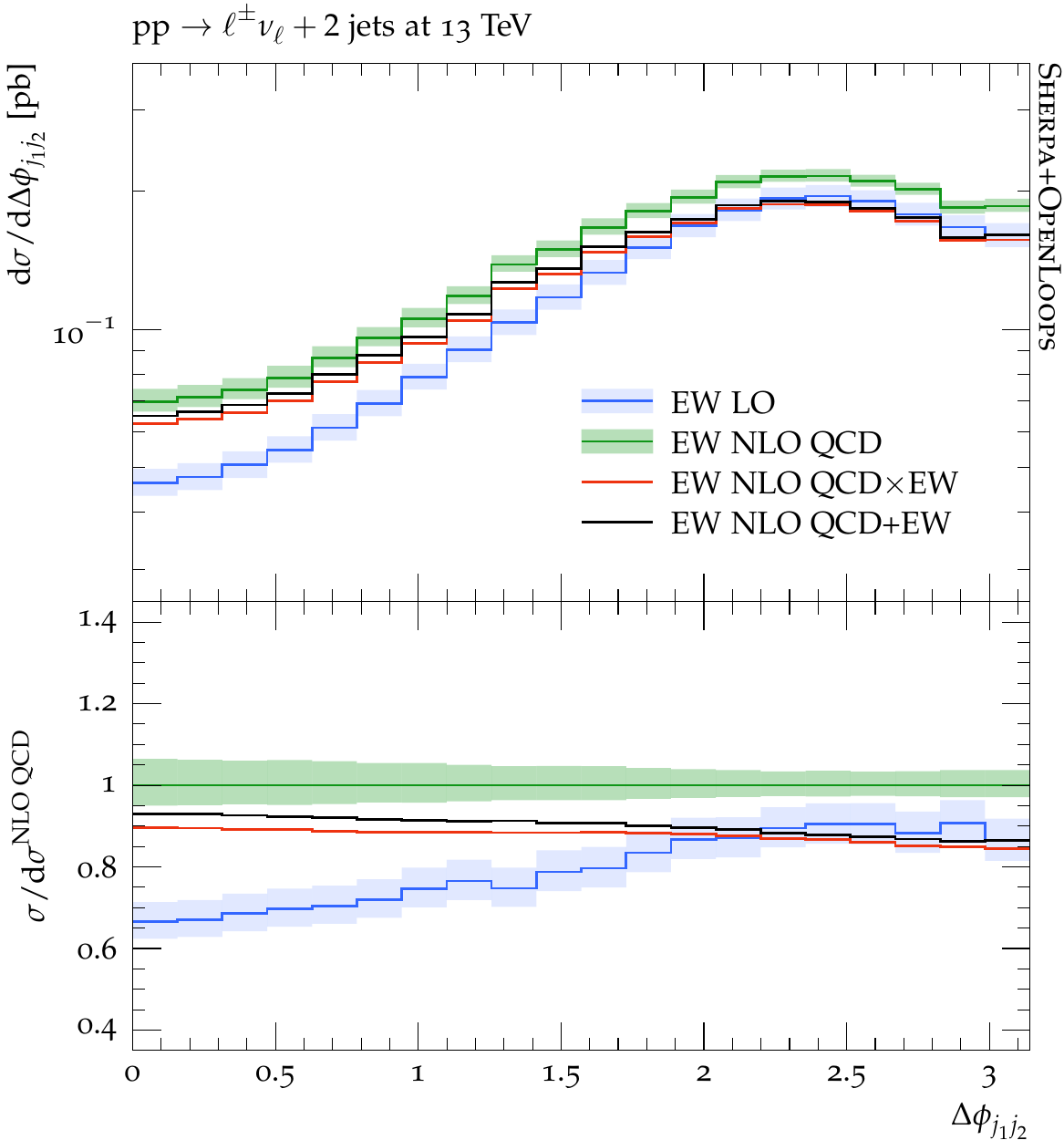}
\caption{Distribution in the azimuthal separation of the two hardest jets, \deltaphijj, 
for EW \ppZnnjj (left) 
and
EW \ppWenjj (right). 
Curves and bands as in Fig. \ref{fig:nom_EW_pTV}.}
\label{fig:nom_EW_dphi}
\end{figure}

In Fig.~\ref{fig:nom_EW_dphi} we plot the differential distribution in the azimuthal separation of the two hardest jets,
\deltaphijj.
 In this observable EW corrections are at the $10\%$ level with hardly any variation across the \deltaphijj range. QCD corrections on the other hand show a mild increase towards smaller
\deltaphijj.
Interestingly, in this region the QCD corrections also show a
non-universality between the two considered processes at the $10\%$ level. 
This non-universality 
can be attributed to the following two mechanisms.
The first one is single-top production, which enters 
only $pp\to$\Wenjj in the form of $s$- and $t$-channel
contributions at LO and also associated $Wt$ production at NLO QCD
(see Sect.~\ref{sec:modes}).
The second mechanism consists of $s$-channel contributions that correspond
to diboson subprocesses of type $q\bar q' \to VV'$, where one of the weak bosons
decays into two jets.
At $\mjj > M_{W,Z}$, such diboson channels 
can contribute through 
hard initial-state radiation, which 
plays the role of one of the two hardest jets.
Their non-universality 
is due to the fact that the QCD corrections to $W^{\pm}Z$
production are much larger as compared to 
$W^+W^-$ and $ZZ$ production.  
Both mechanisms tend to enhance \Wenjj production
at small \deltaphijj, while they tend to be suppressed at larger
\deltaphijj.
The impact of these mild non-universalities is discussed in more detail 
in Sect.~\ref{se:ratios}.



\begin{figure}[t!]
\centering
	\includegraphics[width=0.45\textwidth]{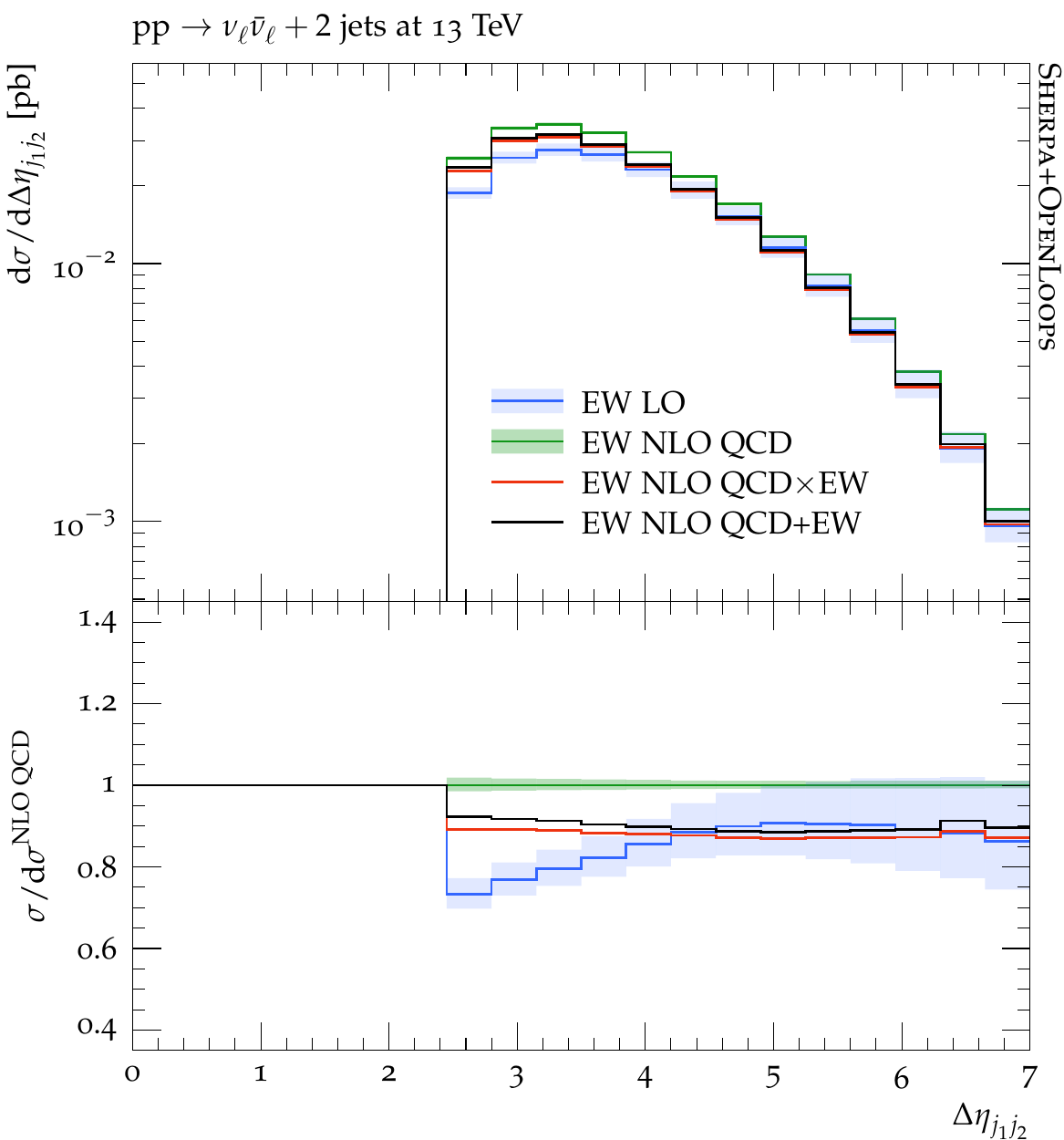}
	\plotsep
	\includegraphics[width=0.45\textwidth]{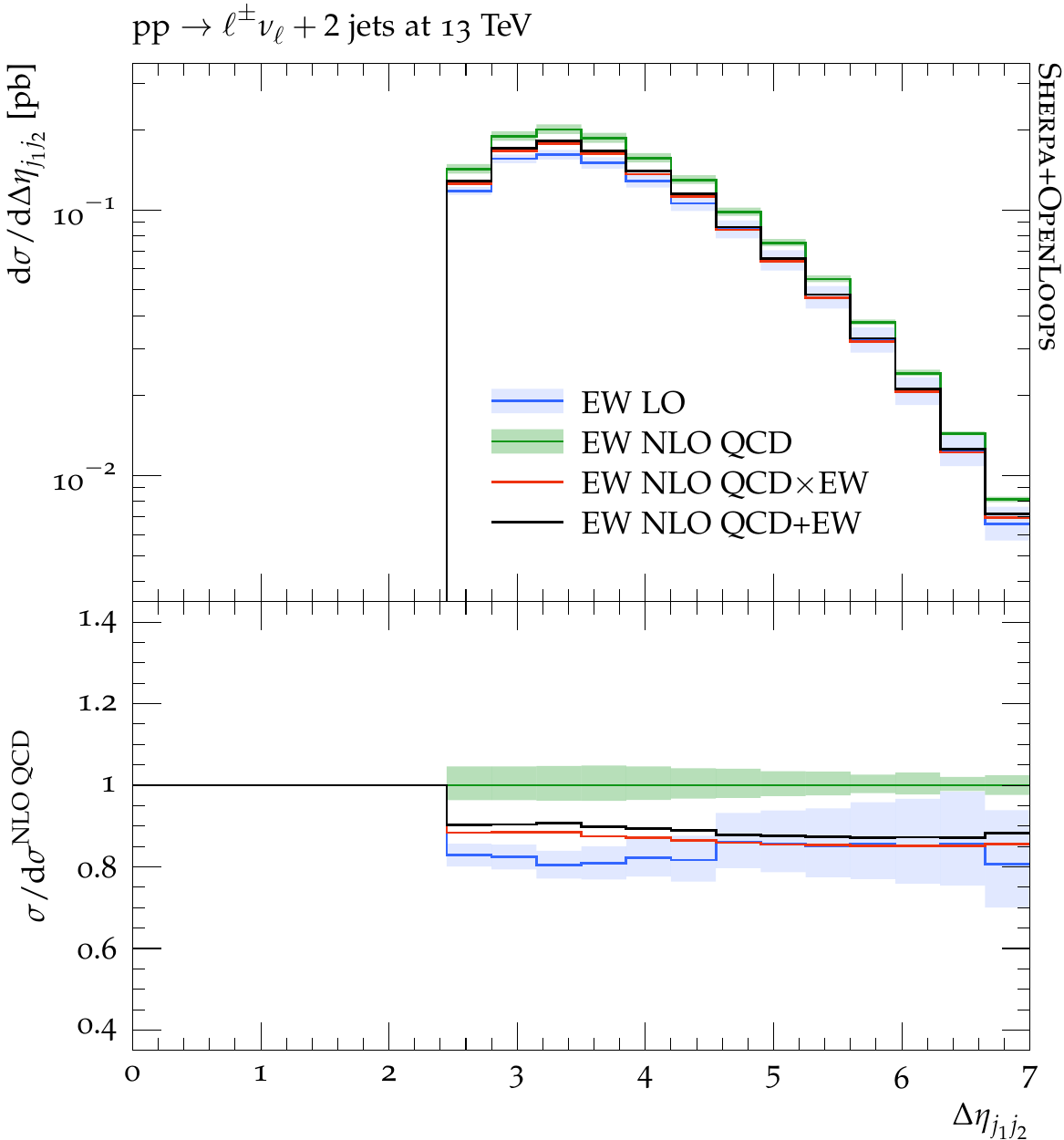}
\caption{Distribution in the rapidity separation of the two hardest jets, \deltaetajj,
for EW \ppZnnjj (left) 
and
EW \ppWenjj (right). 
Curves and bands as in Fig. \ref{fig:nom_EW_pTV}.}
\label{fig:nom_EW_deta}
\end{figure}

Finally, in Fig.~\ref{fig:nom_EW_deta} we consider the distribution in the rapidity  
separation of the two hardest jets, \deltaetajj. 
Also in this case the EW corrections are almost constant and at the level of $10\%$. For the \Znn channel also the QCD corrections are constant and at the level of $20\%$. For the \Wen channel the QCD corrections increase up to $30\%$ for small rapidity separation. In actual analyses for VBF-V production and 
invisible-Higgs searches often tighter requirements on \deltaetajj than the here considered $\deltaetajj > 2.5$ are imposed. This will further increase the level of correlation between the \Wen and \Znn channels. Thus correlation uncertainties derived here and then applied with tighter \deltaetajj requirements can be seen as conservative.

\subsection{Precise predictions and uncertainties for \texorpdfstring{\Vjj}{V+2jet} ratios}
\label{se:ratios}

In this section we present predictions and theoretical uncertainties  
for the ratios of Eq.~\eqref{eq:ratios} between the $\mjj$ distributions 
in \ppZnnjj and \ppWenjj. Numerical predictions for these process ratios
and the related uncertainties can be found 
at~\cite{vjjrepo}, where also additional ratios 
between \ppZlljj and \ppWenjj distributions are available.

The $Z/W$  ratios are the key ingredients of the reweighting procedure defined in Eq.~\eqref{eq:rew}.
As nominal theory predictions 
we take the fixed-order \NLO \QCDtEW ratios
\beq 
\label{eq:nominal}
R^{\ZW,\mode}_{\TH}(x)\,:=\,R^{\ZW,\mode}_{\rm NLO\,\QCDtEW}(x)\,=\,
\frac{\parx\sigma^{Z,\mode}_{\rm NLO\,\QCDtEW}}{\parx\sigma^{W,\mode}_{\rm
NLO\,\QCDtEW}}\,,
\eeq 
where $x=\mjj$ is the dijet invariant mass. 
To describe theory uncertainties we introduce nuisance parameters that 
are directly acting on the $Z/W$ ratios, combining the uncertainties
of the individual processes and their correlations. 
With this approach, our complete predictions with uncertainties 
read
\beq 
\label{eq:fullratiounc}
R^{\ZW,\mode}_{\TH}(x,\vepsTH^{\,Z/W,\mode})\,:=
R^{\ZW,\mode}_{\TH}(x) + \sum_i 
\eps^{\ZW,\mode}_{i,\TH}\,\delta R^{\ZW,\mode}_{i,\TH}(x)\,,
\eeq 
where the nuisance parameters
$\eps^{\ZW,\mode}_{i,\TH}$ are defined as in Eq.~\eqref{eq:nuisancepar}, 
while the $\delta R^{\ZW,\mode}_{i,\TH}(x)$ factors 
embody various sources of theory uncertainty, as defined in the following.
Note that for the two $V+2\,$jet 
production modes ($M$), \ie for QCD and EW
production, we define two independent ratios and uncertainties.


To account for unknown QCD corrections beyond NLO in a conservative way,
we avoid using scale uncertainties and, 
following Ref.~\cite{Lindert:2017olm}, we handle the 
difference between LO QCD and NLO QCD ratios as
uncertainty. 
More precisely, we consider the effect of switching off
NLO QCD corrections in our nominal NLO\,\QCDtEW predictions,
\beq 
\label{eq:QCDuncert}
\delta R^{\ZW,\mode}_{\QCD}(x)\,: =\, \left\vert 
R^{\ZW,\mode}_{\rm NLO\,\EW}(x)
-
R^{\ZW,\mode}_{\rm NLO\,\QCDtEW}(x)
\right\vert\,.
\eeq  
While this approach effectively downgrades the known NLO QCD corrections to 
an uncertainty, the bulk of the QCD corrections cancel in the ratio, and the
uncertainty $\delta R^{\ZW,\mode}_{\QCD}$ remains quite small.

For parton showering and matching at NLO we apply the uncertainty
\beq 
\label{eq:PSuncert}
\delta R^{\ZW,{\mode}}_{\rm PS}(x) \,:=\, \left\vert R^{\ZW,{\mode}}_{\rm NLOPS\, \QCDtEW}(x)-
R^{\ZW,\mode}_{\rm NLO\,\QCDtEW}(x)
\right\vert\,,
\eeq  
\ie the full difference between fixed-order NLO and NLOPS predictions.

To describe the effect of unknown mixed QCD--EW uncertainties beyond NLO we 
intorduce the uncertainty
\beq 
\label{eq:QCDEWmixunc}
\delta R^{\ZW,\mode}_{\rm mix}(x) \,:=\, \left\vert R^{\ZW,\mode}_{\rm NLO\,\QCDpEW}(x)-
R^{\ZW,\mode}_{\rm NLO\,\QCDtEW}(x)
\right\vert\,,
\eeq  
which corresponds to the difference between the additive and multiplicative
combination of NLO QCD and NLO EW corrections. Also this prescription 
can be regarded as a conservative estimate since the multiplicative combination
is expected to provide a correct description of the dominant 
mixed QCD--EW effects beyond NLO.

In case a jet veto is applied to the experimental analysis, also  the following 
uncertainty should be considered,
\beq 
\label{eq:vetounc}
\delta R^{\ZW,\mode}_{\rm veto}(x) \,:=\, \left\vert R^{\ZW,\mode}_{\TH,\rm veto}(x)-
R^{\ZW,\mode}_{\rm NLO\,\QCDtEW}(x)
\right\vert\,,
\eeq  
where the ratio $R^{\ZW,\mode}_{\TH,\rm veto}$ is computed in the presence
of  the  ``theoretical'' veto detailed in Eq.~\eqref{eq:modveto}. Note that the 
effect of the veto cancels to a large extent in the ratio. Thus the 
prescription of Eq.~\eqref{eq:modveto} does not need to be identical to the 
veto that is employed in the experimental analysis.

In order to account for the non-negligible $\deltaphijj$ dependence of 
QCD higher-order effects in the EW production modes (see
Sect.~\ref{se:EWratios}), we split the phase space into the three 
 $\deltaphijj$ bins $\Phi_{i}$ defined in Eq.~\eqref{eq:moddphi}, 
and we define the uncertainty
\beq 
\label{eq:Dphiunc}
\delta R^{\ZW,\mode}_{\Delta\phi}(x) \,:=\sqrt{\, 
\sum\limits_{i=1}^3 \left(
 R^{\ZW,\mode}_{\rm NLO\,\QCDtEW}(x)\Big|_{\Phi_i}
- R^{\ZW,\mode}_{\rm NLO\,\QCDtEW}(x)
\right)^2}\,,
\eeq  
where the first ratio between brackets is restricted to the 
$\Phi_i$ bin.  In other words, the variance of the ratio of Eq.~\eqref{eq:nominal}
in  $\deltaphijj$ space is taken as uncertainty of the
one-dimensional reweighting procedure.

Finally, also PDF uncertainties should be considered. 
In this case, PDF variations in the numerator and denominator of the
$Z/W$ ratio should be correlated.
In the following subsections we present predictions for the
ratios defined in Eq.~\eqref{eq:nominal} and for the various ingredients that enter 
the theoretical uncertainties of Eqs.~\eqref{eq:fullratiounc}--\eqref{eq:Dphiunc}.

\subsubsection{\texorpdfstring{$Z/W$}{Z/W} ratios for the QCD production mode}
\label{se:QCDratios}

\begin{figure}[t!]
\centering
	\includegraphics[width=0.45\textwidth]{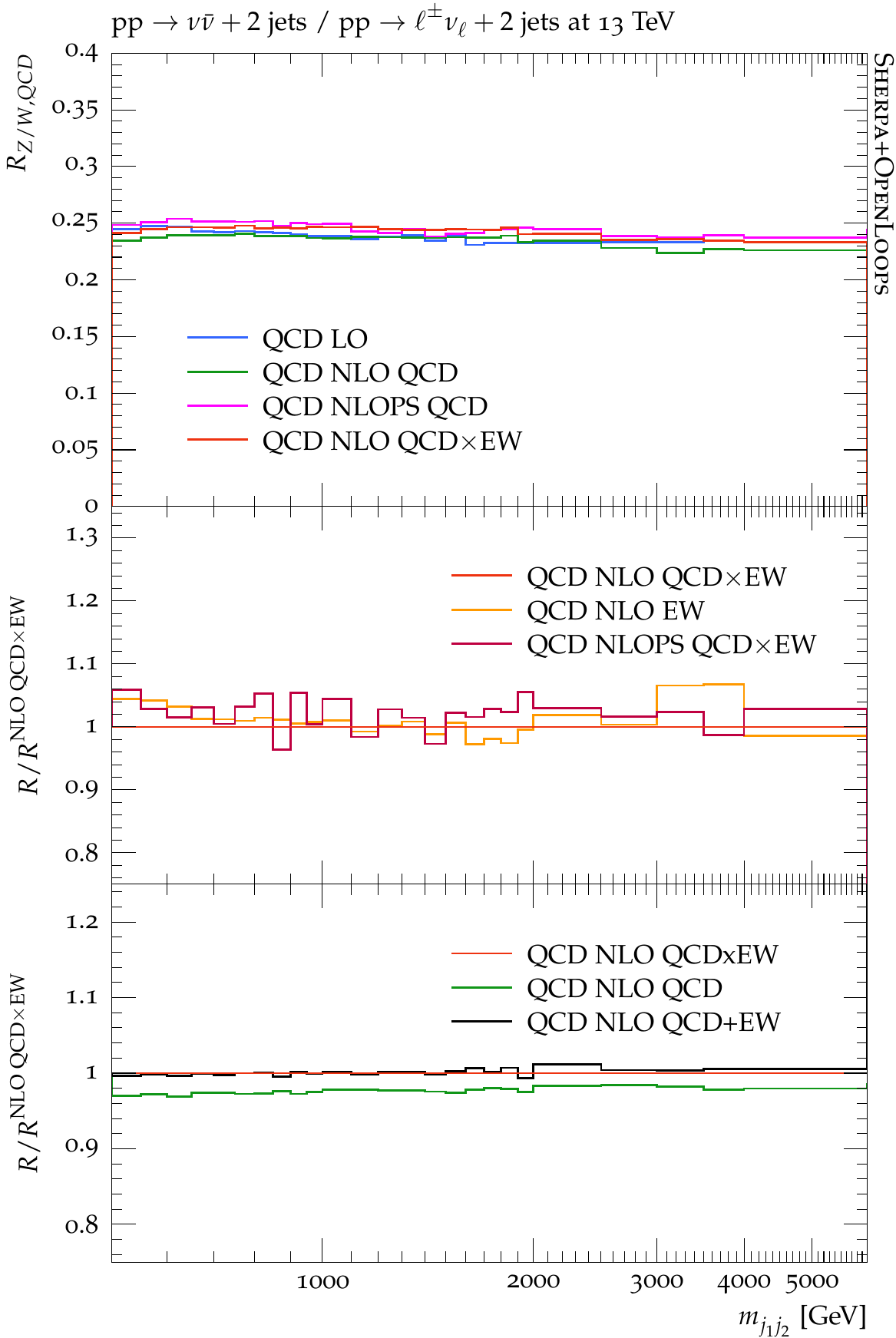}
	\plotsep
	\includegraphics[width=0.45\textwidth]{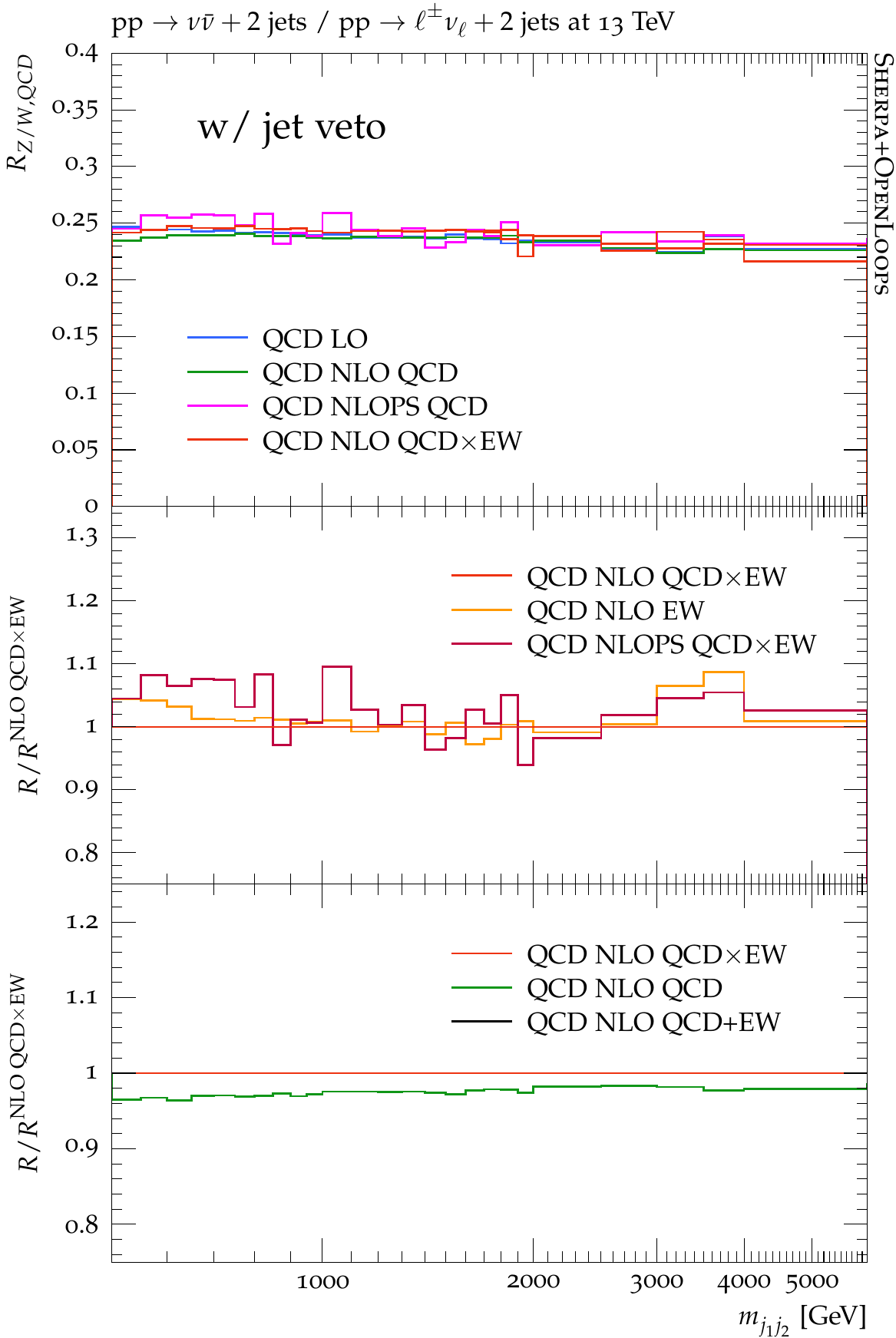}
\caption{%
Ratios of the QCD \ppZnnjj and QCD \ppWenjj 
distributions in \mjj inclusive (left) and in the presence of the
dynamic veto of Eq.~\eqref{eq:modveto} against a third jet (right).
The upper panels compare absolute predictions at LO (blue), NLO QCD (green), NLOPS QCD
(magenta) and NLO QCD$\times$EW (red) accuracy.
The impact of QCD corrections is illustrated in the middle panel, 
which shows the relative variation wrt the nominal 
NLO \QCDtEW prediction (red) when
switching on the parton shower
(NLOPS \QCDtEW, purple) or switching off QCD corrections
(NLO EW, orange).
Similarly, the lowest panel shows the relative effect of switching 
off EW corrections (NLO QCD, green) or 
replacing the multiplicative by the additive combination of 
QCD and EW corrections (NLO QCD+EW, black). 
} 
\label{fig:ratio_QCD_mjj} 
\end{figure}

\begin{figure}[t!]
\centering
	\includegraphics[width=0.45\textwidth]{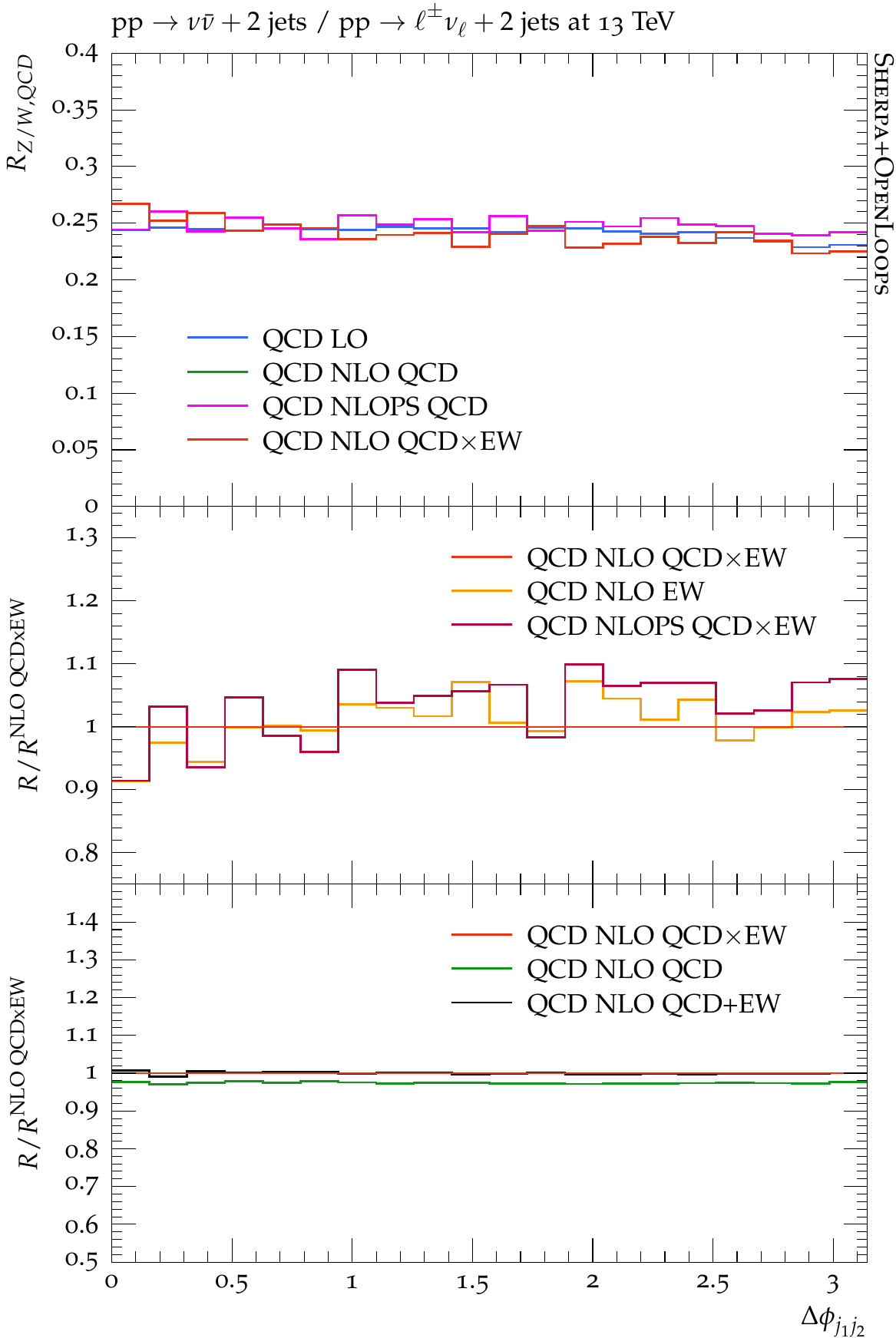}
\caption{%
Ratios of the QCD \ppZnnjj and QCD \ppWenjj 
distributions in \deltaphijj.
The same higher-order predictions and conventions as in
Fig.~\ref{fig:ratio_QCD_mjj} are used.
} 
\label{fig:ratio_QCD_dphijj} 
\end{figure}

As observed in Sect.~\ref{sec:nomQCD}, the QCD and EW corrections to
the QCD production modes of the individual $\Zjj$ and $\Wjj$ processes 
are strongly correlated.
This is confirmed by the smallness of the corrections in the 
$Z/W$ ratios shown in
Figs.~\ref{fig:ratio_QCD_mjj}--\ref{fig:ratio_QCD_dphijj}.

The left and right plots of 
Fig.~\ref{fig:ratio_QCD_mjj} present the ratio of 
$\mjj$-distributions with the inclusive selection cuts, defined in Eq.~\eqref{eq:nominalcuts},
and in the presence of the additional jet veto, defined in Eq.~\eqref{eq:modveto}.
The value of the ratio is around 0.24 and remains 
almost constant in the considered \mjj range from 500 to 5000\,GeV.
The size of the QCD and PS corrections that enter the 
uncertainties of Eqs.~\eqref{eq:QCDuncert}--\eqref{eq:PSuncert}
is shown in the middle panels. In the inclusive selection
NLO~\QCD corrections to the ratio remain below $4-6\%$ in the entire \mjj
range, and PS corrections remain below $6\%$.  
When the jet veto is applied, the $Z/W$ ratio remains stable 
at the percent level. The jet-veto uncertainty of Eq.~\eqref{eq:vetounc}
is thus quite small.
Also QCD and PS corrections to the ratio are
largely insensitive to the jet veto. 

As shown in the lowest frames of Fig.~\ref{fig:ratio_QCD_mjj}, the EW
corrections to the QCD $Z/W$ ratio are around 2\% and almost independent of
\mjj, both for the inclusive selection and including a jet veto. Due to the strong
cancellation of QCD and EW corrections in the ratio, 
the difference between the additive and multiplicative NLO QCD--EW
combinations, which enters
the uncertainty of Eq.~\eqref{eq:QCDEWmixunc}, is completely negligible.

In Fig.~\ref{fig:ratio_QCD_dphijj} we present the QCD $Z/W$
ratio for the distributions in~\deltaphijj without applying a jet veto.
Note that, as a result of the acceptance cuts, Eq.~\refeq{eq:nominalcuts}, 
these~\deltaphijj distributions are dominated by events 
with $500\,\GeV < \mjj < 1500\,\GeV$. The results thus feature a 
very small dependence on~\deltaphijj, both for the nominal ratio, as well as
for the individual corrections. This observation supports the one-dimensional
$\mjj$-reweighting procedure 
proposed in Sect.~\ref{se:reweighting}, 
and the \deltaphijj-uncertainty of Eq.~\eqref{eq:Dphiunc}
can be neglected for the QCD production modes.

\subsubsection{\texorpdfstring{$Z/W$}{Z/W} ratios for the EW production mode}
\label{se:EWratios}

\begin{figure}[t!]
\centering
	\includegraphics[width=0.45\textwidth]{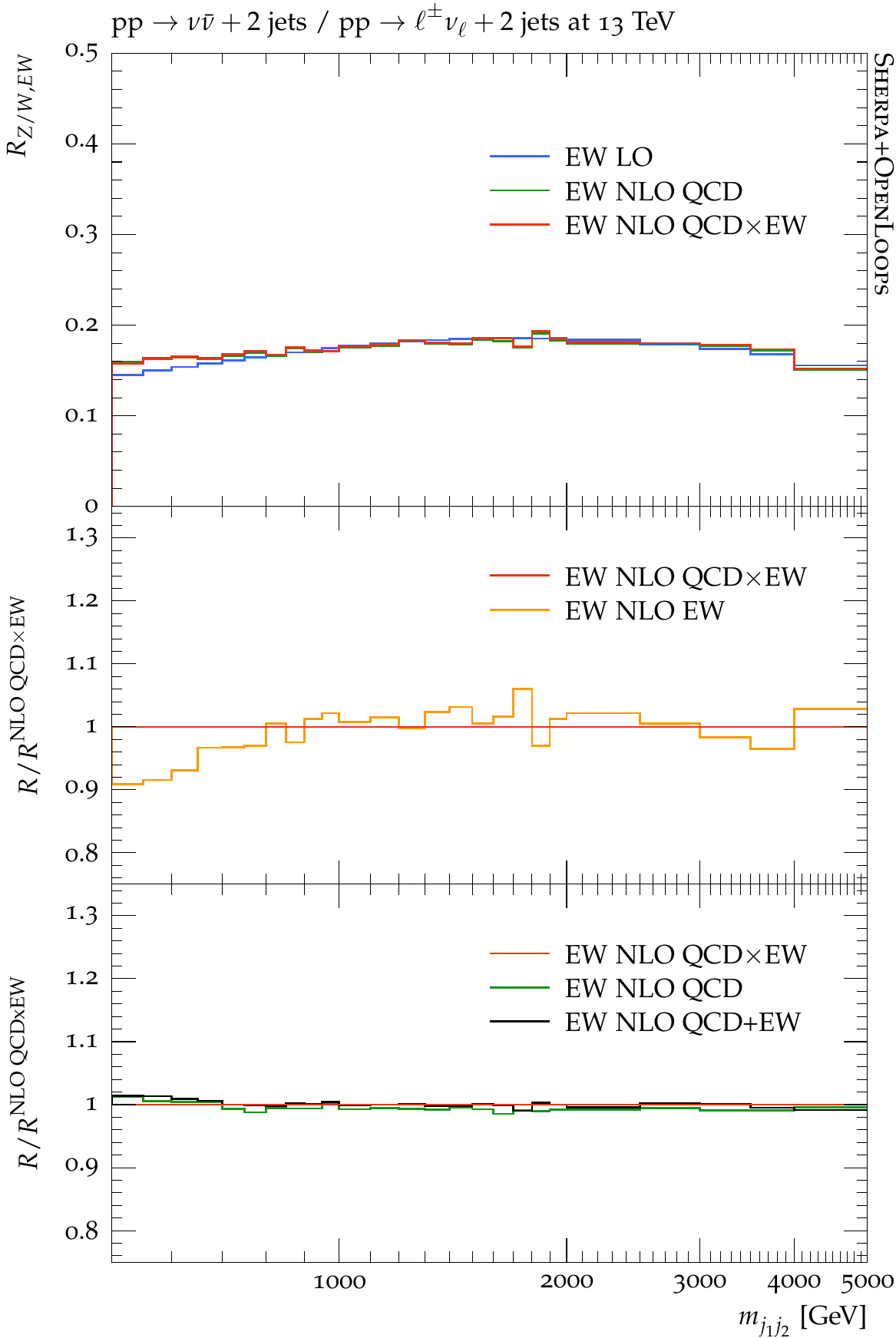}
	\plotsep
	\includegraphics[width=0.45\textwidth]{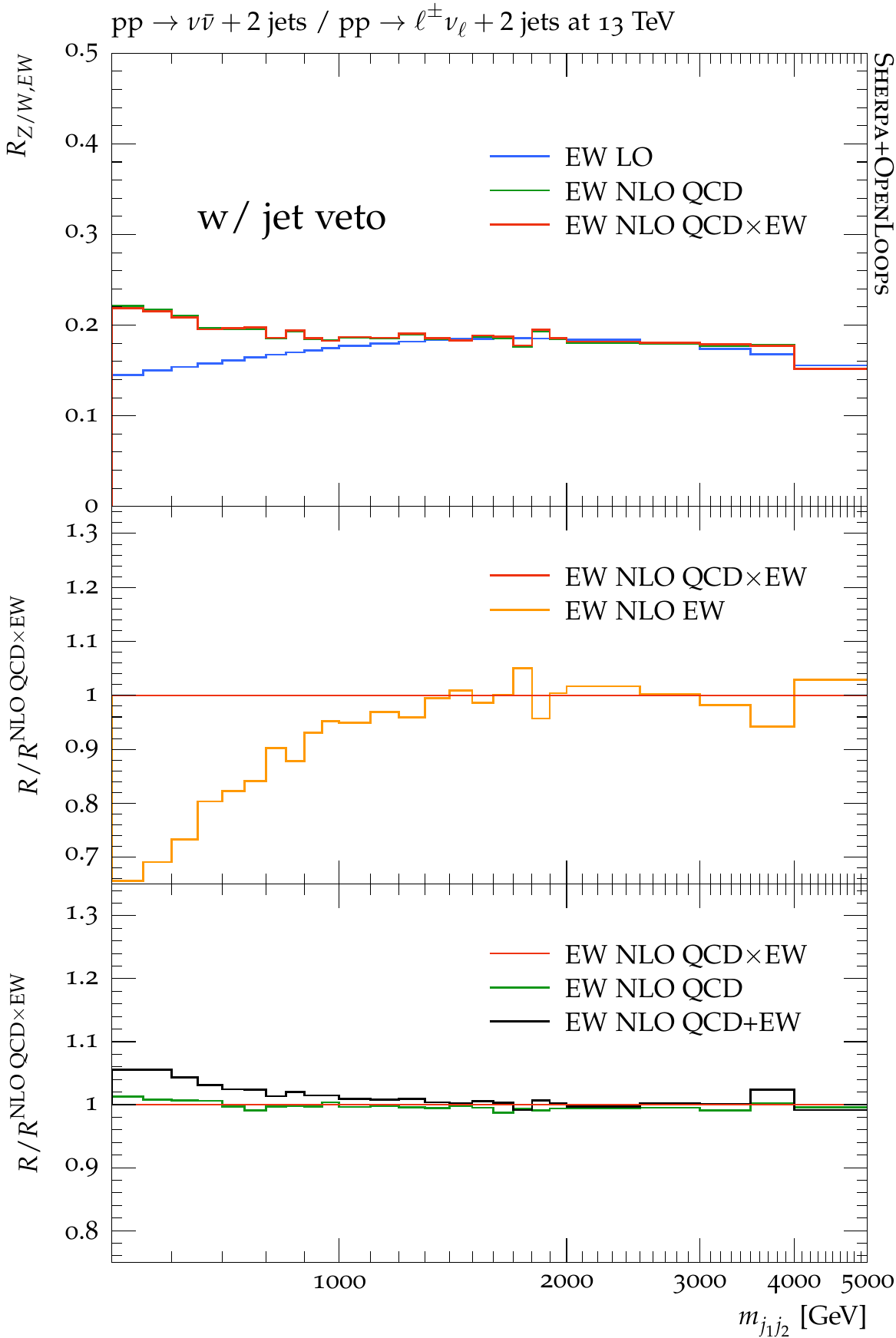}
\caption{%
Ratios of the EW \ppZnnjj and EW \ppWenjj 
distributions in \mjj  inclusive (left) and in the presence of the
dynamic veto of Eq.~\eqref{eq:modveto} against a third jet (right).
Same higher-order predictions and conventions as in
Fig.~\ref{fig:ratio_QCD_mjj}, but without 
matching to the parton shower. 
\\
} 
\label{fig:ratio_EW_mjj}
\end{figure}

\begin{figure}[t!]
\centering
\includegraphics[width=0.45\textwidth]{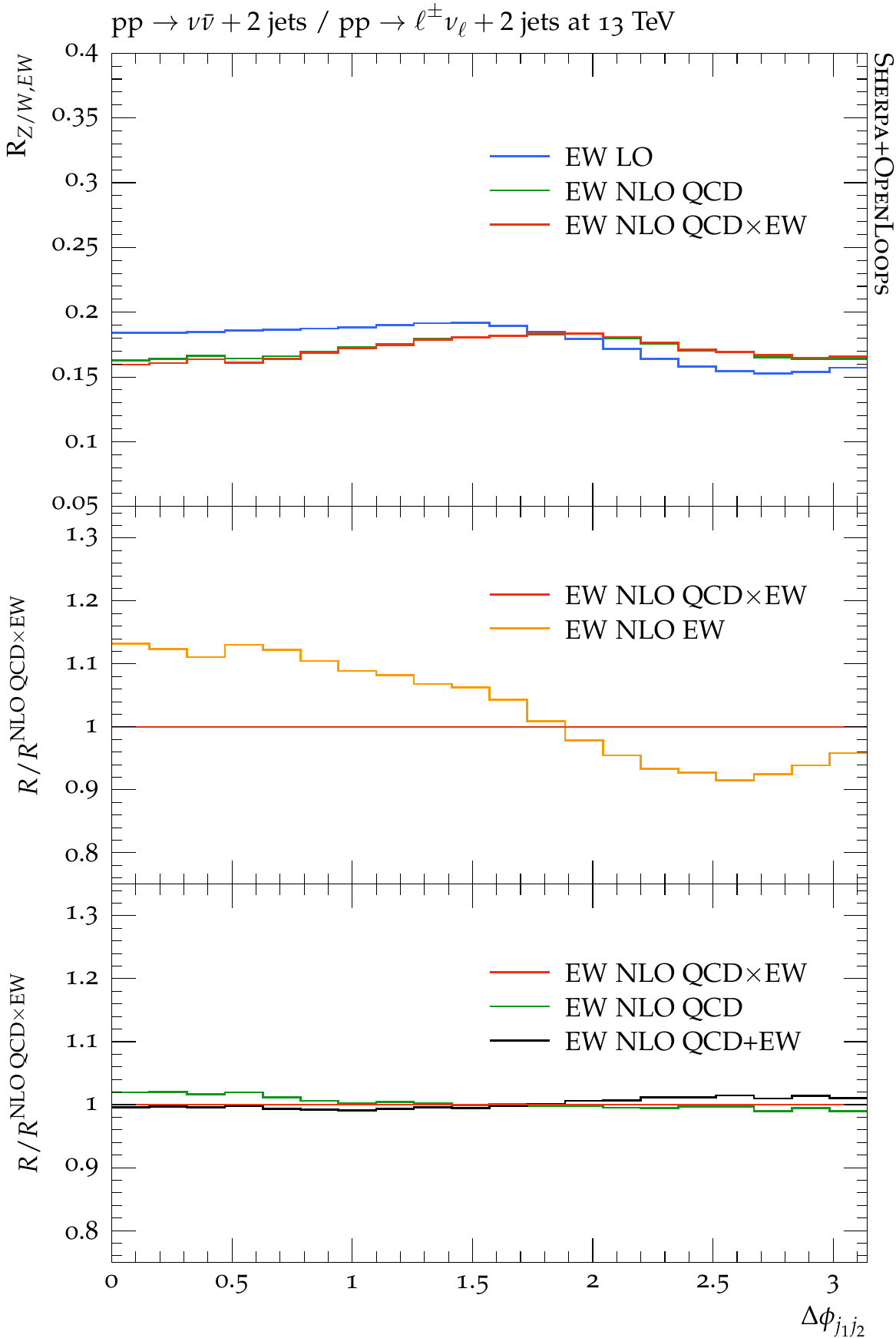}
\caption{%
Ratios of the EW \ppZnnjj and EW \ppWenjj 
distributions in \deltaphijj 
without  jet veto.
Same higher-order predictions and conventions as in
Fig.~\ref{fig:ratio_QCD_mjj}.
}
\label{fig:ratio_EW_dphijj}
\end{figure}


Higher-order predictions for the ratios of distributions in 
EW $\Zjj$ and EW $\Wjj$ production are presented in
Figs.~\ref{fig:ratio_EW_mjj}--\ref{fig:ratio_EW_dphijjbinning}.
The left and right plots of 
Fig.~\ref{fig:ratio_EW_mjj} show the ratio of
$\mjj$-distributions with inclusive selection cuts
and in the presence of the additional jet veto.
The EW $Z/W$ ratio is around 0.15 and remains rather stable 
when $\mjj$ grows from 500\,GeV to 5\,TeV.

In the absence of the jet veto, 
as expected from the findings of Sect.~\ref{sec:nomEW}, 
the ratio is quite stable with respect to higher-order corrections.  
In particluar, for $\mjj>1$\,TeV,  which corresponds to the 
most relevant region for invisible-Higgs searches,
QCD corrections are at the percent level.
Below 1\,TeV the QCD corrections tend to become more significant reaching $+10\%$
at $\mjj=500$\,GeV.
The impact of EW corrections on the inclusive \mjj-ratio does not 
exceed $1\%$ in the plotted \mjj range, and the 
mixed QCD-EW uncertainties of Eq.~\eqref{eq:QCDEWmixunc} 
are negligible.

In the presence of the jet veto, QCD corrections
become rather sizeable below 1\,TeV and reach the level of 
+50\% at 500\,GeV. As a consequence, also mixed QCD--EW 
uncertainties are somewhat enhanced. 
This non-universal behaviour of the QCD corrections 
leads to an enhancement of the QCD uncertainty, as defined in Eq.~\eqref{eq:QCDuncert}.
However, we note that the non-universality of the 
EW production modes at $\mjj<1$\,TeV 
tends to be washed out by the dominance of the
QCD production modes, where all correction effects feature a high degree of 
universality. Moreover, we point out that the prescription of Eq.~\eqref{eq:QCDuncert} 
is very conservative and may be replaced by a more realistic estimate if
QCD uncertainties play a critical role.


Together with their non-universal behaviour at $\mjj<1$\,TeV, 
the QCD corrections to the EW $Z/W$ ratio
feature also a nontrivial dependence on \deltaphijj. This is illustrated 
in Fig.~\ref{fig:ratio_EW_dphijj}, where we plot the 
ratio of the \deltaphijj distributions for 
EW $\Zjj$ and EW $\Wjj$ production. 
The \deltaphijj dependence of this ratio features variations at
the level of $20\%$ at LO and $15\%$ at NLO \QCDtEW.
The EW corrections are very small, and their
dependence on \deltaphijj does not exceed $1\%$.
In contrast, the impact of QCD
corrections on the ratio ranges from $-10\%$
at small \deltaphijj to 
$+10\%$ around $\deltaphijj=2.5$.

\begin{figure}[t!]
	\includegraphics[width=0.33\textwidth]{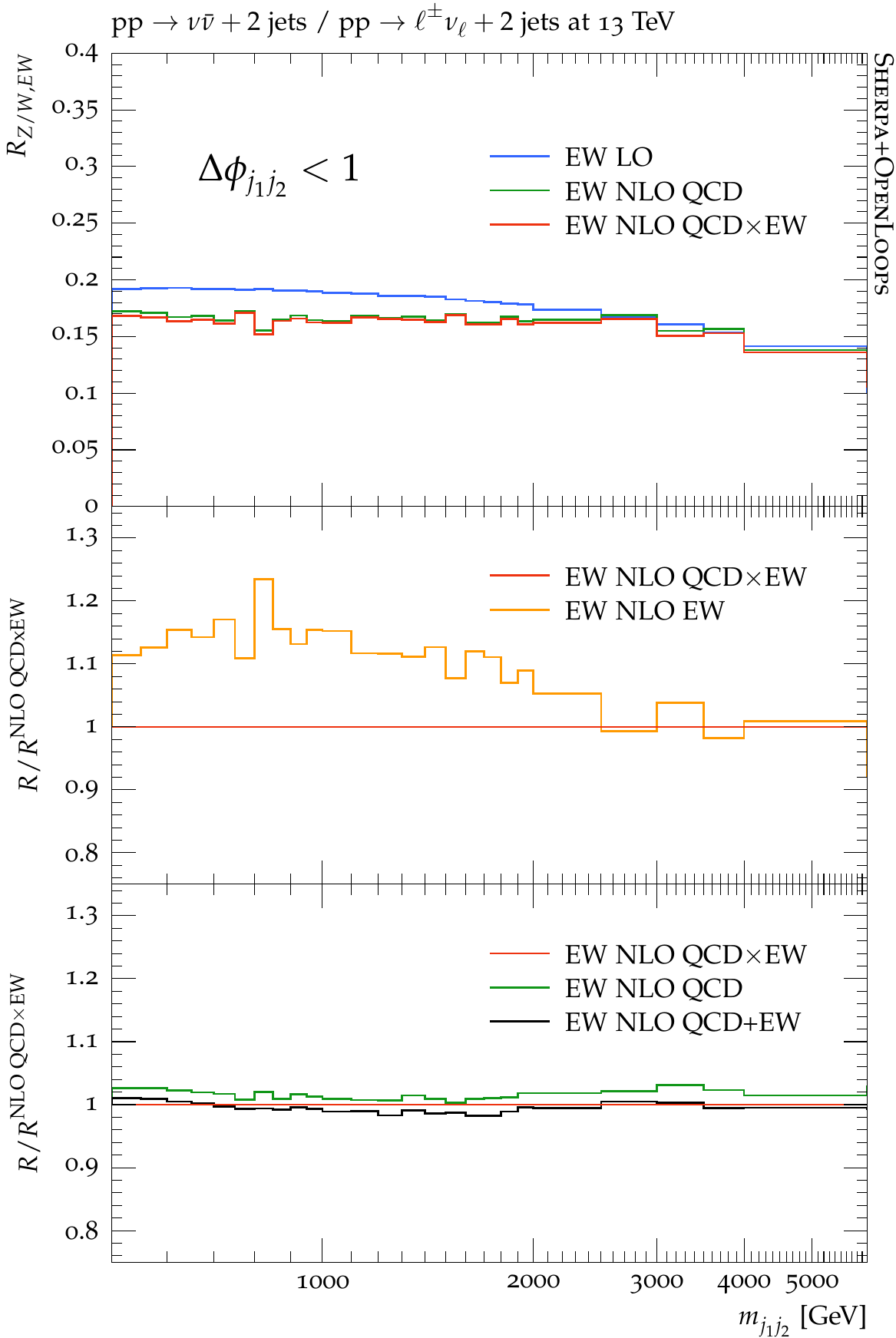}
	\includegraphics[width=0.33\textwidth]{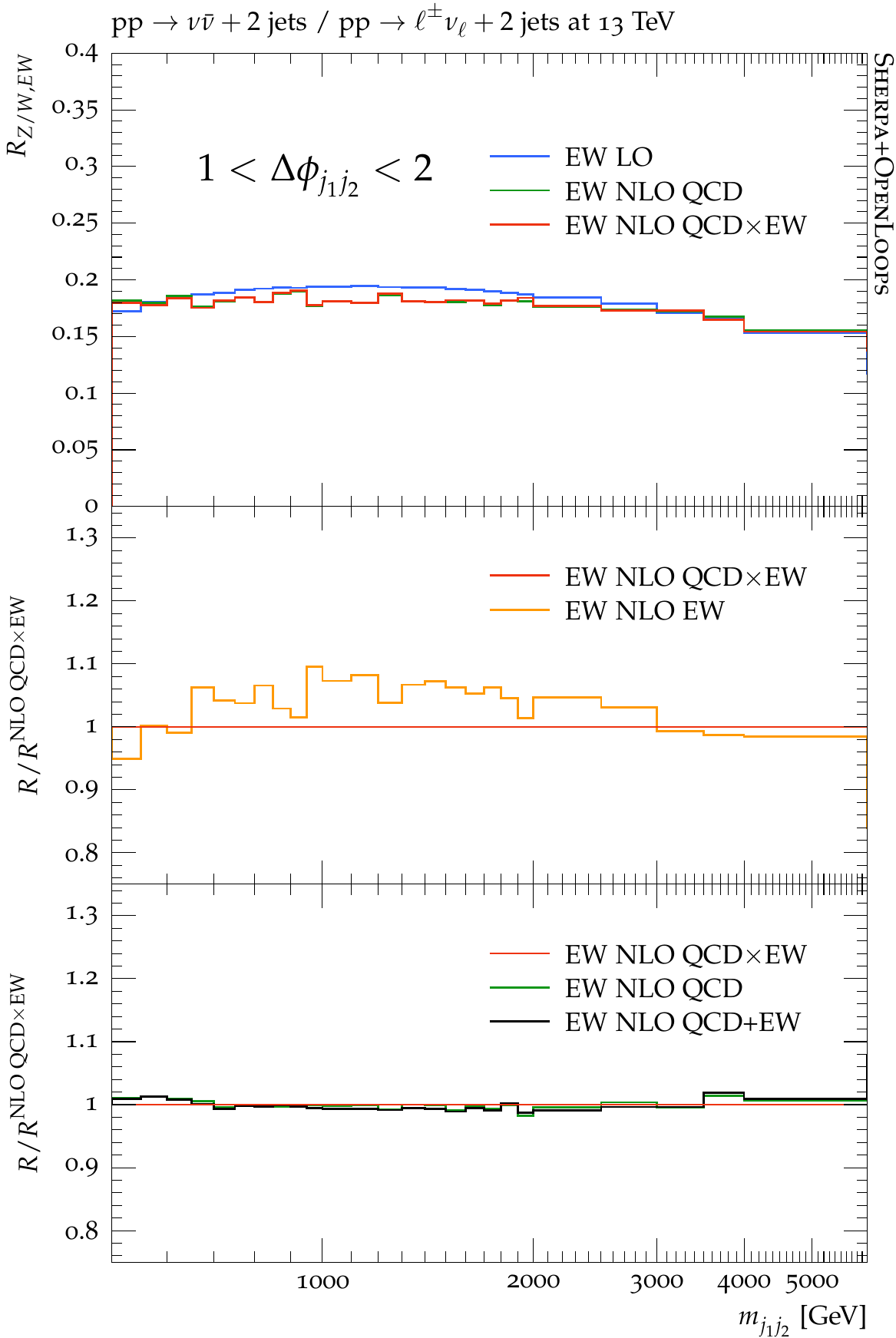}
	\includegraphics[width=0.33\textwidth]{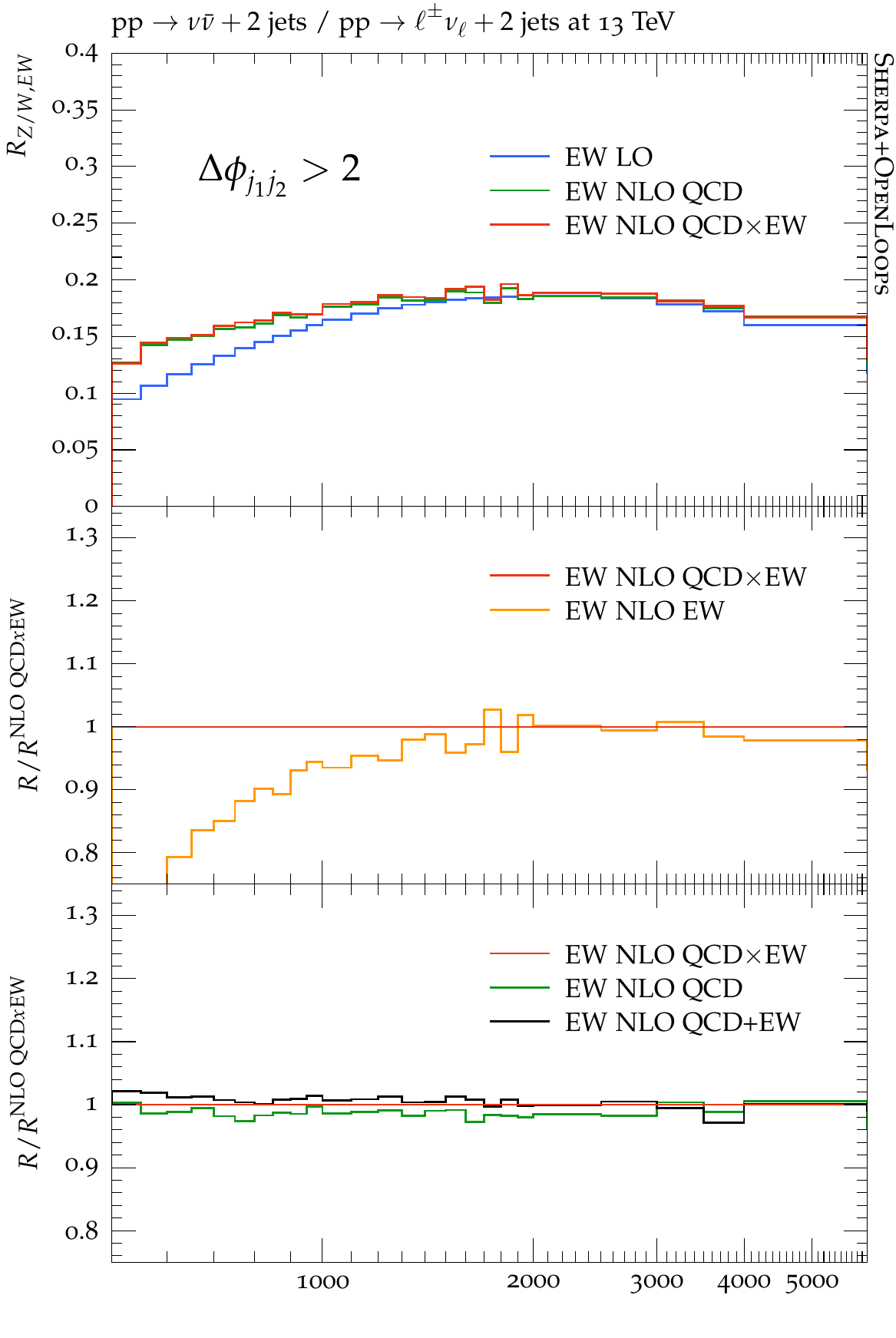} 
\caption{%
Ratios of the EW \ppZnnjj and EW \ppWenjj 
distributions in \mjj without jet veto 
in the regions
$\deltaphijj < 1$ (left), 
$1 < \deltaphijj < 2$ (middle), 
and $\deltaphijj > 2$ (right). 
Same higher-order predictions and conventions as in
Fig.~\ref{fig:ratio_QCD_mjj}, but without 
matching to the parton shower. 
} 
\label{fig:ratio_EW_dphijjbinning}
\end{figure}

In order to account for this \deltaphijj dependence 
in the reweighting of the one-dimensional \mjj distribution
we split the phase space into the three 
\deltaphijj bins defined in Eq.~\eqref{eq:moddphi}. 
The ratios of \mjj distributions for 
EW $\Zjj$ and EW $\Wjj$ production in these three 
\deltaphijj bins are shown in Fig.~\ref{fig:ratio_EW_dphijjbinning}. 
For $\mjj > 2$\,TeV, in all three \deltaphijj-bins we observe very small QCD corrections 
at the one-percent level, consistently with the behaviour of the inclusive \mjj distribution
in~Fig.~\ref{fig:ratio_EW_mjj}. This
is due both to the moderate size of the 
QCD corrections to the individual EW $\Zjj$ and $\Wjj$ cross sections (see
Fig.~\ref{fig:nom_EW_mjj}) 
and to their strong correlation.
%
%
In constrast, for $500\,\GeV < mjj < 2$\,TeV 
the size of the QCD corrections and their dependence on 
$\deltaphijj$ are quite significant.
With decreasing \mjj the impact of the QCD corrections can grow up the level of 
$+10\%$ or $-20\%$, depending on $\deltaphijj$.
Also the nominal NLO\,\QCDtEW ratio features a non-negligible dependence
on $\deltaphijj$. In order to account for the uncertainties 
associated with this nontrivial 
\mjj and \deltaphijj dependence,
the high-order QCD uncertainty
for the inclusive \mjj distribution, defined
in Eq.~\eqref{eq:QCDuncert}, is complemented by the 
additional uncertainty of Eq.~\eqref{eq:Dphiunc},
which accounts for the variation of the nomimal
ratio in the different \deltaphijj bins.
%
%


\section{Conclusions}
\label{sec:conclusions}

The precise control of SM backgrounds is key in order to harness the full
potential of invisible-Higgs searches in the VBF production mode at the LHC. 
Irreducible background contributions to the corresponding signature of 
missing transverse energy 
plus two jets with high invariant mass arise from the SM processes \ppZnnjj and
\ppWenjj, where the lepton is outside of the acceptance region.  
Such backgrounds can be predicted with rather good theoretical accuracy 
in perturbation theory, while the residual theoretical 
uncertainties can be further reduced with a data-driven approach.
In particular, the irreducible  \ppZnnjj background can be
constrained by means of accurate data for 
\ppWenjj with a visible lepton,
in combination with precise theoretical predictions 
for the correlation between
$\Zjj$ and $\Wjj$ production.

In this article we have presented parton-level predictions including
complete NLO 
QCD and EW corrections for all relevant \Vjj processes in the SM.  
These reactions involve 
various perturbative
contributions, which can be split into 
QCD modes, EW modes, and interference
contributions.  
For the
first time we have consistently computed all four perturbative contributions
that arise at NLO\,QCD+EW without applying any approximations.  Based on the
observation that the LO interference between the QCD and EW modes is
very small, the NLO contributions of  
$\ord(\alphaS^3\alpha^2)$ and $\ord(\alphaS^2\alpha^3)$
can be regarded as QCD and EW
corrections to the QCD production mode, 
while $\ord(\alphaS\alpha^4)$ and $\ord(\alpha^5)$
correspond to QCD and EW corrections to the EW
production mode. 
In the signal region for invisible-Higgs searches,
i.e.~at large dijet invariant mass, 
\mjj, the EW $\Vjj$ production mode is dominated by
VBF topologies, but our calculations account for all 
possible $\Vjj$ topologies, including 
contributions that correspond to 
diboson production with semi-leptonic decays,
as well as single-top production and decay 
in the $s$-, $t$- and $Wt$-channels.

The  QCD corrections to the EW modes are small at large \mjj, while the EW
corrections can reach up to $-20\%$.  Both for the QCD and the EW modes, we
have found a very high degree of correlation between the higher-order QCD
and EW corrections to \ppZnnjj and \ppWenjj. 
As a result of this strong correlation, higher-order corrections
and uncertainties cancel to a
large extent in the ratio of \ppZnnjj and \ppWenjj cross sections. 
Based on this observation we have proposed to exploit precise theoretical
predictions for this $Z/W$ ratio in combination with data in order to
control the $\Vjj$ backgrounds to invisible-Higgs searches 
with few-percent precision.
To this end we have provided an explicit recipe, based on 
the reweighting of \mjj distributions, which can be applied 
to the Monte Carlo samples that are used in the 
experimental analyses.
This reweighting is implemented at the level of 
the QCD and EW $Z/W$ ratios, such as to exploit the very small 
theoretical uncertainties in these observables.

In the phase space relevant for invisible-Higgs searches, at $\mjj>1$\,TeV,
the correlation of higher-order corrections in $\Zjj$ and $\Wjj$ production
turns out to be particular strong, and theoretical uncertainties 
in the $Z/W$ ratios are as small as a few percent.  
Moderate decorrelation effects have
been observed at smaller \mjj in the ratio of the EW production
modes. Such effects can reach up to $10\%$ in the ratio.
They are driven by non-universal QCD corrections
to the EW $\Vjj$ production modes, and they originate from 
semileptonic diboson topologies and single-top contributions 
that are not included in the naive VBF approximation.
The $Z/W$ correlation can in principle be further be enhanced separating
these non-universal contributions.  We leave this to future investigation.

Based on the predictions and uncertainties derived in this article
significant sensitivity improvements can be expected in searches for
invisible Higgs decays.  In fact, our predictions and the proposed reweighting procedure
have already  been applied in a recent ATLAS
search~\cite{ATLAS:2022yvh}
yielding an upper limit of $14.5\%$ on the
invisible branching ratio of the Higgs at $95\%$ confidence level. The approach and the theoretical 
predictions presented in this paper 
can also be applied to measurments of $\Vjj$ production via VFB
in order to derive constraints on effective field theories beyond the Standard Model.


\section*{Acknowledgments}
We thank Lorenzo Mai for useful discussions and cross-checks.
We also thank Christian G\"utschow for useful discussions and comments on the manuscript.
J.M.L.\ is supported by the Science and Technology Research Council (STFC) under the Consolidated Grant ST/T00102X/1 and the STFC Ernest Rutherford Fellowship ST/S005048/1.
M.S.\ is supported by the Royal Society through a University Research Fellowship
(URF\textbackslash{}R1\textbackslash{}180549) and an Enhancement Award 
(RGF\textbackslash{}EA\textbackslash{}181033,
 CEC19\textbackslash{}100349, and RF\textbackslash{}ERE\textbackslash{}210397).
The work of S.P.~was supported by the Swiss National Science Foundation (SNSF) 
under contract BSCGI0-157722.
We acknowledge the use of the DiRAC Cumulus HPC facility under Grant No. PPSP226.

\bibliographystyle{JHEP}

\bibliography{vjj}

\providecommand{\href}[2]{#2}\begingroup\raggedright\begin{thebibliography}{10}

\bibitem{Denner:2019fcr}
A.~Denner, S.~Dittmaier, and A.~M\"uck, {\it {PROPHECY4F 3.0: A Monte Carlo
  program for Higgs-boson decays into four-fermion final states in and beyond
  the Standard Model}},  {\em Comput. Phys. Commun.} {\bf 254} (2020) 107336,
  [\href{http://arxiv.org/abs/1912.02010}{{\tt arXiv:1912.02010}}].

\bibitem{Shrock:1982kd}
R.~E. Shrock and M.~Suzuki, {\it {Invisible Decays of Higgs Bosons}},  {\em
  Phys. Lett. B} {\bf 110} (1982) 250.

\bibitem{Choudhury:1993hv}
D.~Choudhury and D.~P. Roy, {\it {Signatures of an invisibly decaying Higgs
  particle at LHC}},  {\em Phys. Lett. B} {\bf 322} (1994) 368--373,
  [\href{http://arxiv.org/abs/hep-ph/9312347}{{\tt hep-ph/9312347}}].

\bibitem{Dominici:2009pq}
D.~Dominici and J.~F. Gunion, {\it {Invisible Higgs Decays from Higgs
  Graviscalar Mixing}},  {\em Phys. Rev. D} {\bf 80} (2009) 115006,
  [\href{http://arxiv.org/abs/0902.1512}{{\tt arXiv:0902.1512}}].

\bibitem{Belanger:2001am}
G.~Belanger, F.~Boudjema, A.~Cottrant, R.~M. Godbole, and A.~Semenov, {\it {The
  MSSM invisible Higgs in the light of dark matter and g-2}},  {\em Phys. Lett.
  B} {\bf 519} (2001) 93--102, [\href{http://arxiv.org/abs/hep-ph/0106275}{{\tt
  hep-ph/0106275}}].

\bibitem{Djouadi:2011aa}
A.~Djouadi, O.~Lebedev, Y.~Mambrini, and J.~Quevillon, {\it {Implications of
  LHC searches for Higgs--portal dark matter}},  {\em Phys. Lett. B} {\bf 709}
  (2012) 65--69, [\href{http://arxiv.org/abs/1112.3299}{{\tt
  arXiv:1112.3299}}].

\bibitem{Baek:2012se}
S.~Baek, P.~Ko, W.-I. Park, and E.~Senaha, {\it {Higgs Portal Vector Dark
  Matter : Revisited}},  {\em JHEP} {\bf 05} (2013) 036,
  [\href{http://arxiv.org/abs/1212.2131}{{\tt arXiv:1212.2131}}].

\bibitem{Calibbi:2013poa}
L.~Calibbi, J.~M. Lindert, T.~Ota, and Y.~Takanishi, {\it {Cornering light
  Neutralino Dark Matter at the LHC}},  {\em JHEP} {\bf 10} (2013) 132,
  [\href{http://arxiv.org/abs/1307.4119}{{\tt arXiv:1307.4119}}].

\bibitem{Beniwal:2015sdl}
A.~Beniwal, F.~Rajec, C.~Savage, P.~Scott, C.~Weniger, M.~White, and A.~G.
  Williams, {\it {Combined analysis of effective Higgs portal dark matter
  models}},  {\em Phys. Rev. D} {\bf 93} (2016), no.~11 115016,
  [\href{http://arxiv.org/abs/1512.06458}{{\tt arXiv:1512.06458}}].

\bibitem{Butter:2015fqa}
A.~Butter, T.~Plehn, M.~Rauch, D.~Zerwas, S.~Henrot-Versill\'e, and R.~Lafaye,
  {\it {Invisible Higgs Decays to Hooperons in the NMSSM}},  {\em Phys. Rev. D}
  {\bf 93} (2016) 015011, [\href{http://arxiv.org/abs/1507.02288}{{\tt
  arXiv:1507.02288}}].

\bibitem{Argyropoulos:2021sav}
S.~Argyropoulos, O.~Brandt, and U.~Haisch, {\it {Collider Searches for Dark
  Matter through the Higgs Lens}},  \href{http://arxiv.org/abs/2109.13597}{{\tt
  arXiv:2109.13597}}.

\bibitem{ATLAS:2017nyv}
{\bf ATLAS} Collaboration, M.~Aaboud et~al., {\it {Search for an invisibly
  decaying Higgs boson or dark matter candidates produced in association with a
  $Z$ boson in $pp$ collisions at $\sqrt{s} =$ 13 TeV with the ATLAS
  detector}},  {\em Phys. Lett. B} {\bf 776} (2018) 318--337,
  [\href{http://arxiv.org/abs/1708.09624}{{\tt arXiv:1708.09624}}].

\bibitem{ATLAS:2018bnv}
{\bf ATLAS} Collaboration, M.~Aaboud et~al., {\it {Search for invisible Higgs
  boson decays in vector boson fusion at $\sqrt{s} = 13$ TeV with the ATLAS
  detector}},  {\em Phys. Lett. B} {\bf 793} (2019) 499--519,
  [\href{http://arxiv.org/abs/1809.06682}{{\tt arXiv:1809.06682}}].

\bibitem{ATLAS:2018nda}
{\bf ATLAS} Collaboration, M.~Aaboud et~al., {\it {Search for dark matter in
  events with a hadronically decaying vector boson and missing transverse
  momentum in $pp$ collisions at $\sqrt{s} = 13$ TeV with the ATLAS detector}},
   {\em JHEP} {\bf 10} (2018) 180, [\href{http://arxiv.org/abs/1807.11471}{{\tt
  arXiv:1807.11471}}].

\bibitem{ATLAS:2019cid}
{\bf ATLAS} Collaboration, M.~Aaboud et~al., {\it {Combination of searches for
  invisible Higgs boson decays with the ATLAS experiment}},  {\em Phys. Rev.
  Lett.} {\bf 122} (2019), no.~23 231801,
  [\href{http://arxiv.org/abs/1904.05105}{{\tt arXiv:1904.05105}}].

\bibitem{CMS:2014gab}
{\bf CMS} Collaboration, S.~Chatrchyan et~al., {\it {Search for invisible
  decays of Higgs bosons in the vector boson fusion and associated ZH
  production modes}},  {\em Eur. Phys. J. C} {\bf 74} (2014) 2980,
  [\href{http://arxiv.org/abs/1404.1344}{{\tt arXiv:1404.1344}}].

\bibitem{CMS:2016dhk}
{\bf CMS} Collaboration, V.~Khachatryan et~al., {\it {Searches for invisible
  decays of the Higgs boson in pp collisions at $\sqrt{s}$ = 7, 8, and 13
  TeV}},  {\em JHEP} {\bf 02} (2017) 135,
  [\href{http://arxiv.org/abs/1610.09218}{{\tt arXiv:1610.09218}}].

\bibitem{CMS:2018yfx}
{\bf CMS} Collaboration, A.~M. Sirunyan et~al., {\it {Search for invisible
  decays of a Higgs boson produced through vector boson fusion in proton-proton
  collisions at $\sqrt{s} =$ 13 TeV}},  {\em Phys. Lett. B} {\bf 793} (2019)
  520--551, [\href{http://arxiv.org/abs/1809.05937}{{\tt arXiv:1809.05937}}].

\bibitem{Lindert:2017olm}
J.~M. Lindert et~al., {\it {Precise predictions for $V+$ jets dark matter
  backgrounds}},  {\em Eur. Phys. J. C} {\bf 77} (2017), no.~12 829,
  [\href{http://arxiv.org/abs/1705.04664}{{\tt arXiv:1705.04664}}].

\bibitem{ATLAS:2021kxv}
{\bf ATLAS} Collaboration, G.~Aad et~al., {\it {Search for new phenomena in
  events with an energetic jet and missing transverse momentum in $pp$
  collisions at $\sqrt {s}$ =13 TeV with the ATLAS detector}},  {\em Phys. Rev.
  D} {\bf 103} (2021), no.~11 112006,
  [\href{http://arxiv.org/abs/2102.10874}{{\tt arXiv:2102.10874}}].

\bibitem{CMS:2021far}
{\bf CMS} Collaboration, A.~Tumasyan et~al., {\it {Search for new particles in
  events with energetic jets and large missing transverse momentum in
  proton-proton collisions at $ \sqrt{s} $ = 13 TeV}},  {\em JHEP} {\bf 11}
  (2021) 153, [\href{http://arxiv.org/abs/2107.13021}{{\tt arXiv:2107.13021}}].

\bibitem{ATLAS:2014sjq}
{\bf ATLAS} Collaboration, G.~Aad et~al., {\it {Measurement of the electroweak
  production of dijets in association with a Z-boson and distributions
  sensitive to vector boson fusion in proton-proton collisions at $\sqrt{s} =$
  8 TeV using the ATLAS detector}},  {\em JHEP} {\bf 04} (2014) 031,
  [\href{http://arxiv.org/abs/1401.7610}{{\tt arXiv:1401.7610}}].

\bibitem{CMS:2016myt}
{\bf CMS} Collaboration, V.~Khachatryan et~al., {\it {Measurement of
  electroweak production of a W boson and two forward jets in proton-proton
  collisions at $ \sqrt{s}=8 $ TeV}},  {\em JHEP} {\bf 11} (2016) 147,
  [\href{http://arxiv.org/abs/1607.06975}{{\tt arXiv:1607.06975}}].

\bibitem{ATLAS:2017nei}
{\bf ATLAS} Collaboration, M.~Aaboud et~al., {\it {Measurement of the
  cross-section for electroweak production of dijets in association with a Z
  boson in pp collisions at $\sqrt {s}$ = 13 TeV with the ATLAS detector}},
  {\em Phys. Lett. B} {\bf 775} (2017) 206--228,
  [\href{http://arxiv.org/abs/1709.10264}{{\tt arXiv:1709.10264}}].

\bibitem{ATLAS:2017luz}
{\bf ATLAS} Collaboration, M.~Aaboud et~al., {\it {Measurements of electroweak
  $Wjj$ production and constraints on anomalous gauge couplings with the ATLAS
  detector}},  {\em Eur. Phys. J. C} {\bf 77} (2017), no.~7 474,
  [\href{http://arxiv.org/abs/1703.04362}{{\tt arXiv:1703.04362}}].

\bibitem{CMS:2017dmo}
{\bf CMS} Collaboration, A.~M. Sirunyan et~al., {\it {Electroweak production of
  two jets in association with a Z boson in proton\textendash{}proton
  collisions at $\sqrt{s}= $ 13 $\,\text {TeV}$}},  {\em Eur. Phys. J. C} {\bf
  78} (2018), no.~7 589, [\href{http://arxiv.org/abs/1712.09814}{{\tt
  arXiv:1712.09814}}].

\bibitem{Campbell:2002tg}
J.~M. Campbell and R.~K. Ellis, {\it {Next-to-Leading Order Corrections to
  $W^+$ 2 jet and $Z^+$ 2 Jet Production at Hadron Colliders}},  {\em Phys.
  Rev. D} {\bf 65} (2002) 113007,
  [\href{http://arxiv.org/abs/hep-ph/0202176}{{\tt hep-ph/0202176}}].

\bibitem{FebresCordero:2006nvf}
F.~Febres~Cordero, L.~Reina, and D.~Wackeroth, {\it {NLO QCD corrections to W
  boson production with a massive b-quark jet pair at the Tevatron p anti-p
  collider}},  {\em Phys. Rev. D} {\bf 74} (2006) 034007,
  [\href{http://arxiv.org/abs/hep-ph/0606102}{{\tt hep-ph/0606102}}].

\bibitem{Campbell:2008hh}
J.~M. Campbell, R.~K. Ellis, F.~Febres~Cordero, F.~Maltoni, L.~Reina,
  D.~Wackeroth, and S.~Willenbrock, {\it {Associated Production of a $W$ Boson
  and One $b$ Jet}},  {\em Phys. Rev. D} {\bf 79} (2009) 034023,
  [\href{http://arxiv.org/abs/0809.3003}{{\tt arXiv:0809.3003}}].

\bibitem{Berger:2009zg}
C.~F. Berger, Z.~Bern, L.~J. Dixon, F.~Febres~Cordero, D.~Forde, T.~Gleisberg,
  H.~Ita, D.~A. Kosower, and D.~Maitre, {\it {Precise Predictions for $W$ + 3
  Jet Production at Hadron Colliders}},  {\em Phys. Rev. Lett.} {\bf 102}
  (2009) 222001, [\href{http://arxiv.org/abs/0902.2760}{{\tt
  arXiv:0902.2760}}].

\bibitem{Ellis:2009zw}
R.~K. Ellis, K.~Melnikov, and G.~Zanderighi, {\it {Generalized unitarity at
  work: first NLO QCD results for hadronic $W^+$ 3jet production}},  {\em JHEP}
  {\bf 04} (2009) 077, [\href{http://arxiv.org/abs/0901.4101}{{\tt
  arXiv:0901.4101}}].

\bibitem{Ellis:2009zyy}
R.~K. Ellis, K.~Melnikov, and G.~Zanderighi, {\it {W+3 jet production at the
  Tevatron}},  {\em Phys. Rev. D} {\bf 80} (2009) 094002,
  [\href{http://arxiv.org/abs/0906.1445}{{\tt arXiv:0906.1445}}].

\bibitem{Berger:2009ep}
C.~F. Berger, Z.~Bern, L.~J. Dixon, F.~Febres~Cordero, D.~Forde, T.~Gleisberg,
  H.~Ita, D.~A. Kosower, and D.~Maitre, {\it {Next-to-Leading Order QCD
  Predictions for W+3-Jet Distributions at Hadron Colliders}},  {\em Phys. Rev.
  D} {\bf 80} (2009) 074036, [\href{http://arxiv.org/abs/0907.1984}{{\tt
  arXiv:0907.1984}}].

\bibitem{Berger:2010zx}
C.~F. Berger, Z.~Bern, L.~J. Dixon, F.~Febres~Cordero, D.~Forde, T.~Gleisberg,
  H.~Ita, D.~A. Kosower, and D.~Maitre, {\it {Precise Predictions for W + 4 Jet
  Production at the Large Hadron Collider}},  {\em Phys. Rev. Lett.} {\bf 106}
  (2011) 092001, [\href{http://arxiv.org/abs/1009.2338}{{\tt
  arXiv:1009.2338}}].

\bibitem{Bern:2013gka}
Z.~Bern, L.~J. Dixon, F.~Febres~Cordero, S.~H\"oche, H.~Ita, D.~A. Kosower,
  D.~Ma\^\i{}tre, and K.~J. Ozeren, {\it {Next-to-Leading Order $W + 5$-Jet
  Production at the LHC}},  {\em Phys. Rev. D} {\bf 88} (2013), no.~1 014025,
  [\href{http://arxiv.org/abs/1304.1253}{{\tt arXiv:1304.1253}}].

\bibitem{Anger:2017nkq}
F.~R. Anger, F.~Febres~Cordero, S.~H\"oche, and D.~Ma\^\i{}tre, {\it {Weak
  vector boson production with many jets at the LHC $\sqrt{s}= 13$ TeV}},  {\em
  Phys. Rev. D} {\bf 97} (2018), no.~9 096010,
  [\href{http://arxiv.org/abs/1712.08621}{{\tt arXiv:1712.08621}}].

\bibitem{Badger:2021nhg}
S.~Badger, H.~B. Hartanto, and S.~Zoia, {\it {Two-Loop QCD Corrections to Wbb
  Production at Hadron Colliders}},  {\em Phys. Rev. Lett.} {\bf 127} (2021),
  no.~1 012001, [\href{http://arxiv.org/abs/2102.02516}{{\tt
  arXiv:2102.02516}}].

\bibitem{Abreu:2021asb}
S.~Abreu, F.~F. Cordero, H.~Ita, M.~Klinkert, B.~Page, and V.~Sotnikov, {\it
  {Leading-Color Two-Loop Amplitudes for Four Partons and a W Boson in QCD}},
  \href{http://arxiv.org/abs/2110.07541}{{\tt arXiv:2110.07541}}.

\bibitem{Re:2012zi}
E.~Re, {\it {NLO corrections merged with parton showers for Z+2 jets production
  using the POWHEG method}},  {\em JHEP} {\bf 10} (2012) 031,
  [\href{http://arxiv.org/abs/1204.5433}{{\tt arXiv:1204.5433}}].

\bibitem{Alwall:2014hca}
J.~Alwall, R.~Frederix, S.~Frixione, V.~Hirschi, F.~Maltoni, O.~Mattelaer,
  H.~S. Shao, T.~Stelzer, P.~Torrielli, and M.~Zaro, {\it {The automated
  computation of tree-level and next-to-leading order differential cross
  sections, and their matching to parton shower simulations}},  {\em JHEP} {\bf
  07} (2014) 079, [\href{http://arxiv.org/abs/1405.0301}{{\tt
  arXiv:1405.0301}}].

\bibitem{Sherpa:2019gpd}
{\bf Sherpa} Collaboration, E.~Bothmann et~al., {\it {Event Generation with
  Sherpa 2.2}},  {\em SciPost Phys.} {\bf 7} (2019), no.~3 034,
  [\href{http://arxiv.org/abs/1905.09127}{{\tt arXiv:1905.09127}}].

\bibitem{Frederix:2015eii}
R.~Frederix, S.~Frixione, A.~Papaefstathiou, S.~Prestel, and P.~Torrielli, {\it
  {A study of multi-jet production in association with an electroweak vector
  boson}},  {\em JHEP} {\bf 02} (2016) 131,
  [\href{http://arxiv.org/abs/1511.00847}{{\tt arXiv:1511.00847}}].

\bibitem{Hoeche:2012yf}
S.~H{\"o}che, F.~Krauss, M.~Sch{\"o}nherr, and F.~Siegert, {\it {QCD matrix
  elements + parton showers: The NLO case}},  {\em JHEP} {\bf 04} (2013) 027,
  [\href{http://arxiv.org/abs/1207.5030}{{\tt arXiv:1207.5030}}].

\bibitem{Gehrmann:2012yg}
T.~Gehrmann, S.~H{\"o}che, F.~Krauss, M.~Sch{\"o}nherr, and F.~Siegert, {\it
  {NLO QCD matrix elements + parton showers in $e^+e^-$ ---\ensuremath{>}
  hadrons}},  {\em JHEP} {\bf 01} (2013) 144,
  [\href{http://arxiv.org/abs/1207.5031}{{\tt arXiv:1207.5031}}].

\bibitem{Lonnblad:2012ix}
L.~L\"onnblad and S.~Prestel, {\it {Merging Multi-leg NLO Matrix Elements with
  Parton Showers}},  {\em JHEP} {\bf 03} (2013) 166,
  [\href{http://arxiv.org/abs/1211.7278}{{\tt arXiv:1211.7278}}].

\bibitem{Frederix:2012ps}
R.~Frederix and S.~Frixione, {\it {Merging meets matching in MC@NLO}},  {\em
  JHEP} {\bf 12} (2012) 061, [\href{http://arxiv.org/abs/1209.6215}{{\tt
  arXiv:1209.6215}}].

\bibitem{Andersen:2012gk}
J.~R. Andersen, T.~Hapola, and J.~M. Smillie, {\it {W Plus Multiple Jets at the
  LHC with High Energy Jets}},  {\em JHEP} {\bf 09} (2012) 047,
  [\href{http://arxiv.org/abs/1206.6763}{{\tt arXiv:1206.6763}}].

\bibitem{Andersen:2016vkp}
J.~R. Andersen, J.~J. Medley, and J.~M. Smillie, {\it {$Z/\gamma$ plus multiple
  hard jets in high energy collisions}},  {\em JHEP} {\bf 05} (2016) 136,
  [\href{http://arxiv.org/abs/1603.05460}{{\tt arXiv:1603.05460}}].

\bibitem{Andersen:2020yax}
J.~R. Andersen, J.~A. Black, H.~M. Brooks, E.~P. Byrne, A.~Maier, and J.~M.
  Smillie, {\it {Combined subleading high-energy logarithms and NLO accuracy
  for W production in association with multiple jets}},  {\em JHEP} {\bf 04}
  (2021) 105, [\href{http://arxiv.org/abs/2012.10310}{{\tt arXiv:2012.10310}}].

\bibitem{Denner:2014ina}
A.~Denner, L.~Hofer, A.~Scharf, and S.~Uccirati, {\it {Electroweak corrections
  to lepton pair production in association with two hard jets at the LHC}},
  {\em JHEP} {\bf 01} (2015) 094, [\href{http://arxiv.org/abs/1411.0916}{{\tt
  arXiv:1411.0916}}].

\bibitem{Kallweit:2014xda}
S.~Kallweit, J.~M. Lindert, P.~Maierh{\"o}fer, S.~Pozzorini, and
  M.~Sch{\"o}nherr, {\it {NLO electroweak automation and precise predictions
  for W+multijet production at the LHC}},  {\em JHEP} {\bf 04} (2015) 012,
  [\href{http://arxiv.org/abs/1412.5157}{{\tt arXiv:1412.5157}}].

\bibitem{Kallweit:2015dum}
S.~Kallweit, J.~M. Lindert, P.~Maierh{\"o}fer, S.~Pozzorini, and
  M.~Sch{\"o}nherr, {\it {NLO QCD+EW predictions for V + jets including
  off-shell vector-boson decays and multijet merging}},  {\em JHEP} {\bf 04}
  (2016) 021, [\href{http://arxiv.org/abs/1511.08692}{{\tt arXiv:1511.08692}}].

\bibitem{Chiesa:2015mya}
M.~Chiesa, N.~Greiner, and F.~Tramontano, {\it {Automation of electroweak
  corrections for LHC processes}},  {\em J. Phys. G} {\bf 43} (2016), no.~1
  013002, [\href{http://arxiv.org/abs/1507.08579}{{\tt arXiv:1507.08579}}].

\bibitem{Oleari:2003tc}
C.~Oleari and D.~Zeppenfeld, {\it {QCD corrections to electroweak nu(l) j j and
  l+ l- j j production}},  {\em Phys. Rev. D} {\bf 69} (2004) 093004,
  [\href{http://arxiv.org/abs/hep-ph/0310156}{{\tt hep-ph/0310156}}].

\bibitem{Jager:2010aj}
B.~Jager, {\it {Next-to-leading order QCD corrections to photon production via
  weak-boson fusion}},  {\em Phys. Rev. D} {\bf 81} (2010) 114016,
  [\href{http://arxiv.org/abs/1004.0825}{{\tt arXiv:1004.0825}}].

\bibitem{Jager:2012xk}
B.~Jager, S.~Schneider, and G.~Zanderighi, {\it {Next-to-leading order QCD
  corrections to electroweak Zjj production in the POWHEG BOX}},  {\em JHEP}
  {\bf 09} (2012) 083, [\href{http://arxiv.org/abs/1207.2626}{{\tt
  arXiv:1207.2626}}].

\bibitem{Schissler:2013nga}
F.~Schissler and D.~Zeppenfeld, {\it {Parton Shower Effects on W and Z
  Production via Vector Boson Fusion at NLO QCD}},  {\em JHEP} {\bf 04} (2013)
  057, [\href{http://arxiv.org/abs/1302.2884}{{\tt arXiv:1302.2884}}].

\bibitem{Cacciari:2008gp}
M.~Cacciari, G.~P. Salam, and G.~Soyez, {\it {The anti-$k_t$ jet clustering
  algorithm}},  {\em JHEP} {\bf 04} (2008) 063,
  [\href{http://arxiv.org/abs/0802.1189}{{\tt arXiv:0802.1189}}].

\bibitem{Schonherr:2017qcj}
M.~Sch\"onherr, {\it {An automated subtraction of NLO EW infrared
  divergences}},  {\em Eur. Phys. J. C} {\bf 78} (2018), no.~2 119,
  [\href{http://arxiv.org/abs/1712.07975}{{\tt arXiv:1712.07975}}].

\bibitem{Frixione:2002ik}
S.~Frixione and B.~R. Webber, {\it {Matching NLO QCD computations and parton
  shower simulations}},  {\em JHEP} {\bf 06} (2002) 029,
  [\href{http://arxiv.org/abs/hep-ph/0204244}{{\tt hep-ph/0204244}}].

\bibitem{Hoeche:2011fd}
S.~H{\"o}che, F.~Krauss, M.~Sch{\"o}nherr, and F.~Siegert, {\it {A critical
  appraisal of NLO+PS matching methods}},  {\em JHEP} {\bf 09} (2012) 049,
  [\href{http://arxiv.org/abs/1111.1220}{{\tt arXiv:1111.1220}}].

\bibitem{Schumann:2007mg}
S.~Schumann and F.~Krauss, {\it {A Parton shower algorithm based on
  Catani-Seymour dipole factorisation}},  {\em JHEP} {\bf 03} (2008) 038,
  [\href{http://arxiv.org/abs/0709.1027}{{\tt arXiv:0709.1027}}].

\bibitem{Krauss:2001iv}
F.~Krauss, R.~Kuhn, and G.~Soff, {\it {AMEGIC++ 1.0: A Matrix element generator
  in C++}},  {\em JHEP} {\bf 02} (2002) 044,
  [\href{http://arxiv.org/abs/hep-ph/0109036}{{\tt hep-ph/0109036}}].

\bibitem{Gleisberg:2007md}
T.~Gleisberg and F.~Krauss, {\it {Automating dipole subtraction for QCD NLO
  calculations}},  {\em Eur. Phys. J. C} {\bf 53} (2008) 501--523,
  [\href{http://arxiv.org/abs/0709.2881}{{\tt arXiv:0709.2881}}].

\bibitem{Gleisberg:2008ta}
T.~Gleisberg, S.~H{\"o}che, F.~Krauss, M.~Sch{\"o}nherr, S.~Schumann,
  F.~Siegert, and J.~Winter, {\it {Event generation with SHERPA 1.1}},  {\em
  JHEP} {\bf 02} (2009) 007, [\href{http://arxiv.org/abs/0811.4622}{{\tt
  arXiv:0811.4622}}].

\bibitem{Buccioni:2019sur}
F.~Buccioni, J.-N. Lang, J.~M. Lindert, P.~Maierh\"ofer, S.~Pozzorini,
  H.~Zhang, and M.~F. Zoller, {\it {OpenLoops 2}},  {\em Eur. Phys. J. C} {\bf
  79} (2019), no.~10 866, [\href{http://arxiv.org/abs/1907.13071}{{\tt
  arXiv:1907.13071}}].

\bibitem{Cascioli:2011va}
F.~Cascioli, P.~Maierhofer, and S.~Pozzorini, {\it {Scattering Amplitudes with
  Open Loops}},  {\em Phys. Rev. Lett.} {\bf 108} (2012) 111601,
  [\href{http://arxiv.org/abs/1111.5206}{{\tt arXiv:1111.5206}}].

\bibitem{Buccioni:2017yxi}
F.~Buccioni, S.~Pozzorini, and M.~Zoller, {\it {On-the-fly reduction of open
  loops}},  {\em Eur. Phys. J. C} {\bf 78} (2018), no.~1 70,
  [\href{http://arxiv.org/abs/1710.11452}{{\tt arXiv:1710.11452}}].

\bibitem{Denner:2016kdg}
A.~Denner, S.~Dittmaier, and L.~Hofer, {\it {Collier: a fortran-based Complex
  One-Loop LIbrary in Extended Regularizations}},  {\em Comput. Phys. Commun.}
  {\bf 212} (2017) 220--238, [\href{http://arxiv.org/abs/1604.06792}{{\tt
  arXiv:1604.06792}}].

\bibitem{vanHameren:2010cp}
A.~van Hameren, {\it {OneLOop: For the evaluation of one-loop scalar
  functions}},  {\em Comput. Phys. Commun.} {\bf 182} (2011) 2427--2438,
  [\href{http://arxiv.org/abs/1007.4716}{{\tt arXiv:1007.4716}}].

\bibitem{Denner:2005fg}
A.~Denner et~al., {\it {Electroweak corrections to charged-current $e^+ e^-\to$
  4 fermion pro results}},  {\em Nucl.Phys.} {\bf B724} (2005) 247--294,
  [\href{http://arxiv.org/abs/hep-ph/0505042}{{\tt hep-ph/0505042}}].

\bibitem{Manohar:2016nzj}
A.~Manohar, P.~Nason, G.~P. Salam, and G.~Zanderighi, {\it {How bright is the
  proton? A precise determination of the photon PDF}},
  \href{http://arxiv.org/abs/1607.04266}{{\tt arXiv:1607.04266}}.

\bibitem{Hoeche:2009rj}
S.~Hoeche, F.~Krauss, S.~Schumann, and F.~Siegert, {\it {QCD matrix elements
  and truncated showers}},  {\em JHEP} {\bf 05} (2009) 053,
  [\href{http://arxiv.org/abs/0903.1219}{{\tt arXiv:0903.1219}}].

\bibitem{vjjrepo}
J.~M. Lindert, {\it {Public repository with NLO QCD+EW theoretical predictions
  and uncertainties for $V+$2\,jet ratios}},  {\em
  {\url{https://gitlab.com/Lindert/vjj.git}}}.

\bibitem{ATLAS:2022yvh}
{\bf ATLAS} Collaboration, G.~Aad et~al., {\it {Search for invisible
  Higgs-boson decays in events with vector-boson fusion signatures using 139
  $\text{fb}^{-1}$ of proton-proton data recorded by the ATLAS experiment}},
  \href{http://arxiv.org/abs/2202.07953}{{\tt arXiv:2202.07953}}.

\end{thebibliography}\endgroup

\end{document}